\documentclass[aps,twocolumn,prd,amssymb,amsmath,showpacs]{revtex4}

\usepackage[10pt]{type1ec}  
\usepackage[T1]{fontenc}
\usepackage{CJKutf8}
\usepackage[overlap, CJK]{ruby}
\usepackage{CJKulem}
\newenvironment{SChinese}{%
  \CJKfamily{gbsn}%
  \CJKtilde
  \CJKnospace}{}

\usepackage{graphicx,stackengine,xcolor}
\usepackage{multirow}
\usepackage[breaklinks,colorlinks,urlcolor=blue,citecolor=blue]{hyperref}
\usepackage{mathrsfs}
\usepackage{soul}
\usepackage{color}

\usepackage[utf8]{inputenc}
\usepackage{bm}                      
\usepackage{dcolumn}                 

\newcommand{\be}{\beta}

\newcommand{\de}{\delta}
\newcommand{\De}{\Delta}
\newcommand{\ep}{\varepsilon}

\newcommand{\la}{\lambda}
\newcommand{\La}{\Lambda}

\newcommand{\ptrans}{p_{\rm trans}}
\newcommand{\ntrans}{n_{\rm trans}}
\newcommand{\nb}{n_{\rm B}}
\newcommand{\etrans}{\varepsilon_{\rm trans}}
\newcommand{\Mtrans}{M_{\rm trans}}

\newcommand{\Msolar}{{\rm M}_{\odot}}
\newcommand{\Mchirp}{{\mathcal M}}
\newcommand{\pcent}{p_{\rm cent}}

\newcommand{\cQM}{{c^{\phantom{1}}_{\rm QM}}}

\newcommand{\cQMsq}{c^2_{\rm QM}}
\newcommand{\cNMsq}{c^2_{\rm NM}}

\newcommand{\Mmax}{M_{\rm max}}
\newcommand{\Rtyp}{R_{1.4}}

\newcommand{\beq}{\begin{equation}}
\newcommand{\eeq}{\end{equation}}
\newcommand{\ba}{\begin{array}}
\newcommand{\ea}{\end{array}}
\newcommand{\bea}{\begin{eqnarray}}
\newcommand{\eea}{\end{eqnarray}}
\newcommand{\bc}{\begin{center}}
\newcommand{\ec}{\end{center}}

\newcommand{\dsp}{\displaystyle}
\newcommand\eqn[1]{(\ref{#1})}      
\newcommand\Eqn[1]{Eq.~(\ref{#1})}  

\newcommand{\nn}{\nonumber \\}

\begin{document}

\title{Tidal deformability with sharp phase transitions in (binary) neutron stars}
\author{Sophia Han (\begin{CJK}{UTF8}{}\begin{SChinese}韩 君\end{SChinese}\end{CJK})$^{1,3,4}$} 
\author{Andrew W. Steiner$^{1,2}$} 

\affiliation{$^{1}$Department of Physics and Astronomy, University of
  Tennessee, Knoxville, TN 37996, USA}
\affiliation{$^{2}$Physics Division, Oak Ridge National Laboratory, Oak
  Ridge, TN 37831, USA}
\affiliation{$^{3}$Department of Physics and Astronomy, Ohio University, Athens, OH~45701, USA}  
\affiliation{$^{4}$Department of Physics, University of California Berkeley, Berkeley, CA~94720, USA}

\date{April 8, 2019} 

\begin{abstract}
The neutron star tidal deformability is a critical parameter which determines the pre-merger gravitational-wave signal in a neutron star merger. In this article, we show how neutron star tidal deformabilities behave in the presence of one or two sharp phase transition(s). We characterize how the tidal deformability changes when the properties of these phase transitions are modified in dense matter equation of state (EoS). Sharp phase transitions lead to the smallest possible tidal deformabilities and also induce discontinuities in the relation between tidal deformability and gravitational mass. These results are qualitatively unmodified by a modest softening of the phase transition. Finally, we test two universal relations involving the tidal deformability and show that their accuracy is limited by sharp phase transitions.
\end{abstract}

\pacs{97.60.Jd, 95.30.Cq, 26.60.-c}

\maketitle
\section{Introduction}
\label{sec:intro}

During the late stage (last few orbits) of binary neutron star inspiral, while tidal effects are the largest to measure, higher-order effects and nonlinear hydrodynamics become important that full numerical simulations or sophisticated waveform models are essential to best model the phase evolution. Flanagan and Hinderer \cite{Flanagan:2007ix,Hinderer:2007mb} pointed out that in the early part of the phase evolution, a small but clean signature is also measurable, which can be characterized by the EoS-dependent tidal deformability $\la$ and Love number $k_2$; the tidal parameter $\la$ remains the dominant source of EoS-dependent effects throughout the entire inspiral. A recent binary neutron star (BNS) merger event GW170817 detected by the LIGO-Virgo collaboration \cite{LIGO:2017qsa} placed the very first constraint on the dimensionless tidal deformability $\La$ for a $1.4\,\Msolar$ neutron star by assuming a linear expansion of $\La (m)$, however in that analysis the EoS dependence of two individual neutron stars were treated uncorrelated. Employing the quasi-universal relation $\La_{a} (\La_{s}, q)$ \cite{Yagi:2016bkt,Chatziioannou:2018vzf}, Ref.~\cite{LIGO:2018exr} reanalyzed the data and claimed improved limits e.g. on the $\La$ value for a $1.4\,\Msolar$ neutron star to be $70\leq\La_{1.4}\leq580$ (at $90\%$ confidence level) for low spin priors, but possible phase transitions were not taken into account.

It has been confirmed that assuming individual neutron stars obey the same normal nuclear matter EoS, the weighted-average tidal deformability $\tilde{\La}$ in the binary as a function of the chirp mass $\Mchirp=(m_{1}m_{2})^{3/5}/(m_1+m_2)^{1/5}$ which can be accurately determined during the inspiral \cite{LIGO:2017qsa,LIGO:2018wiz}, is relatively insensitive to the unknown mass ratio $q=m_2/m_1$~\cite{Radice:2017lry, Raithel:2018ncd}. In this paper, we investigate how the degeneracy is altered when a sharp first-order phase transition from normal nuclear matter to quark matter takes place in the interior of neutron stars, and determine the most sensitive phase transition parameter to tidal deformation in the binary.

In Sec.~\ref{sec:input} we describe properties of candidate EoSs applied in this work and their distinctive behavior in the speed of sound, followed by calculations of Love number and tidal deformability in Sec.~\ref{sec:tidal_calc}. We present in detail the formulae for the case where discontinuity in energy density (and the speed of sound) exists. Results of $k_2$ and $\La$ as a function of mass for candidate EoSs are given in Sec.~\ref{sec:results}, and we also expand discussions on the relevance to binary observables. In Sec.~\ref{sec:cmp} we compare our work with previous work. Our conclusions are summarized in Sec.~\ref{sec:con}.

Finally, exemplary plots depicting abrupt changes in tidal Love number and tidal deformability close to the onset of first-order phase transition can be found in Appendix~\ref{app:tidal_k2_pts}. Appendix~\ref{app:css} contains the setup of the generic EoS template that features possible phase transition at supra-nuclear density to quark matter.

We use units in which $\hbar=G=c=1$.

\section{physics input/theoretical framework}
\label{sec:input}

We consider three representative classes of equations of state: i) normal nuclear matter equations of state, ii) equations of state with sharp first-order transition from nuclear matter to quark matter at high density, and iii) equations of state for self-bound strange quark stars. In case ii) we also evaluate possible effects of smoothing near the phase boundary.

\subsection{Nuclear matter equations of state}
\label{sec:nucl_eos}

\begin{figure}[htb]
\includegraphics[width=\hsize]{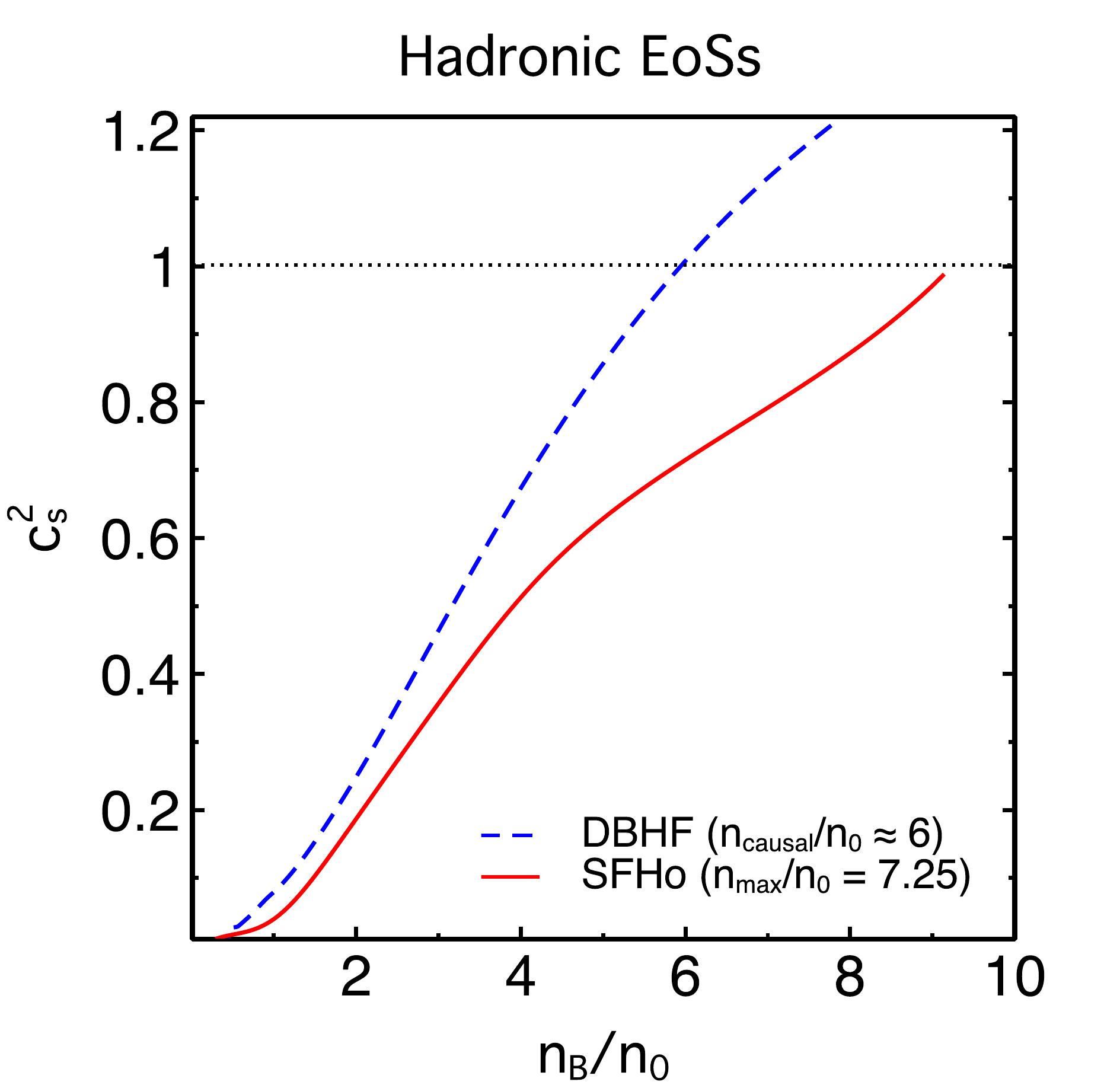}
\caption{(Color online) Speed-of-sound squared as a function of the baryon number density $\cNMsq(n_{\rm B})$ in nuclear matter for the SFHo and DBHF EoS; the DBHF EoS goes superluminal around the central density of the maximum-mass neutron star it supports.}
\label{fig:c2-dbhf-sfho}
\end{figure}

In order to show how our results depend on the nuclear matter equation of state used to describe low-density matter, we employ both the SFHo~\cite{Steiner:2012rk} and DBHF~\cite{GrossBoelting:1998jg}~EoSs. These are representative of EoSs with small or large values for the density derivative of the symmetry energy ($L$), respectively, and both of them are compatible with recent constraints inferred from the unitary Fermi gas~\cite{Kolomeitsev:2016sjl}. Some basic properties of matter at the nuclear saturation density and of neutron stars without a phase transition are summarized in Table~\ref{tab:nucl_eos}. The speed of sound, a measure of ``stiffness'', as a function of baryon number density is shown in Fig.~\ref{fig:c2-dbhf-sfho}. We assume the neutron star has a hadronic crust as described in \citet{Baym71tg} and \citet{Negele73ns}. 

\subsection{Template for hybrid EoSs with phase transitions}

\begin{figure}[htb]
\includegraphics[width=\hsize]{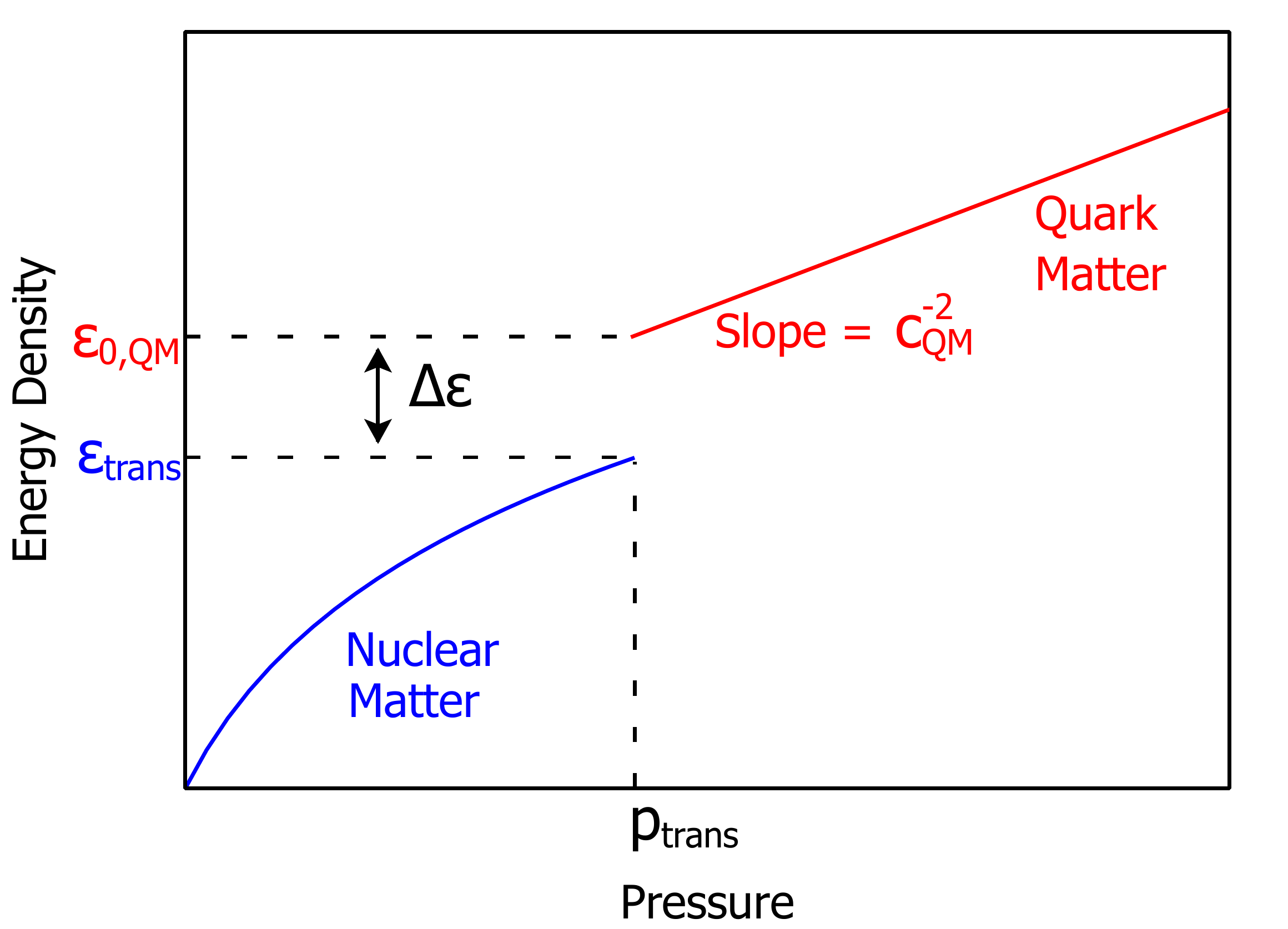}
\caption{(Color online) Equation of state $\ep(p)$ for dense matter with sharp first-order phase transition. The quark matter EoS is specified by the transition pressure $\ptrans$, the energy density discontinuity $\De\ep$, and the speed of sound in quark matter $\protect\cQM$ (assumed density-independent).}
\label{fig:eos-css}
\end{figure}

One can write the high-density EoSs with a first-order phase
transition using the generic constant-sound-speed (CSS)
parametrization~\cite{Alford:2013aca} in terms of three parameters:
the transition pressure $\ptrans$, the discontinuity in energy density
at the transition $\De\ep$, and the speed of sound in quark matter
$\cQM$ which is assumed density-independent. For a given nuclear matter
EoS, the full CSS EoS is then (schematically see
Fig.~\ref{fig:eos-css}) \beq \ep(p) = \left\{\!
\begin{array}{ll}
\ep_{\rm NM}(p) & p<\ptrans \\
\ep_{\rm NM}(\ptrans)+\De\ep+c_{\rm QM}^{-2} (p-\ptrans) & p>\ptrans
\end{array}
\right.\ 
\label{eqn:CSS_EoS}
\eeq
The CSS form can be viewed as the lowest-order terms of a Taylor
expansion of the high-density EoS about the transition pressure.
Following Ref.~\cite{Alford:2013aca}, we express the three parameters
in dimensionless form, as $\ptrans/\etrans$, $\De\ep/\etrans$ (equal
to $\la-1$ in the notation of Ref.~\cite{Haensel:1983}) and $\cQMsq$,
where $\etrans \equiv \ep_{\rm NM}(\ptrans)$.

We use the term ``hybrid star'' to refer to stars with central
pressures above the critical value $\ptrans$ where the hadronic phase
becomes energetically disfavored, and therefore contain a quark core. The part of the mass-radius relation that arises from
such stars is the hybrid branch. For a sharp hadron/quark phase transition (Maxwell construction), there are four topologies of the
mass-radius curve for compact stars: the hybrid branch may be
connected to the hadronic branch (C), or disconnected (D), or both are present (B) or absent (A); see Fig.~\ref{fig:MR-De} in Appendix~\ref{app:css} taken from Ref.~\cite{Alford:2013aca}.

\begin{table}[htb]
\begin{center}
\begin{tabular}{c | c@{\quad} c}
\hline \\[-2ex]
Property & SFHo  & DBHF \\[0.5ex]
\hline \\[-2ex]
Saturation baryon density $n_0 \rm (fm^{-3})$   & 0.16 & 0.18  \\
Binding energy/baryon $E/A$ (MeV)   &  -16.17   &  -16.15  \\
Compressibility $K_0$ (MeV)           & 245.2 &  230  \\
Symmetry energy $S_0 $ (MeV)          & 31.2  &  34.4  \\
$L = 3 n_0  \, [ dS_0/dn ]_{n_0}$ (MeV)    & 45.7  &  69.4 \\
Maximum mass of star ($\Msolar$)          & 2.06 & 2.31 \\
Radius of the heaviest star (km)  & 10.38 &11.26 \\
Radius of $M=1.4\,\Msolar$ star (km)  & 11.97 &13.41 \\
\hline
\end{tabular}
\end{center}
\caption{Calculated properties of symmetric nuclear matter for the
  SFHo and DBHF nuclear equations of state used here. SFHo is softer,
  and DBHF is stiffer (see Sec.~\ref{sec:nucl_eos}).}
\label{tab:nucl_eos}
\end{table}  

\begin{table}[htb]
\begin{center}
\begin{tabular}{c|ccc}
\hline \\[-2ex]
& \multicolumn{3}{c}{$\rm{CSS} \,(\ntrans/n_0, \De\ep/\etrans, \cQMsq)$} \\[0.5ex]
\hline \\[-2ex]
Nuclear EoS &  Set I & Set II & Set III \\[0.5ex]
\hline \\[-2ex]
SFHo  & (2.0, 0.9, 1) & (1.0, 1.8, 1) &\\
 & (2.5, 0.6, 1) & (1.0, 1.5, 1) & \\
 & (3.0, 0.3, 1) & (1.5, 1.2, 1) & \\
 & (3.5, 0.2, 1) & (1.5, 0.9, 1) & \\
 & (1.0, 3.1, 1) &(2.0, 0.6, 1) &\\[0.5ex]
 \hline \\[-2ex]
DBHF & (2.0, 1.0, 1)& (1.5, 1.3, 1) & \\
& (2.5, 0.7, 1) &(1.5, 1.0, 1) & \\
& (3.0, 0.4, 1) &(1.5, 0.7, 1) &  (2.5, 0.6, 1)\\
& (3.5, 0.4, 1) &(2.0, 0.7, 1)&(3.0, 0.6, 1)\\
 &(1.0, 3.2, 1)&(2.0, 0.4, 1) &(3.5, 0.5, 1) \\[0.5ex]
\hline
\end{tabular}
\end{center}
\caption{Categories of hybrid EoS parameters applied in this work. Set I has broad variation of onset density for phase transition $\ntrans$, set II focuses on $\ntrans$ between $1\sim2$ times nuclear saturation density, and set III is intended for better manifestation of both hybrid branches (connected and disconnected) on the mass-radius curve; see on the CSS phase diagram Fig.~\ref{fig:diag-dbhf-sfho-c2-1}.
}
\label{tab:hyb_EoS}
\end{table} 

\begin{figure*}[htb]
\parbox{0.5\hsize}{
\includegraphics[width=\hsize]{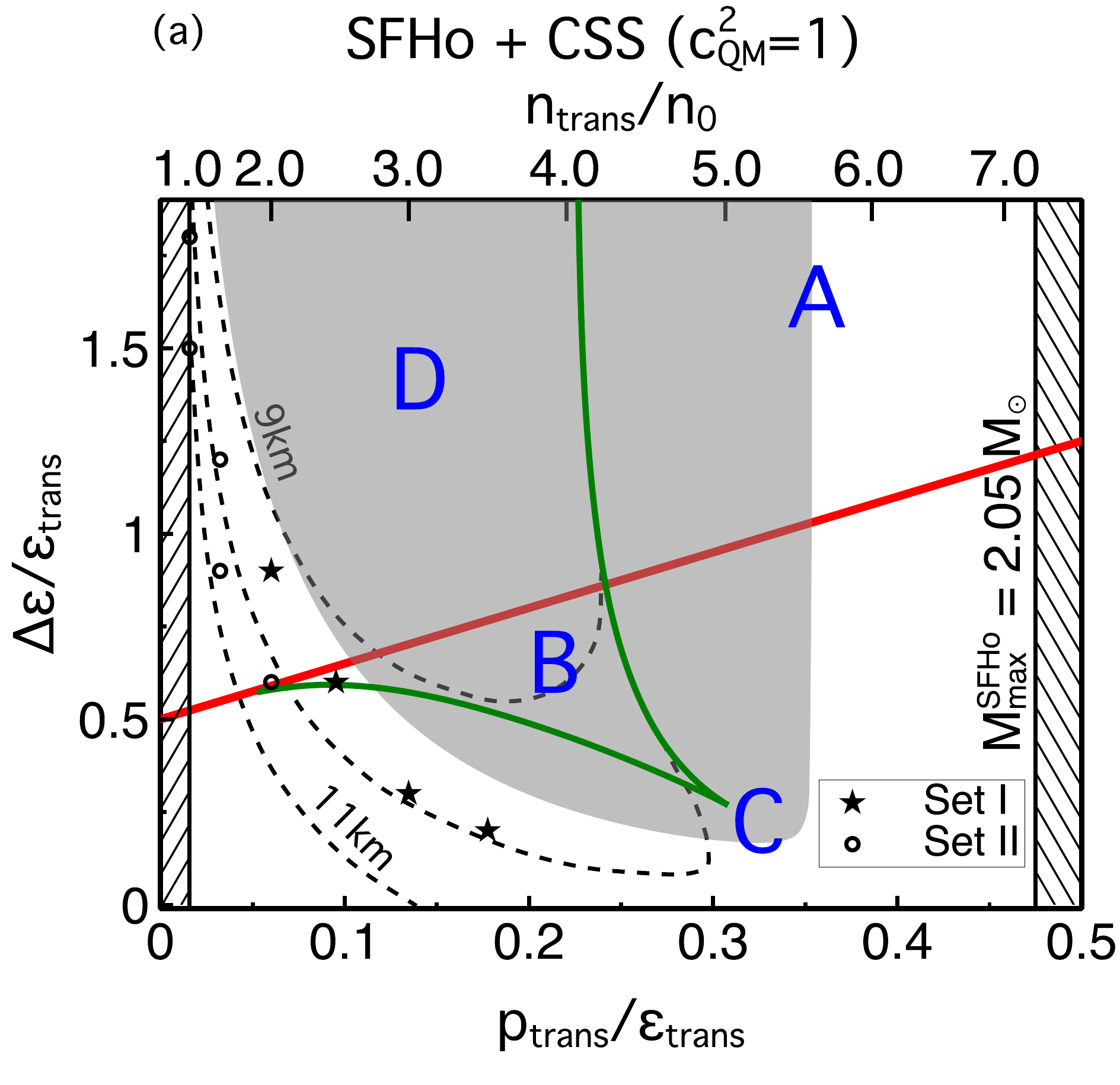}\\[-2ex]
}\parbox{0.5\hsize}{
\includegraphics[width=\hsize]{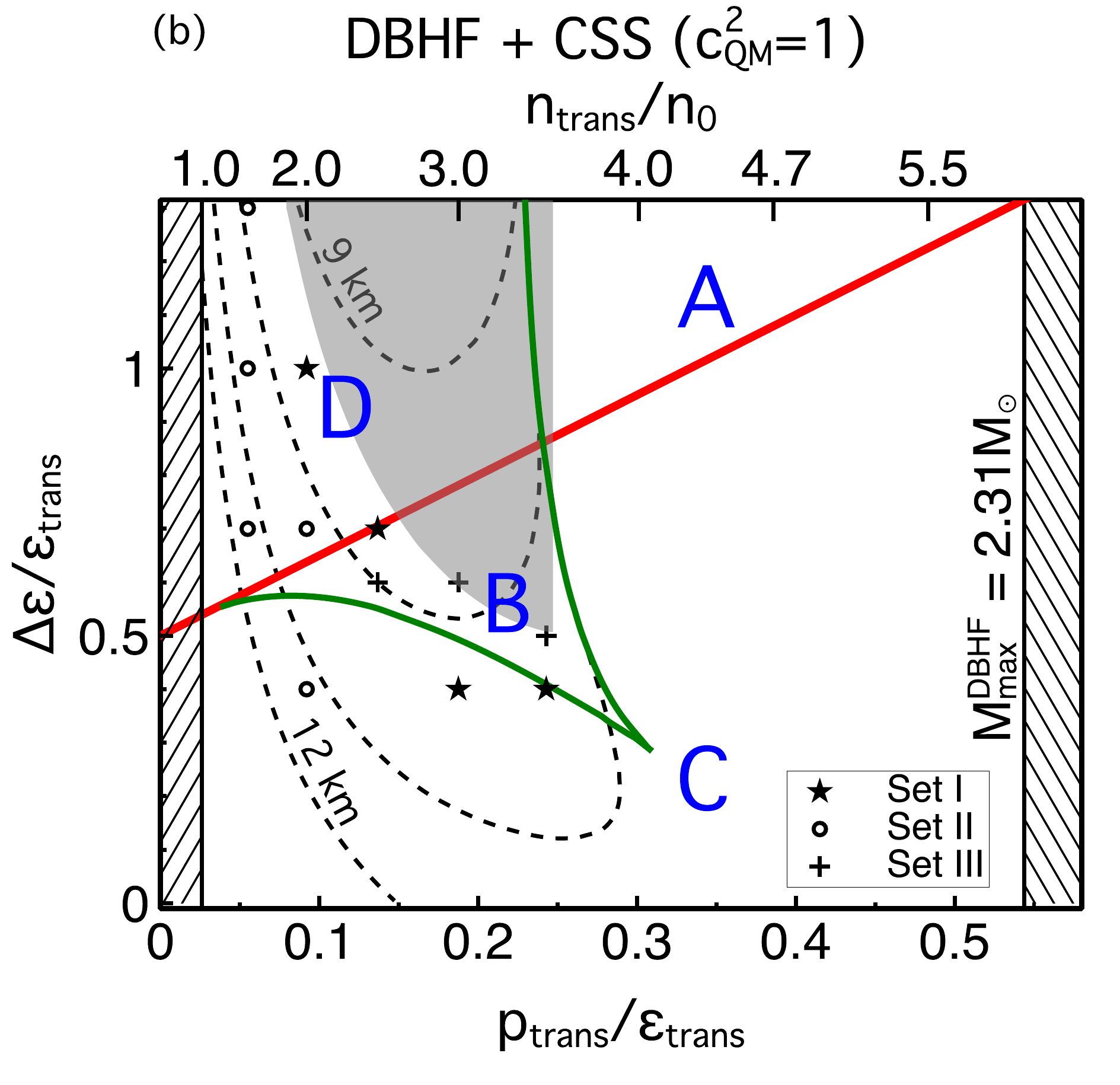}\\[-2ex]
}
\caption{(Color online) Topology of the mass-radius relation for hybrid star
  configurations in the CSS framework, with SFHo/DBHF as the hadronic
  part of the equation of state. We fix $\cQMsq$ (to be unity) and
  vary $\ptrans/\etrans$ and $\De\ep/\etrans$. The four regions are
  (A) no hybrid branch (``absent''); (B) both connected and
  disconnected hybrid branches; (C) connected hybrid branch only; and
  (D) disconnected hybrid branch only. The gray region refers to EoSs
  that are excluded by the $2\,\Msolar$ constraint; black dashed
  curves are radius contours of the maximum-mass star on the hybrid
  branch (connected or disconnected). Set I, set II and set III of
  phase transition parameters from Table~\ref{tab:hyb_EoS} are
  denoted by asterisks, open circles and line crosses. }
\label{fig:diag-dbhf-sfho-c2-1}
\end{figure*}

\begin{figure*}[htb]
\parbox{0.5\hsize}{
\includegraphics[width=\hsize]{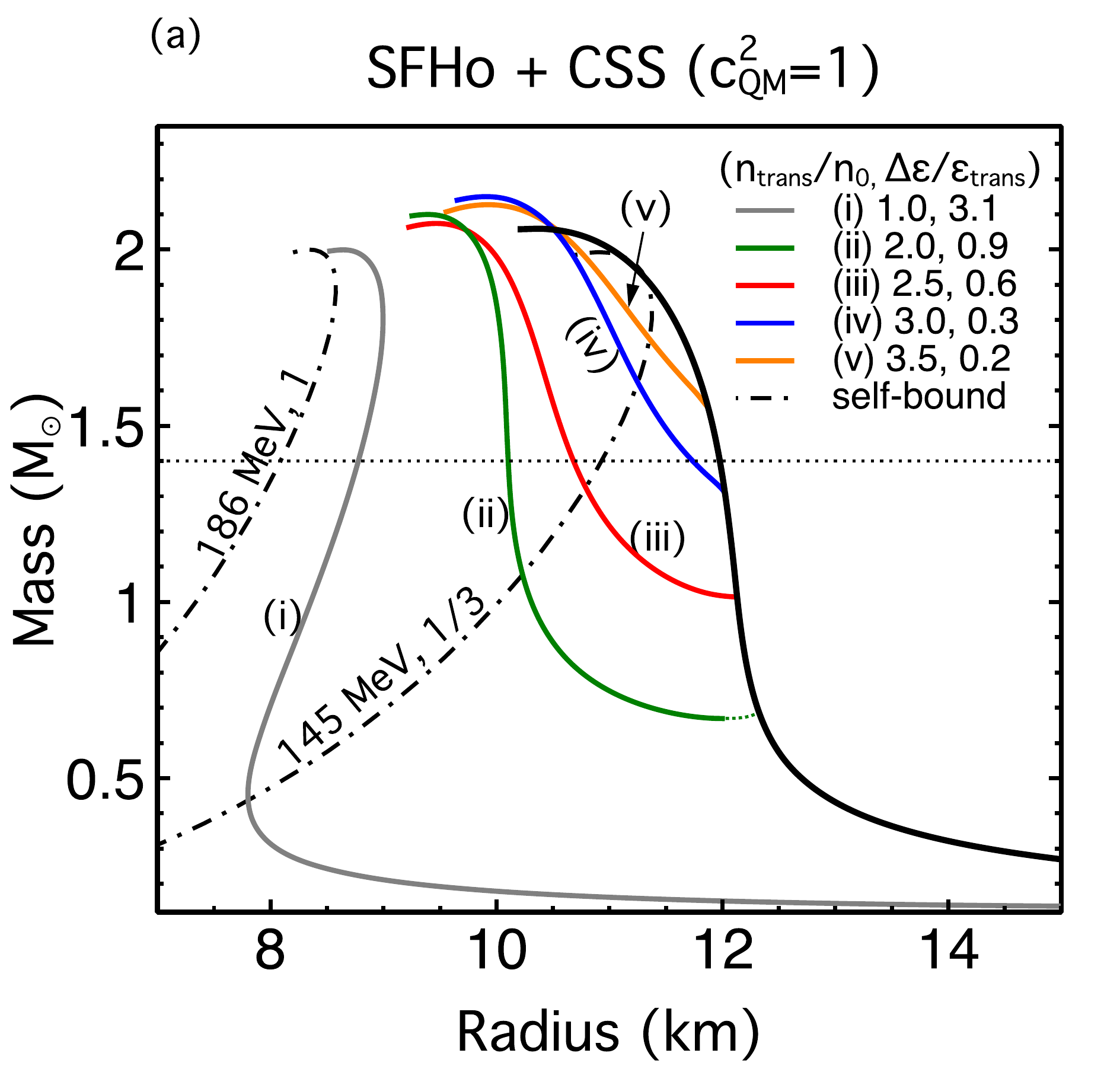}\\[-2ex]
}\parbox{0.5\hsize}{
\includegraphics[width=\hsize]{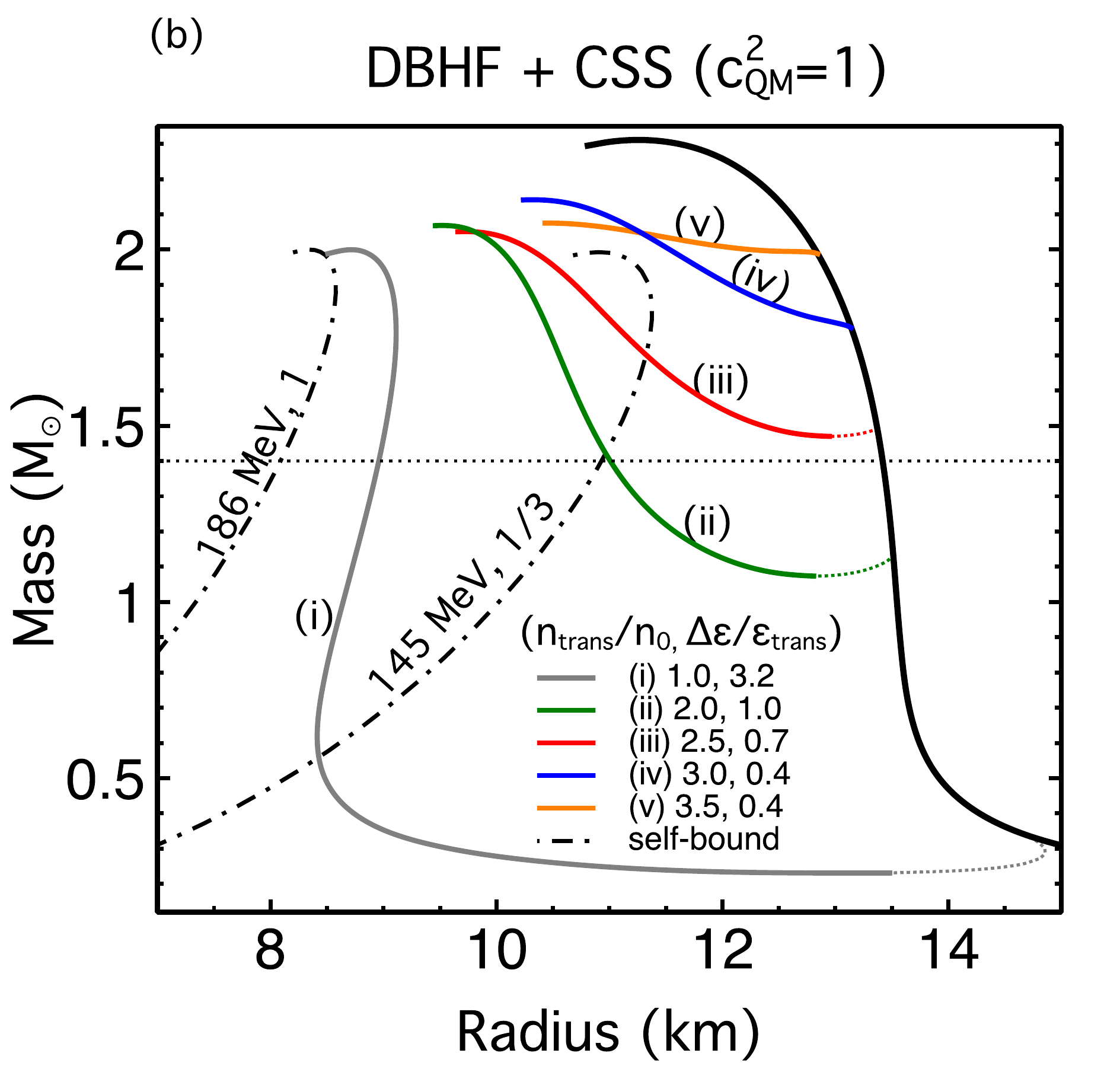}\\[-2ex]
}\\[2ex]
\parbox{0.5\hsize}{
\includegraphics[width=\hsize]{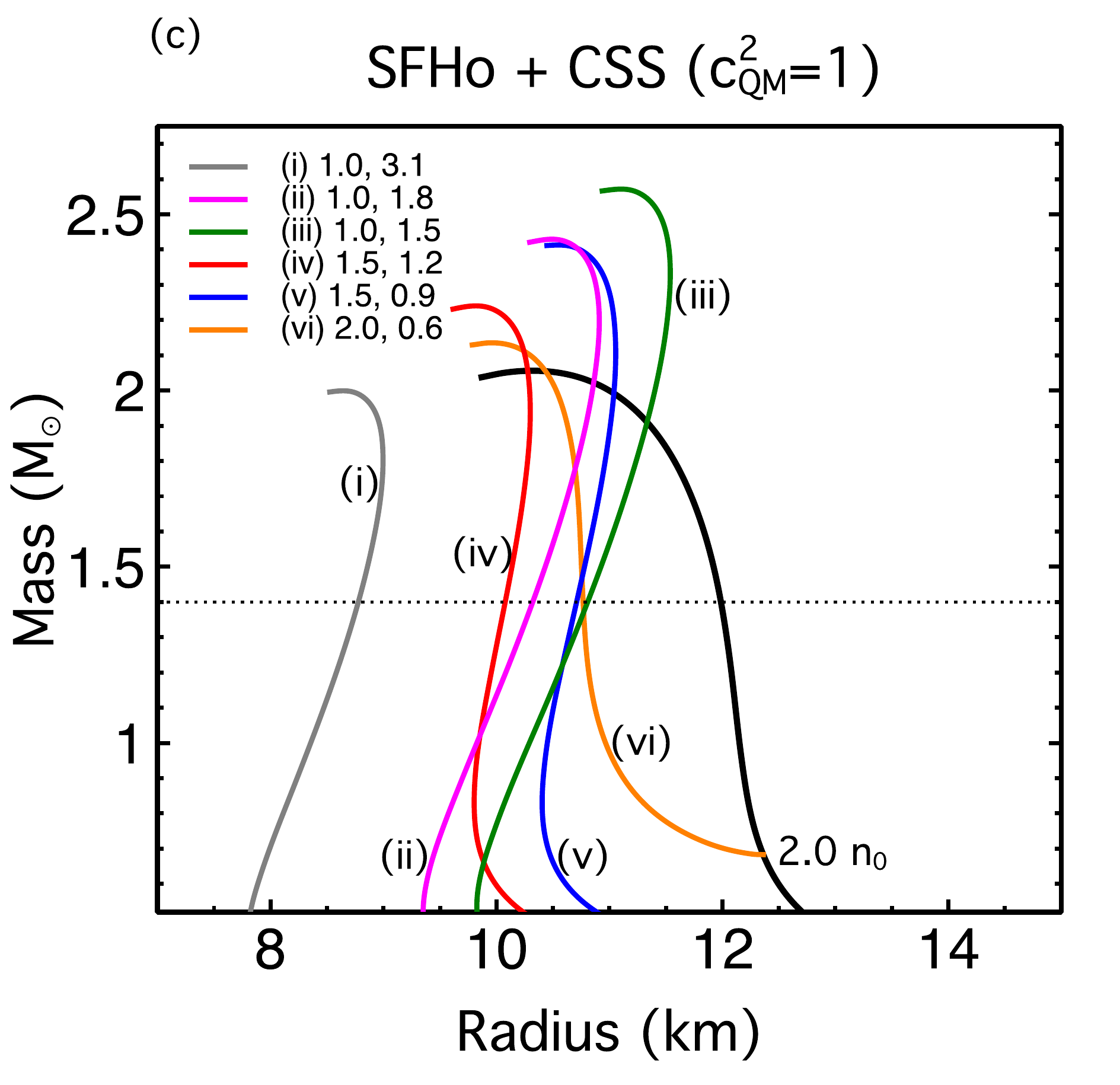}\\[-2ex]
}\parbox{0.5\hsize}{
\includegraphics[width=\hsize]{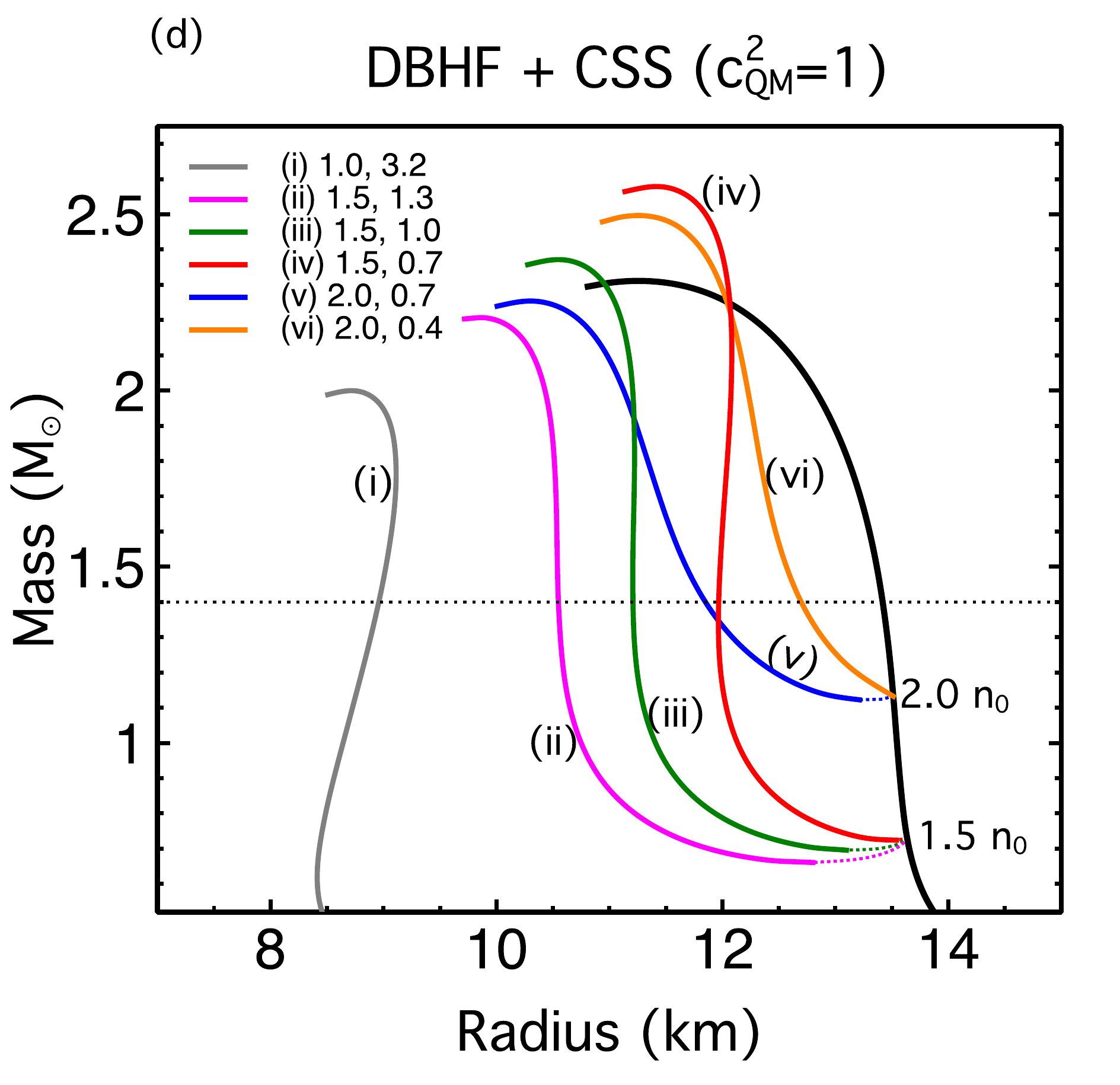}\\[-2ex]
}
\caption{(Color online) Mass-radius diagrams for SFHo/DBHF + CSS parameter sets I and II listed in Table~\ref{tab:hyb_EoS}. Black curves refer to purely-hadronic stars, and dash-dotted curves the strange quark stars specified by the linear EoS parameters $B$ and $\cQMsq$ in \Eqn{eqn:eos_bare}; $M=1.4\,\Msolar$ is indicated by the horizontal dotted line. The last EoSs of set I represented by gray curves (i) lead to the smallest radius in all hybrid configurations that are compatible with $\Mmax\geq\,2\Msolar$, which requires the phase transition to occur right above saturation density $\ntrans=n_0$ with a huge latent heat $\De\ep/\etrans\approx3$.}
\label{fig:MR-curves-1}
\end{figure*}

To survey the impact that a single sharp phase transition has on the neutron star structure, we select 10 SFHo parameterizations and 13 DBHF parameterizations, all of which fulfill the $2\,\Msolar$ maximum-mass constraint \cite{Demorest:2010bx,Antoniadis:2013pzd} from pulsar observations. The values of the CSS parameters for these 23 EoSs are given in Table~\ref{tab:hyb_EoS}. Although at asymptotically high densities all EoS should approach the QCD limit $c_s^2 \approx 1/3$ \cite{Borsanyi:2012cr,Kurkela:2014vha,Bedaque:2014sqa}, uncertainties of the speed of sound behavior at intermediate densities relevant are still large. In order to be compatible with the observational constraint $\Mmax\geq2\,\Msolar$, strong first-order phase transitions which substantially soften the transition region require a large speed of sound elsewhere in the star. Typically $\cQMsq \gtrsim 0.5$ is needed when the hadronic EoS is soft and $\cQMsq \gtrsim 0.4$ when the hadronic EoS is stiff, and for $\cQMsq \approx 1/3$ even if the hadronic EoS is quite stiff almost no \textit{detectable} hybrid configurations are present \cite{Alford:2013aca,Alford:2015gna}. In this paper, we principally assume that quark matter is maximally stiff ($c_s^2 =1$), aiming to derive a conservative limit.

Fig.~\ref{fig:diag-dbhf-sfho-c2-1} shows the selected EoSs lead to three 
topologies described above (except ``A'' where stable hybrid branch is absent), depending on the value of $\ptrans/\etrans$ (or equivalently $\ntrans/n_0$) and $\De\ep/\etrans$. Fig.~\ref{fig:MR-curves-1} shows corresponding mass-radius relationship. Small radii and sufficiently high maximum masses
are ordinarily obtained for hybrid EoS parameters in region ``D'' or
``B'', where low values of tidal response are anticipated due to its
high sensitivity to the stellar radius $R$ (see \Eqn{eqn:k2-lam}).

In our results below, we also examine the impact that two sharp phase transitions have on the neutron star structure and the tidal deformability. Such an EoS might be realized in nature if nuclear matter is replaced by two-flavor color-superconducting (2SC) quark matter at moderate densities and color-flavor-locked (CFL) quark matter at the highest densities in the core~\cite{Alford:2017qgh}. The presence of an additional phase transition introduces three new parameters, in which the analytic form reads

\beq
\ep(p) = \left\{\!
\begin{array}{ll}
\ep_{\rm NM}(p) & p<p_{1} \\
\ep_1+\De\ep_1+s_1^{-1} (p-p_1) & p_1<p<p_2 \\
\ep_2+\De\ep_2+s_2^{-1} (p-p_2) & p>p_2.
\end{array}
\right.\ 
\label{eqn:seq_EoS}
\eeq

where $p_{1}\equiv p_{\rm NM} (\nb = \nb^{\rm trans, 1})$, and $\ep_1\equiv\ep_{\rm NM}(\ptrans)$.

\subsection{Bare (crustless) strange quark matter EoS}

To study pure strange quark stars that are self-bound, we adopt a
linearized EoS (\Eqn{eqn:eos_bare}) where the pressure is zero below a
few times nuclear saturation density \beq \varepsilon\left(p\right) =
4B^4+\frac{1}{\cQMsq} p. \;
\label{eqn:eos_bare}
\eeq
In bag models of very weakly interacting quarks, $\cQMsq\approx1/3$ and $B$ is the bag constant. We choose $\cQMsq$ between $1/3$ (characteristic of very weakly interacting massless quarks) and $1$ (maximal value consistent with causality) in order to scan extreme range of outcomes for quark matter.

\section{Calculation of the Love number and tidal deformability}
\label{sec:tidal_calc}

To linear order, the tidal deformability $\la$ that characterizes the response of the neutron star to external disturbance is defined by the ratio of the induced mass-quadrupole moment $\mathit{Q}_{ij}$ and the applied tidal field $\mathcal{E}_{ij}$
\beq
\mathit{Q}_{ij}=-\la\,\mathcal{E}_{ij}.
\label{eqn:lam}
\eeq

It is related to the $l=2$ dimensionless tidal Love number $k_2$
\bea
\la&=&\frac{2}{3}k_2 R^{5},\nn
\La&=&\la/M^{5},
\label{eqn:k2-lam}
\eea
where $R$ and $M$ are the radius and mass of the star, and $\La$ is the dimensionless tidal deformability.

\begin{figure*}[htb]
\parbox{0.48\hsize}{
\includegraphics[width=\hsize]{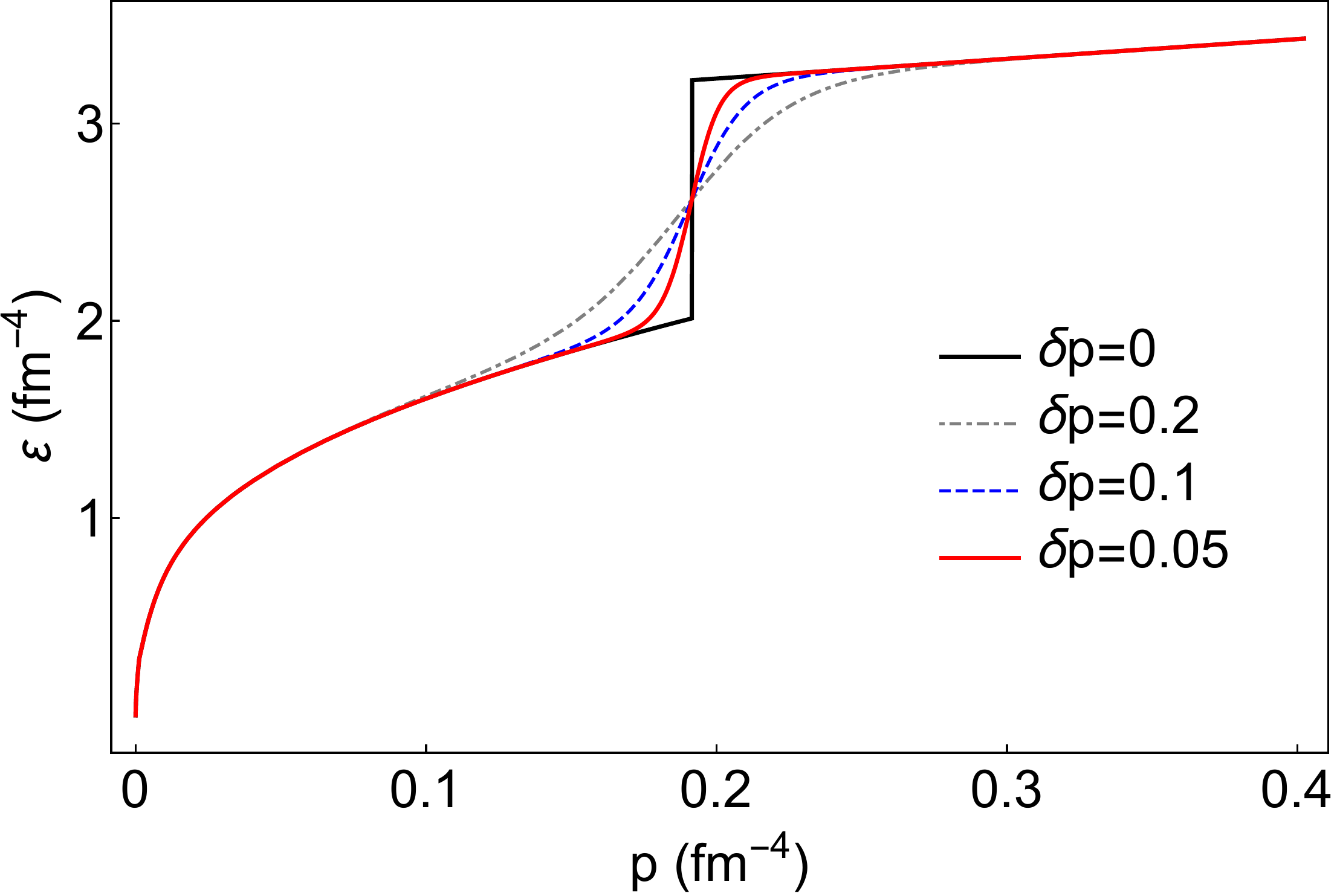}\\
}\parbox{0.51\hsize}{
\includegraphics[width=\hsize]{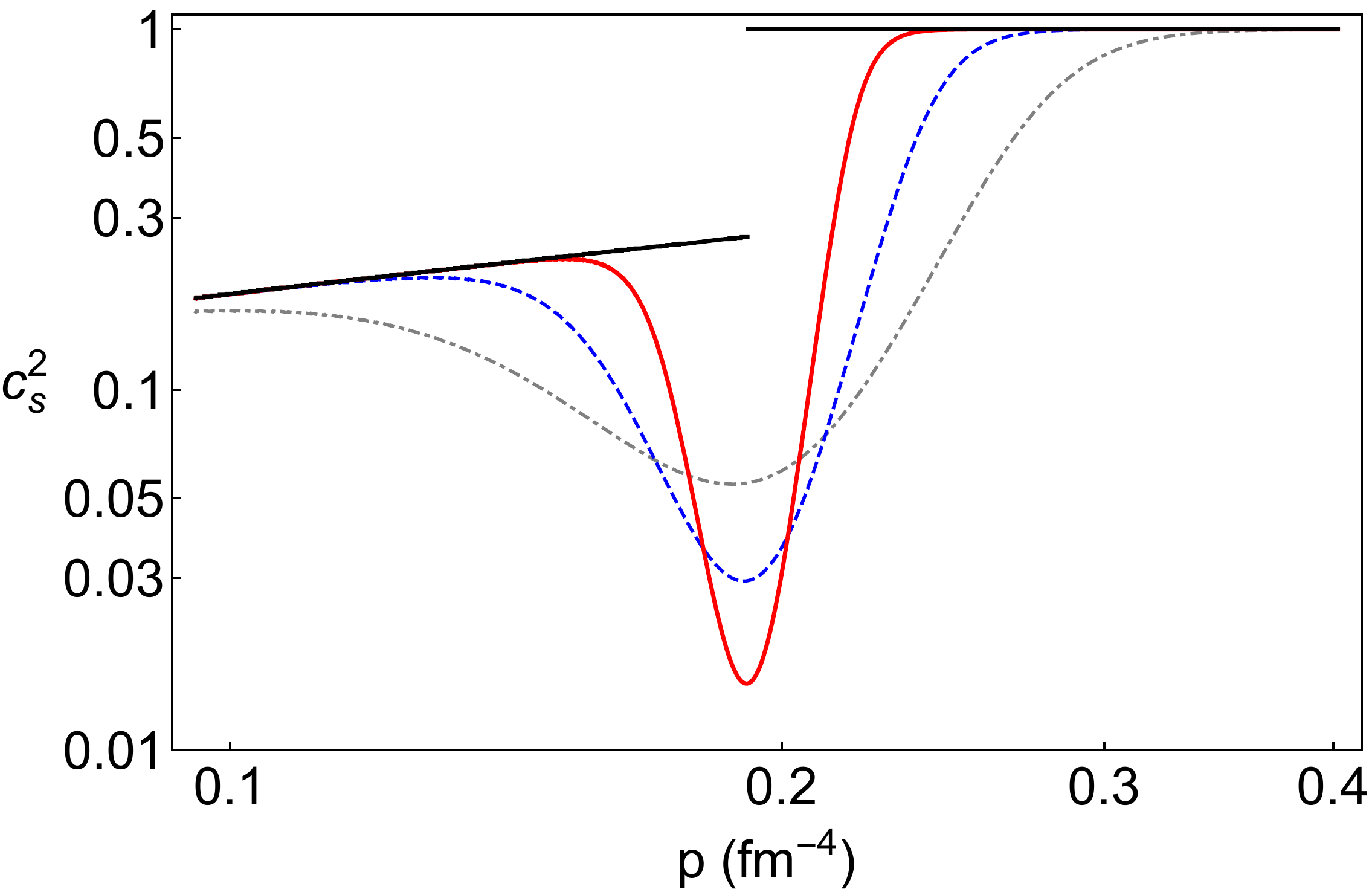}\\
}
\caption{(Color online) LHS: $\ep(p)$ for SFHo + CSS hybrid EoS with sharp hadron/quark interface (black solid), specified by $\ntrans/n_0=2.5$, $\De\ep/\etrans=0.6$ and $\cQMsq=1$, and three smoothed EoSs with crossover region (red solid, blue dashed and gray dash-dotted), specified by the width $\de p = 0.05$, 0.1, and 0.2 respectively. RHS: sound-speed squared as a function of the pressure $c_s^2(p)$ for each EoS. Note that in the sharp transition case, $1/c_s^2(p)$ encounters a singularity at $\ptrans$, not a finite discontinuity.}
\label{fig:eos-csq-smt}
\end{figure*}

We follow the method developed
in~\cite{Hinderer:2007mb,Damour:2009vw},
the quantity $k_2$ can be expressed in terms of the compactness parameter $\be=M/R$
\bea
k_2&=&\frac{8\be^2}{5}(1-2\be)^2[2+2\be(y-1)-y] \nn
&&\times \{2\be[6-3y+3\be(5y-8)] \nn
&&+4\be^3[13-11y+\be(3y-2) \nn
&&+2\be^2(1+y)]+3(1-2\be)^2[2-y \nn
&&+2\be(y-1)]\ln(1-2\be)\}^{-1}
\label{eqn:k2-beta}
\eea
where the quantity $y$ is defined as $y\equiv y(r) |_{r=R}$. The function $y(r)$ satisfies the first-order differential equation \cite{Postnikov:2010yn}
\bea
ry'(r)+y(r)^2+y(r)e^{\la(r)}\left\{1+4\pi r^2 \right.\nn
\left.\left[p(r)-\ep(r)\right]\right\}+r^2 Q(r)&=&0,
\label{eqn:y_r}
\eea
\bea
Q(r)&=&4\pi e^{\la(r)}\left[5\ep(r)+9p(r)+\frac{\ep(r)+p(r)}{dp/d\ep}\right] \nn
&&-6\frac{e^{\la(r)}}{r^2}-(\nu'(r))^2,
\label{eqn:Q_r}
\eea
with the boundary condition $y(0)=2$. The metric coefficients $\la(r)$ and $\nu(r)$ are given by
\bea
e^{\la(r)}&=&\left[1-\frac{2m(r)}{r}\right]^{-1},\nn
\frac{d\nu}{dr}&=&\frac{2}{r}\left[\frac{m(r)+4\pi p(r) r^3}{r-2m(r)}\right].
\label{eqn:metric}
\eea

For spherically symmetric configurations, by choosing the equation of
state $\ep(p)$ and the value of central pressure $\pcent$, one can
obtain the Love number and tidal deformability by solving
simultaneously \Eqn{eqn:y_r} and the Tolman-Oppenheimer-Volkoff (TOV)
equations \cite{Tolman:1939jz,Oppenheimer:1939ne}.

\begin{figure*}[htb]
\parbox{0.48\hsize}{
\includegraphics[width=\hsize]{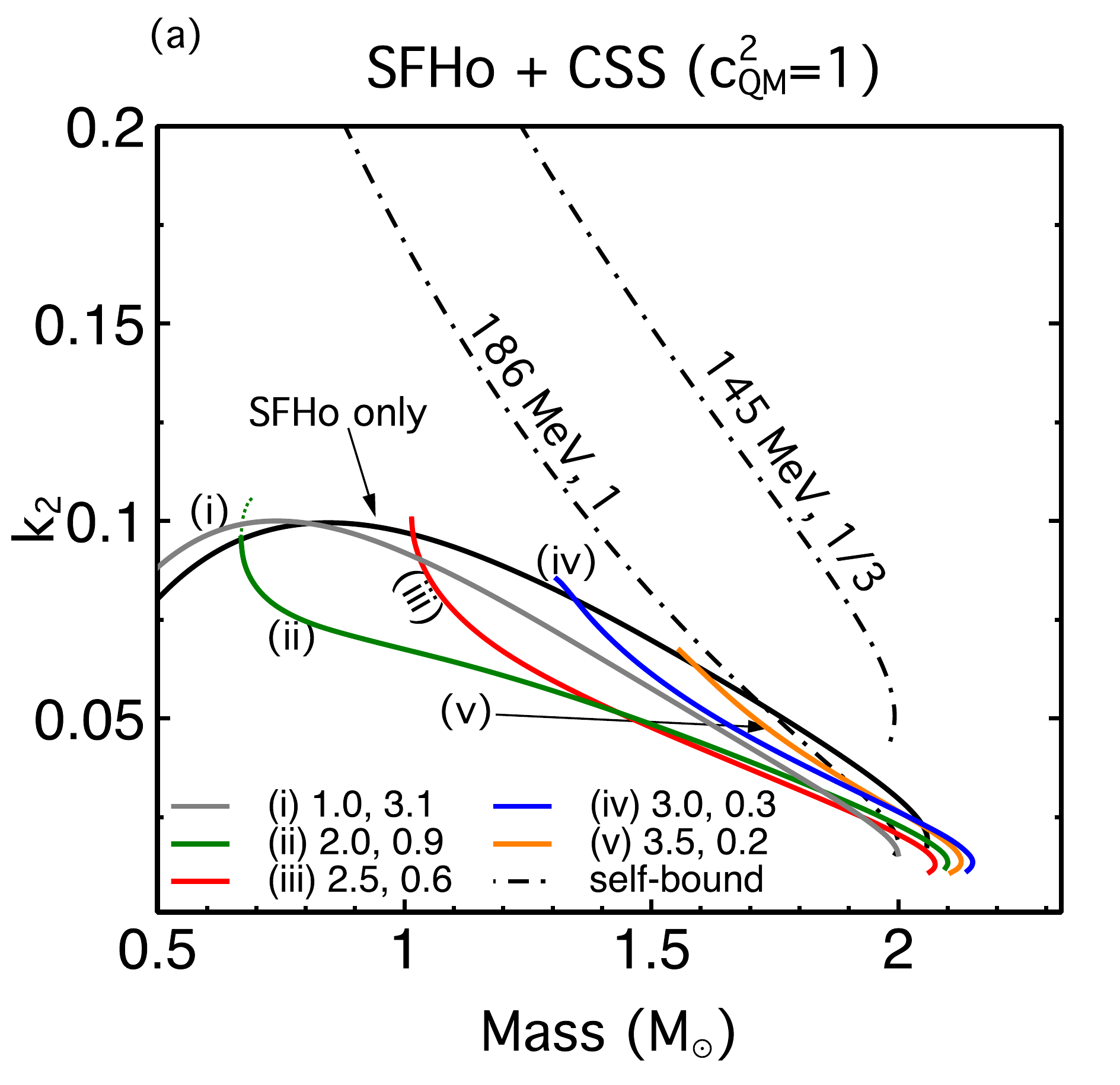}\\[-2ex]
}\parbox{0.48\hsize}{
\includegraphics[width=\hsize]{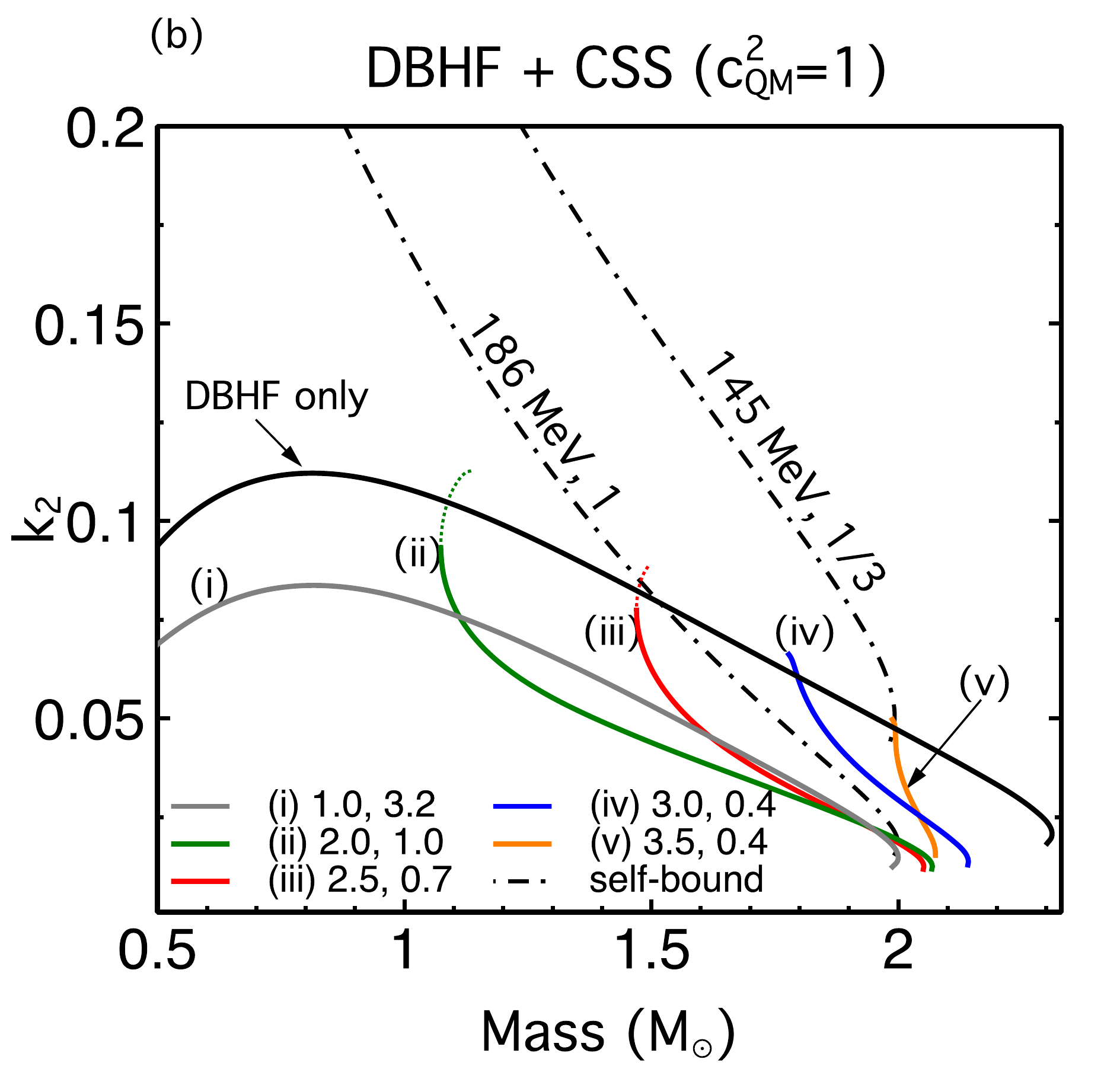}\\[-2ex]
}\\[2ex]
\parbox{0.5\hsize}{
\includegraphics[width=\hsize]{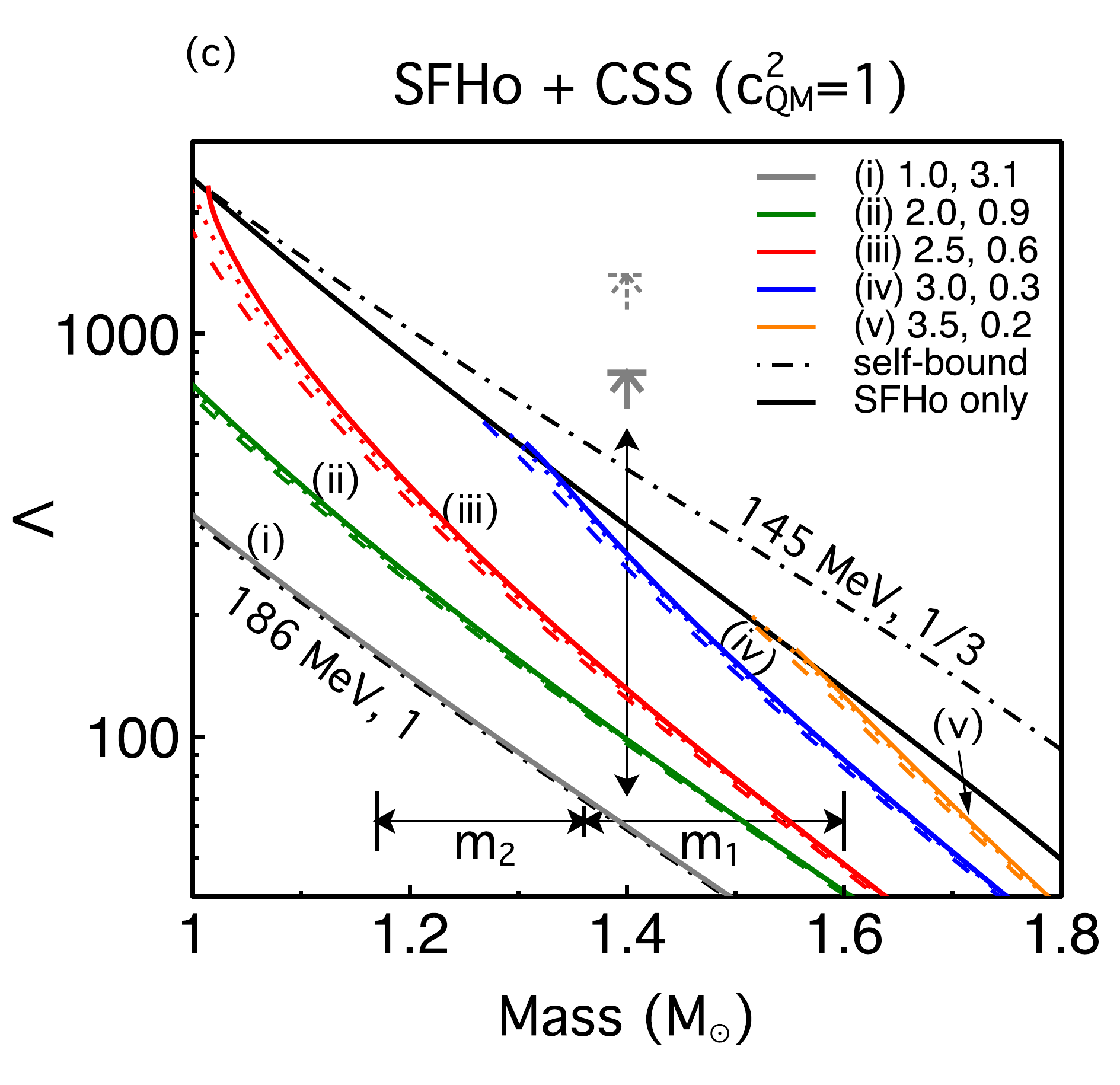}\\[-2ex]
}\parbox{0.5\hsize}{
\includegraphics[width=\hsize]{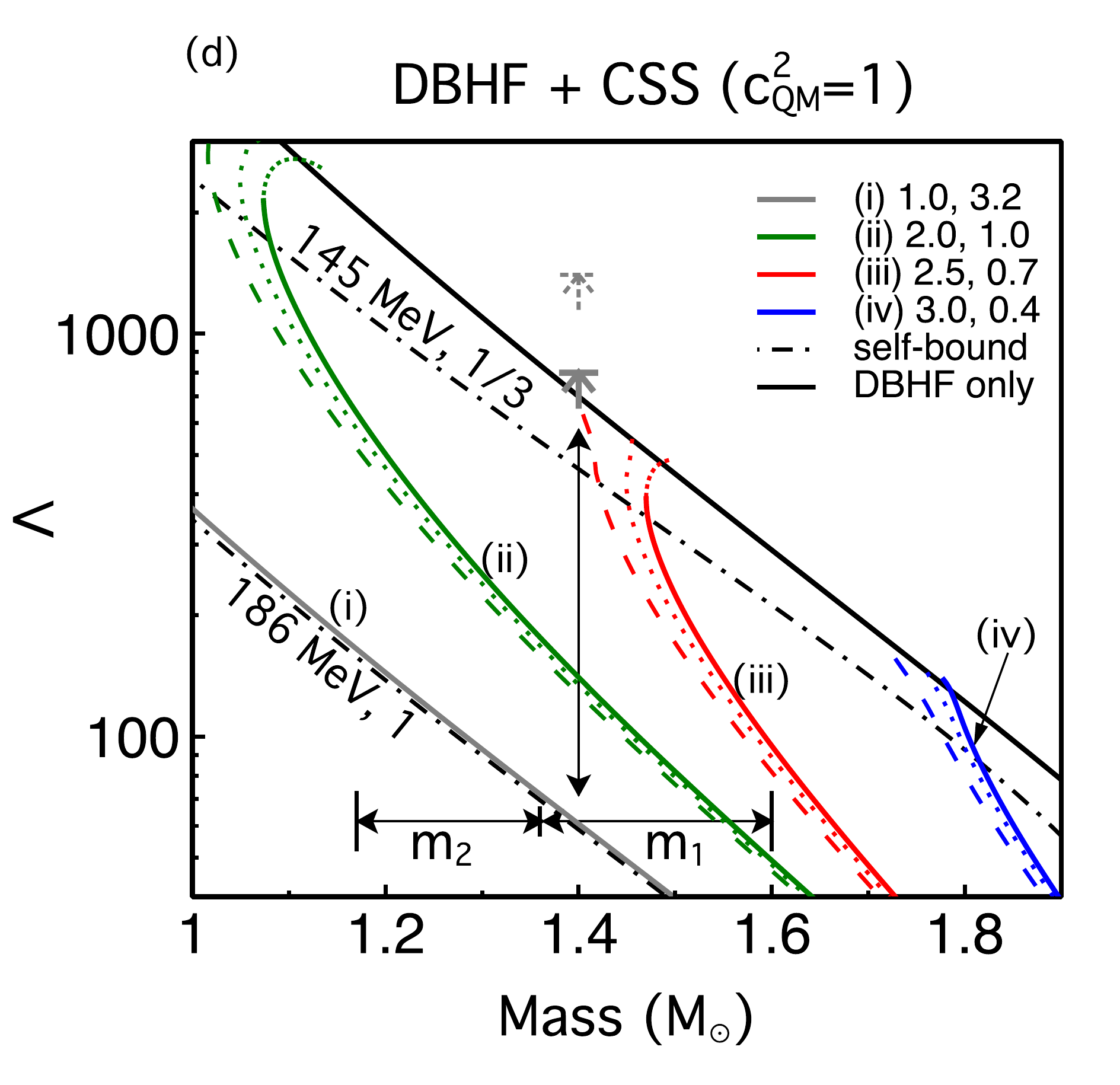}\\[-2ex]
}
\caption{(Color online) Love number $k_2$ and dimensionless tidal deformability $\La$ based on SFHo/DBHF + CSS (set I) hybrid EoSs with sharp first-order phase transitions that the onset density varies extensively ($\ntrans=1\sim3.5 \,n_0$); dotted segments on the solid colored curves represent unstable stellar configurations. Black solid curves represent purely-hadronic EoSs, and dash-dotted lines exemplify self-bound strange quark stars (SQSs) specified by the linear EoS \Eqn{eqn:eos_bare}. In lower panels $\La(M)$ for crossover phase transitions are also plotted, which lie below corresponding curves for sharp phase transitions; the widths of crossover are $\de p=0.2$ (dashed) and $\de p=0.1$ (dotted) respectively. Two gray arrows $\La=1400, 800$ correspond to the original upper bounds on the dimensionless tidal deformability for $1.4\,\Msolar$ stars, with high (dashed) or low (solid) spin priors from gravitational-wave detection GW170817~\cite{LIGO:2017qsa}, while the vertical line with double arrowheads represents updated limits $\La_{1.4}=190_{-120}^{+390}$ given in most recent work \cite{LIGO:2018exr}; in both analyses the possibility of sharp phase transition was not taken into account.
}
\label{fig:k2-Lam-set1}
\end{figure*}

\subsubsection{First-order phase transitions and the role of sound speed}

In the presence of a finite energy density discontinuity, the last term in \Eqn{eqn:y_r} involves a singularity $\propto (\ep+p)/c_s^2$, where $c_s^2=dp/d\ep$ is the sound speed squared. This was first discussed by Damour \& Nagar~\cite{Damour:2009vw} addressing the surface/vacuum discontinuity for incompressible stars. Postnikov et al. \cite{Postnikov:2010yn} extended this approach to the context of possible first-order transitions inside the stars. To capture the delta-function behavior across the point of discontinuity 
\beq
1/c_s^2=(d\ep/dp) |_{p \neq p_d}+\de (p-p_d) \De \ep,
\label{eqn:csq_disc}
\eeq
one needs to properly match the solutions of $y(r)$ at $r_d^{-}$ and $r_d^{+}$
\beq
y(r_d^{+})=y(r_d^{-})-\frac{4\pi r_d^3 \,\De \ep}{m(r_d)},
\label{eqn:y_r_disc}
\eeq
where $\De \ep=\ep(r_d^{-})-\ep(r_d^{+})$. 

In the incompressible limit where the energy density is uniform ($\ep=const=\ep_0$) inside the star and vanishes outside the star, \Eqn{eqn:y_r_disc} is reduced to the matching condition at the surface $y(R^{+})=y(R^{-})-3$. As a result, the energy density discontinuity at the sharp surface of bare self-bound strange quark stars (SQSs) leads to drastic change in the value of tidal Love number $k_2$~\cite{Damour:2009vw,Hinderer:2009ca}.

For normal hadronic matter models, the $p(\ep)$ curve is assumed to be smooth, except for an energy discontinuity at the crust/core interface of neutron stars. For strange quark stars which are surrounded by a thin nuclear crust, the exact nature of phase transition is unknown. Thus strange quark stars are good candidates to examine the effect on $\la$ and $k_2$ values from discontinuous energy density at an interface. Postnikov et al. \cite{Postnikov:2010yn} first studied the crust/core transition in this case, which is characteristic of enormous $\De\ep/\etrans$ ($=30\sim 10^5$). They find that such a phase transition leads to an order of magnitude variation in $\la$, according to the treatment of the discontinuity (sharp or smoothed). They also find that the Love number of strange quark stars can be smaller or larger than that for hadronic stars. In this work, we elaborate on the generic behavior of the tidal deformability in the presence of hybrid stars, where phase transitions arise in the core region transforming normal nuclear matter into quark matter at supranuclear densities, with density discontinuity mostly being of order one ($\De\ep/\etrans\lesssim 3$), and compare the results with both purely-hadronic stars and self-bound strange quark stars. In Sec.~\ref{sec:tidal} we show that the smaller radii and tidal Love numbers obtained in sharp first-order transition to a stiff quark phase ($c_s^2\approx1$) predict a new lower bound on the dimensionless tidal deformability of neutron stars.

\subsubsection{Smoothing into a crossover}
The discontinuity in $c_s^2=dp/d\ep$ that enters the differential equation for tidal deformability calculations is essential for EoSs with sharp phase transitions, and here we introduce a regularized form that describes a crossover \cite{Alford:2017vca}

\bea
\ep(p)&=&\frac{1}{2}\left[1-\tanh \left(\frac{p-\ptrans}{\de p}\right) \right] \ep_{\rm NM}(p) \nn
&&+\frac{1}{2}\left[1+\tanh \left(\frac{p-\ptrans}{\de p}\right) \right] \ep_{\rm QM}(p),
\label{eqn:CSS_EoS_smt}
\eea
where the nuclear matter EoS $\ep_{\rm NM}(p)$ and quark matter EoS $\ep_{\rm QM}(p)$ take the same form as in standard CSS parameterization \Eqn{eqn:CSS_EoS}. In the vicinity of the crossover region where $p\in [\ptrans -\de p, \,\ptrans+\de p]$, the sound speed is changing rapidly.  Applying the common calculation method of Love number and tidal deformability for the smooth EoS describing a crossover, one can numerically evaluate the solutions and observe the trend when the width of crossover $\de p$ is varied. 

The limiting case $\de p \to 0$ restores a true density discontinuity (sharp transition) in the EoS, for which matching condition at the phase boundary in \Eqn{eqn:y_r_disc} is necessary. To illustrate this with an example, we plot in Fig.~\ref{fig:eos-csq-smt} the hybrid EoS $\ep(p)$ with sharp hadron/quark interface specified by $\ntrans/n_0=2.5$, $\De/\etrans=0.6$ and $\cQMsq=1$, along with its sound-speed squared as a function of the pressure $c_s^2(p)$. The SFHo EoS is chosen as the hadronic part. We then compare them to three regulated smooth EoSs (crossover) setting $\de p = 0.2, 0.1, 0.05$ in \Eqn{eqn:CSS_EoS_smt} respectively. In practice, a substantial (but still continuous) change in $c_s^2$ similar to the RHS of Fig.~\ref{fig:eos-csq-smt} can be a precursor of phase transition in ``realistic'' quark models. We present results of the tidal parameters for rapid crossovers that mimic a discontinuity, and find that the generic trend exhibited does not change appreciably; see plots and discussion in Sec.~\ref{sec:tidal}.

\section{Results and discussion}
\label{sec:results}

\subsection{Love number and tidal deformability}
\label{sec:tidal}

As masses and tidal deformabilities are the measurable macroscopic quantities during binary inspirals from gravitational-wave observations, in Fig.~\ref{fig:k2-Lam-set1} we plot tidal Love number $k_2$ and dimensionless tidal deformability $\La$ as a function of the neutron star mass, for the three classes of equations of state studied in this work.

For purely-hadronic EoSs, there is a small spread of $k_2$ values, and the
trend with varying the mass are similar. For self-bound strange quark stars, the value of $k_2$ is larger, but tidal deformability is smaller because of their much smaller radii (see Fig.~\ref{fig:MR-curves-1}). In both cases, $\La(M)$ is a smooth, monotonically decreasing function. 

To analyze the effects of sharp first-order phase transition, we calculate and plot $k_2(M)$ and $\La(M)$ for the SFHo/DBHF+ CSS set I hybrid EoSs. When the central pressure of the star reaches above the critical pressure of phase transition $\ptrans$, $k_2$ and $\La$ decrease below the values which are obtained in a purely-hadronic star. This decrease is large enough that hadronic stars and hybrid stars will be distinguishable in future BNS merger events (see discussion in Sec.~\ref{sec:bns-obs}). The behavior of $k_2$ and $\La$ in the DBHF-2SC-CFL and SFHo-2SC-CFL models discussed later in Sec.~\ref{sec:correlation} shows the same trend, with two decreases which occur when the central pressure reaches the two phase transitions.

\begin{figure*}[htb]
\parbox{0.5\hsize}{
\includegraphics[width=\hsize]{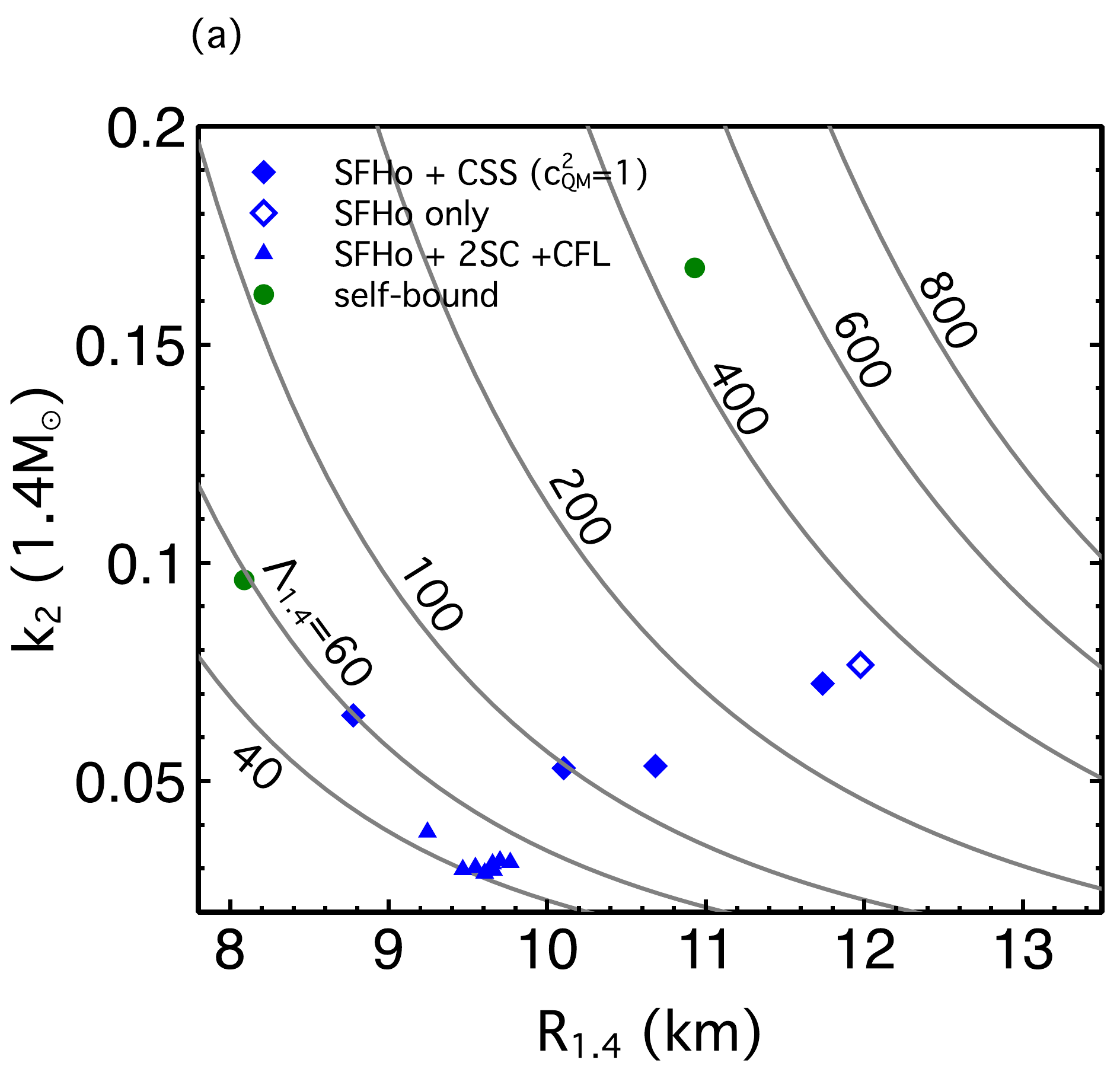}\\[-2ex]
}\parbox{0.5\hsize}{
\includegraphics[width=\hsize]{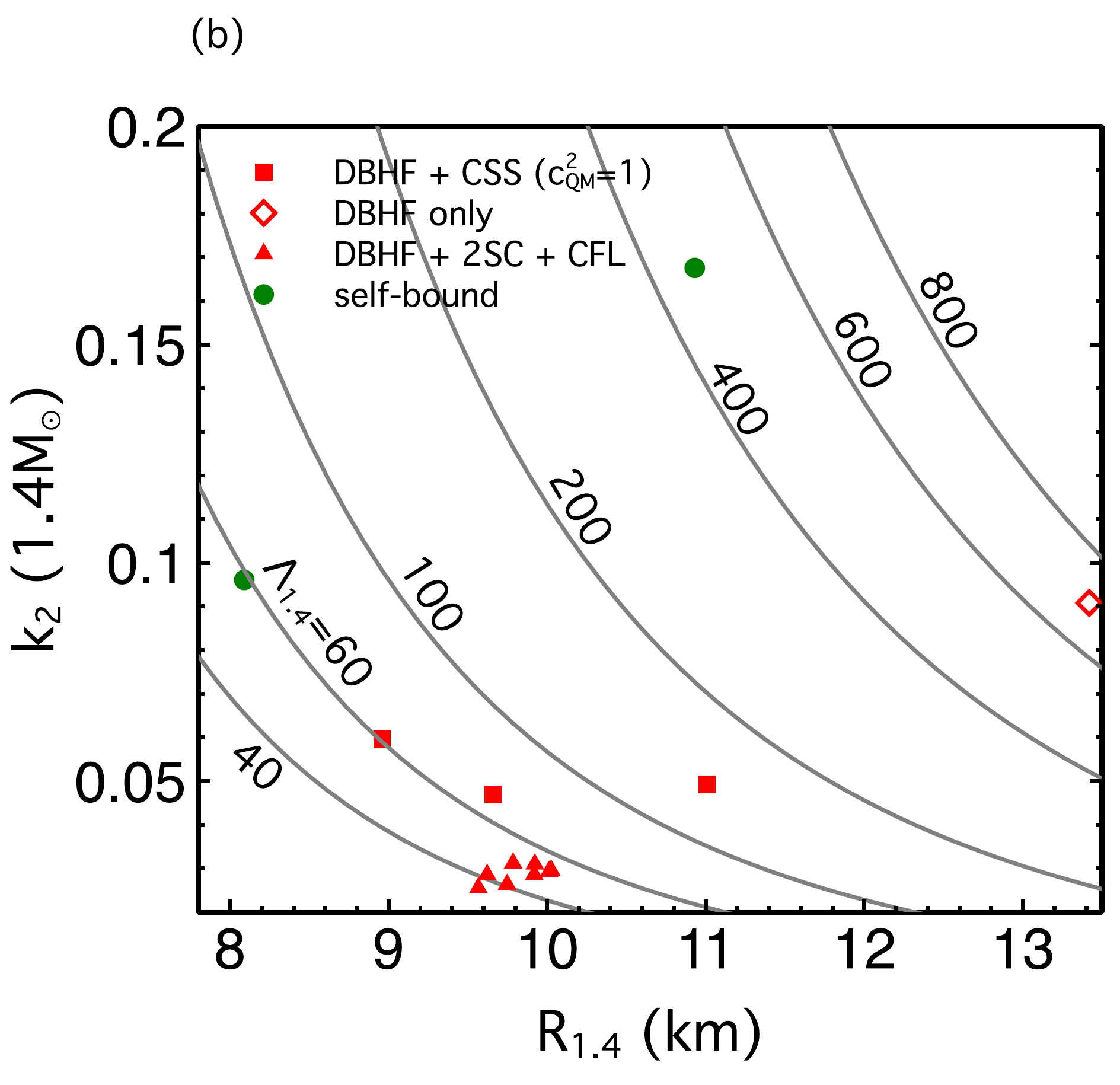}\\[-2ex]
}\\[5ex]
\parbox{0.5\hsize}{
\includegraphics[width=\hsize]{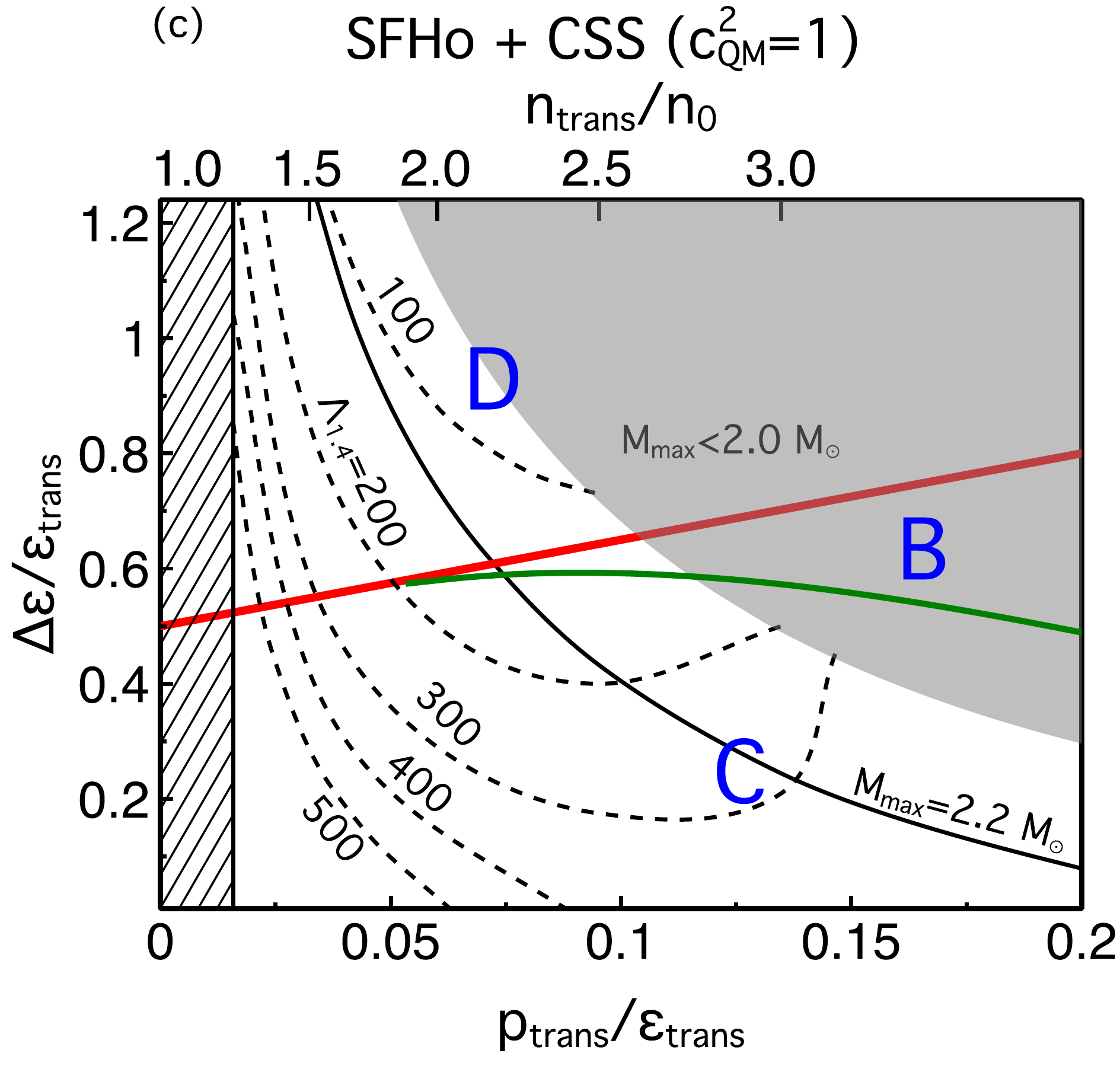}\\[-2ex]
}\parbox{0.5\hsize}{
\includegraphics[width=\hsize]{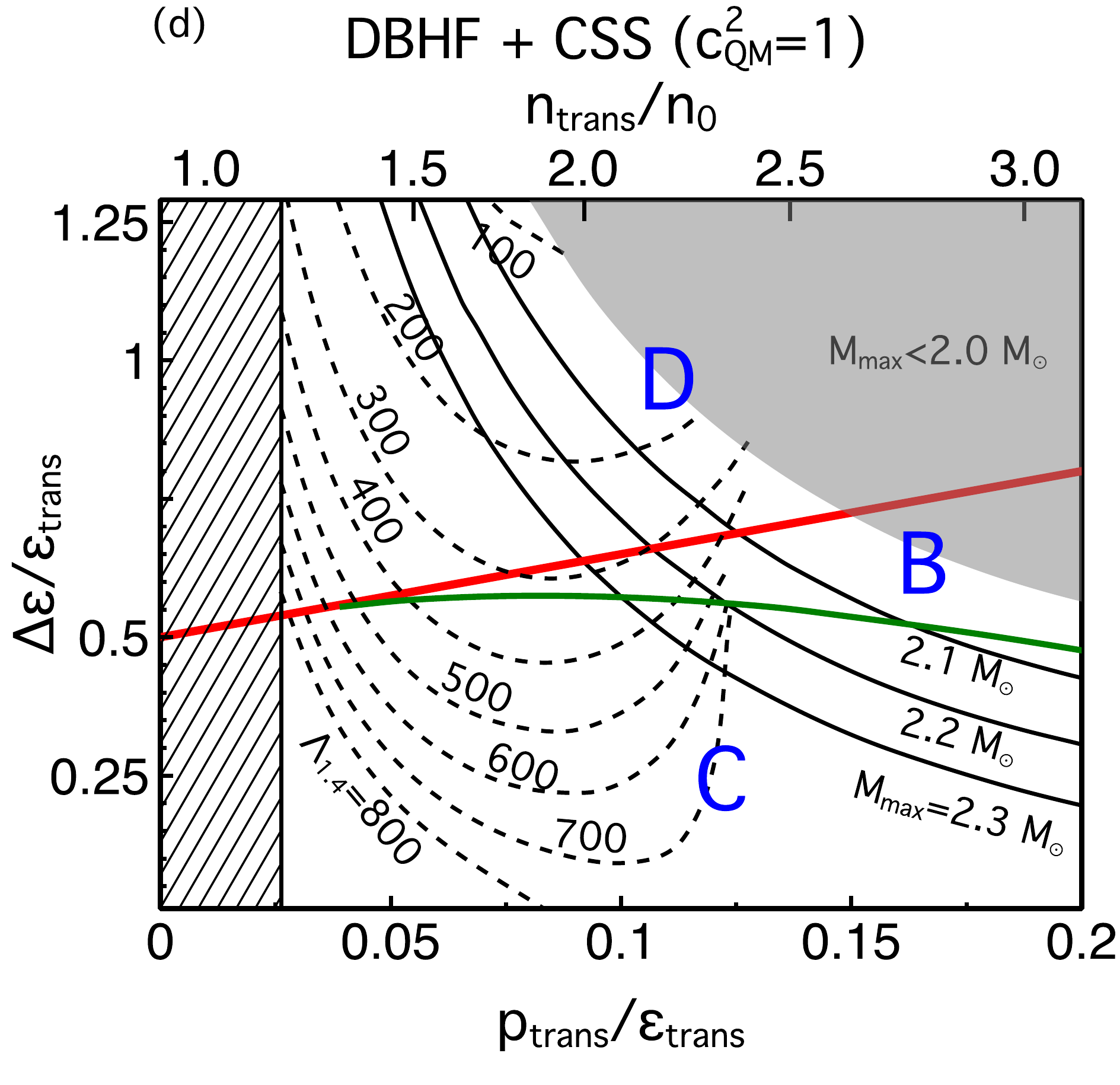}\\[-2ex]
}
\caption{(Color online) Panels (a) and (b): values of $k_2$ and $R$ for $1.4\,\Msolar$ purely-hadronic stars, self-bound quark stars and hybrid stars. Lowest $\La_{1.4} $ are associated with a strong nuclear $\to$ 2SC phase transition around $1\sim2$ times nuclear saturation density followed by a weak 2SC $\to$ CFL phase transition at higher densities (see \Eqn{eqn:seq_EoS} and text). Panels (c) and (d): constraints from $\Mmax$ and $\La_{1.4}$ on the quark matter phase space, assuming a single phase transition based on SFHo/DBHF + CSS parametrization. 
}
\label{fig:lam_bar-contour}
\end{figure*}

\begin{figure*}[htb]
\parbox{0.5\hsize}{
\includegraphics[width=\hsize]{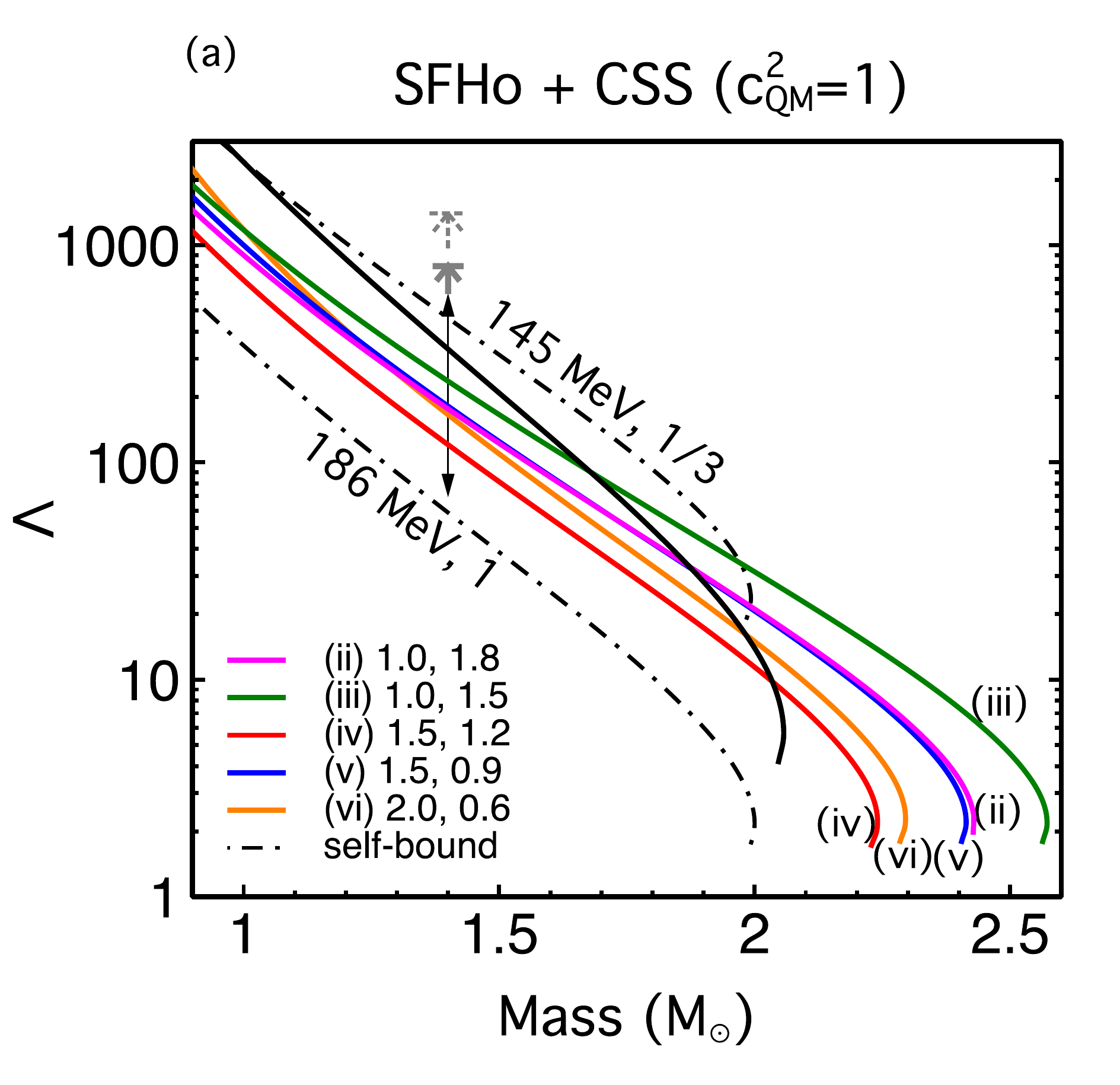}\\[-2ex]
}\parbox{0.5\hsize}{
\includegraphics[width=\hsize]{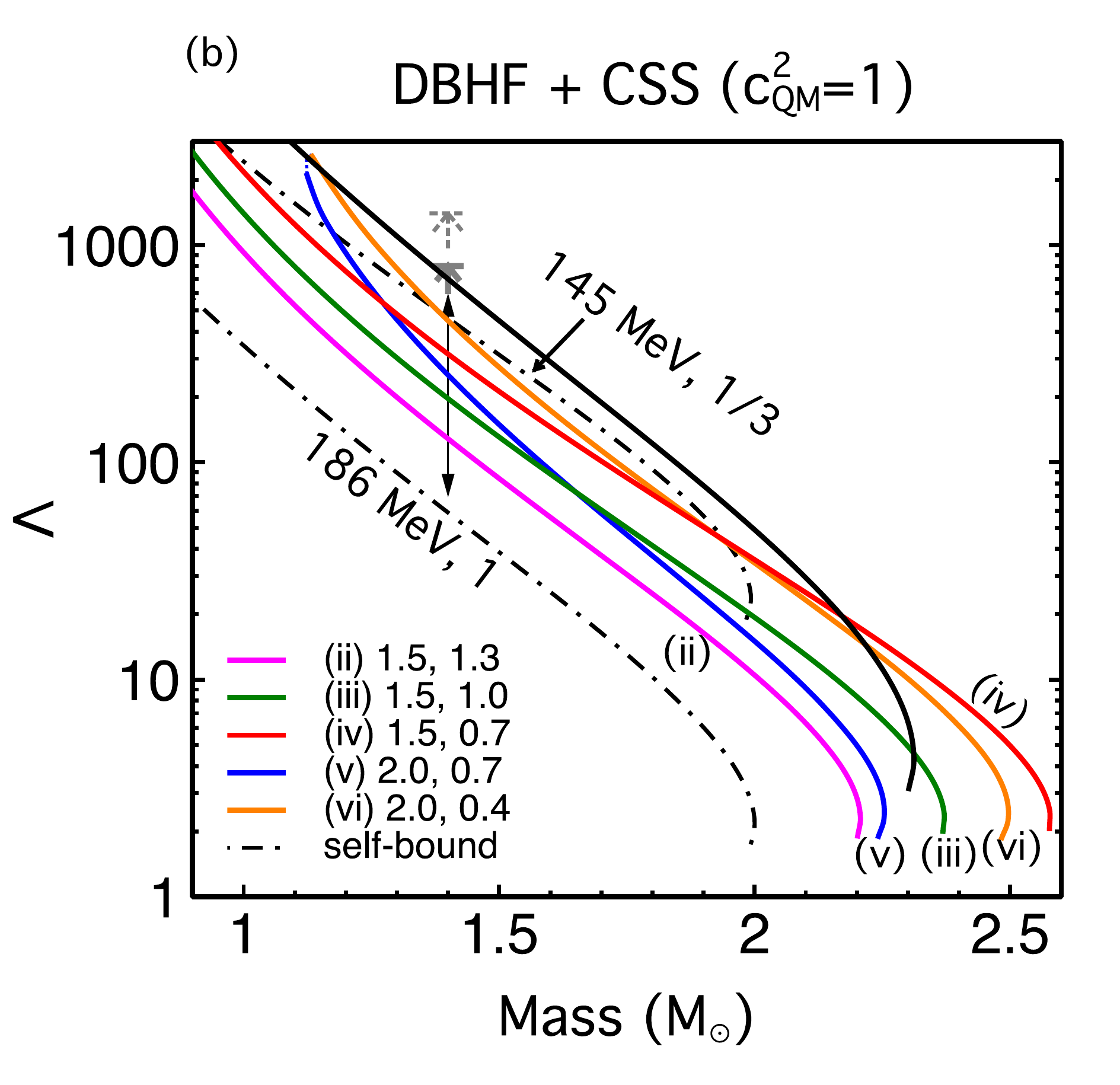}\\[-2ex]
}
\caption{(Color online) Dimensionless tidal deformability $\La$ based on SFHo/DBHF + CSS (set II) hybrid EoSs, for which phase transitions set in fairly early ($\ntrans=1\sim\,2 n_0$) and the onset mass for quark cores to appear is relatively low ($\Mtrans\lesssim1.2\Msolar$). Black solid curves represent purely-hadronic EoSs, and dash-dotted lines exemplify self-bound strange quark stars (SQSs) specified by the linear EoS \Eqn{eqn:eos_bare}. Two gray arrows $\La=1400, 800$ correspond to the original upper bounds on the dimensionless tidal deformability for $1.4\,\Msolar$ stars, with high (dashed) or low (solid) spin priors from gravitational-wave detection GW170817~\cite{LIGO:2017qsa}, while the vertical line with double arrowheads represents updated limits $\La_{1.4}=190_{-120}^{+390}$ given in most recent work \cite{LIGO:2018exr}; in both analyses the possibility of sharp phase transition was not taken into account.
}
\label{fig:Lam-M-set2}
\end{figure*}

The rapid crossover that approximates a true discontinuity in the EoS gives rise to even lower values of $\La$ around the phase transition region (see dashed and dotted curves in Fig.~\ref{fig:k2-Lam-set1}, panels (c) and (d)), but not as noticeable when the central density is much higher. This is due to the fact that for massive hybrid stars with large quark cores, global quantities such as tidal deformability become insensitive to the exact nature of phase transition being sharp first-order or rapid crossover. Also, for smoothed EoSs with very short range of the crossover e.g. $\de p=0.05$ in Fig.~\ref{fig:eos-csq-smt}, the $\La(M)$ relation is almost identical to those with sharp transitions and the curves are hardly visible, hence we choose not to show them on the plots. 

Panels (a) and (b) in Fig.~\ref{fig:lam_bar-contour} display a scatter of data points on the $(k_2, R)$ plane for $1.4\,\Msolar$ stars. Filled diamonds are obtained from SFHo/DBHF + CSS (set I) hybrid EoSs, lying on $\La_{1.4}$ contours with lower values compared to the purely-hadronic ones (open diamonds) chiefly due to the decrease in both $k_2$ and $R$. For self-bound SQSs, as mentioned above, higher value of $k_2$ are compensated by smaller radii (see Figs.~\ref{fig:k2-Lam-set1} and \ref{fig:MR-curves-1}), resulting in $\La_{1.4}$ that ranges from $60 \sim 500$ which is sensitive to its stiffness $\cQMsq$ (if $\Mmax=2\,\Msolar$ is fixed). In most cases radii play the dominant role, except particular sequential nuclear $\to$ 2SC $\to$ CFL phase transitions (labeled with filled triangles) that reduce the value of $k_2$ at the canonical mass $1.4\,\Msolar$ to 0.03--0.04. For the EoS template of two sharp transitions, see \Eqn{eqn:seq_EoS} in Sec.~\ref{sec:correlation}.

Repeating the calculation for different values of $\ntrans$ and $\De\ep/\etrans$, we show in panels (c) and (d) the $\La_{1.4}$ contours for all hybrid EoSs with a single phase transition that can have quark cores in $1.4\,\Msolar$ stars while $\Mmax\geq2\,\Msolar$. As can be seen from the plots, the minimal $\La_{1.4}$ is associated with $\ntrans =n_0$ and extremely large $\De\ep$ (the last EoSs of set I in Table~\ref{tab:hyb_EoS}). Also shown are the contours for maximum mass, indicating more severe constraints on quark matter if neutron stars heavier than $2\,\Msolar$ were to be detected. 

If the phase transition take place at low densities, e.g. $1\leq\ntrans/n_0\leq 2$ for set II EoSs in Table~\ref{tab:hyb_EoS}, most of the neutron stars we observe are likely to be above the onset mass $\Mtrans$ for phase transition, and accordingly the $\La(M)$ relation on the hadronic branch is less relevant (see Fig.~\ref{fig:Lam-M-set2}).

\subsection{Binary merger observables}
\label{sec:bns-obs}

The weighted-average dimensionless tidal deformability is given by

\beq
\tilde{\La}=\frac{16}{13}\frac{(m_1+12m_2)m_1^4\,\La_1+(m_2+12m_1)m_2^4\,\La_2}{(m_1+m_2)^5}
\label{eqn:La_wt}
\eeq
where $m_1$ and $m_2$ are the component masses in the binary system ($m_1\geq m_2$). A specific combination of $m_1$ and $m_2$, the chirp mass $\Mchirp=(m_{1}m_{2})^{3/5}/(m_1+m_2)^{1/5}$, can be well determined from gravitational-wave signals detected~\cite{LIGO:2017qsa} and is relatively insensitive to the mass ratio $q=m_{2}/m_1$ \cite{Radice:2017lry, Raithel:2018ncd}.

\begin{figure*}[htb]
\parbox{0.5\hsize}{
\includegraphics[width=\hsize]{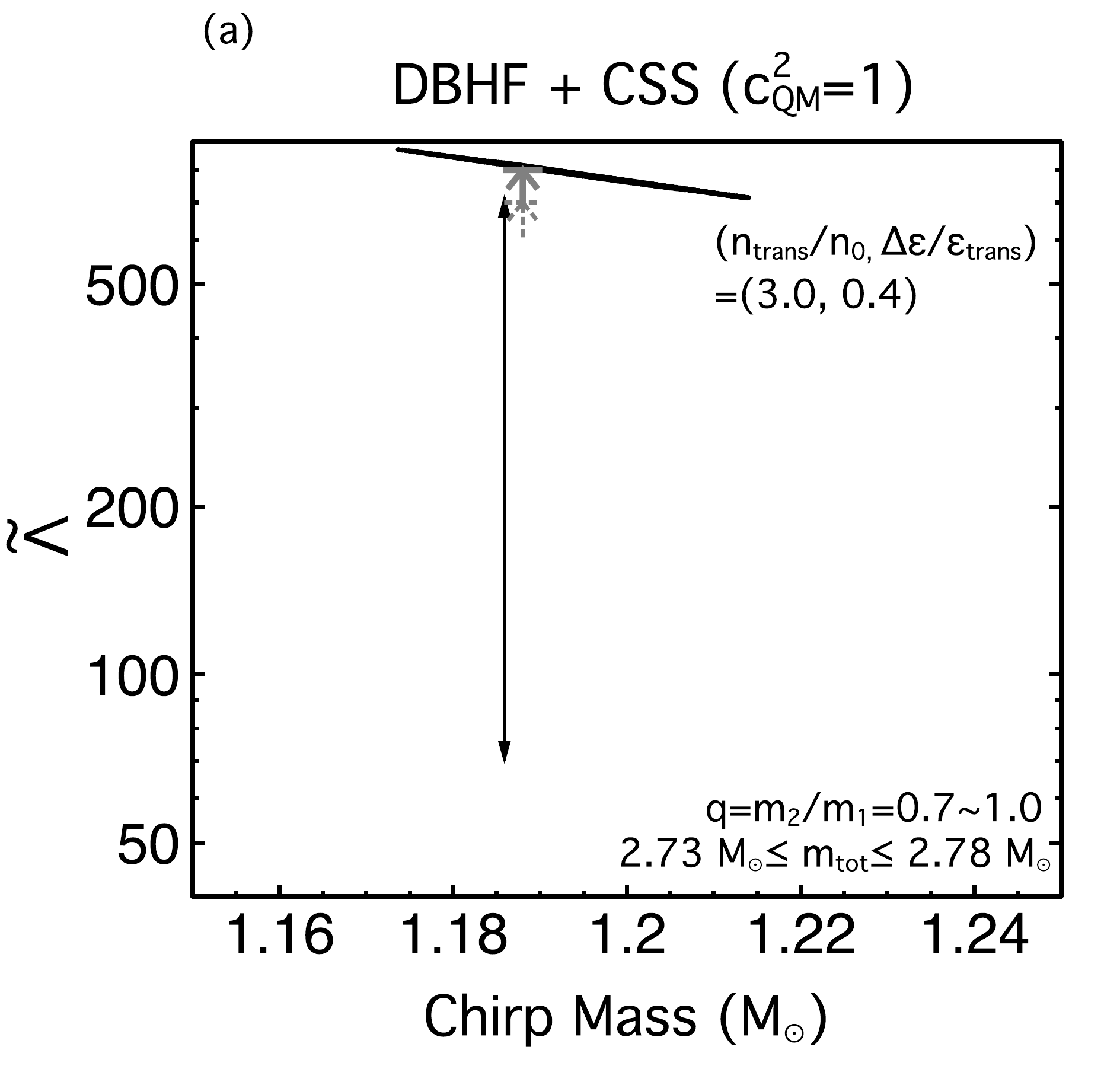}\\[-2ex]
}\parbox{0.5\hsize}{
\includegraphics[width=\hsize]{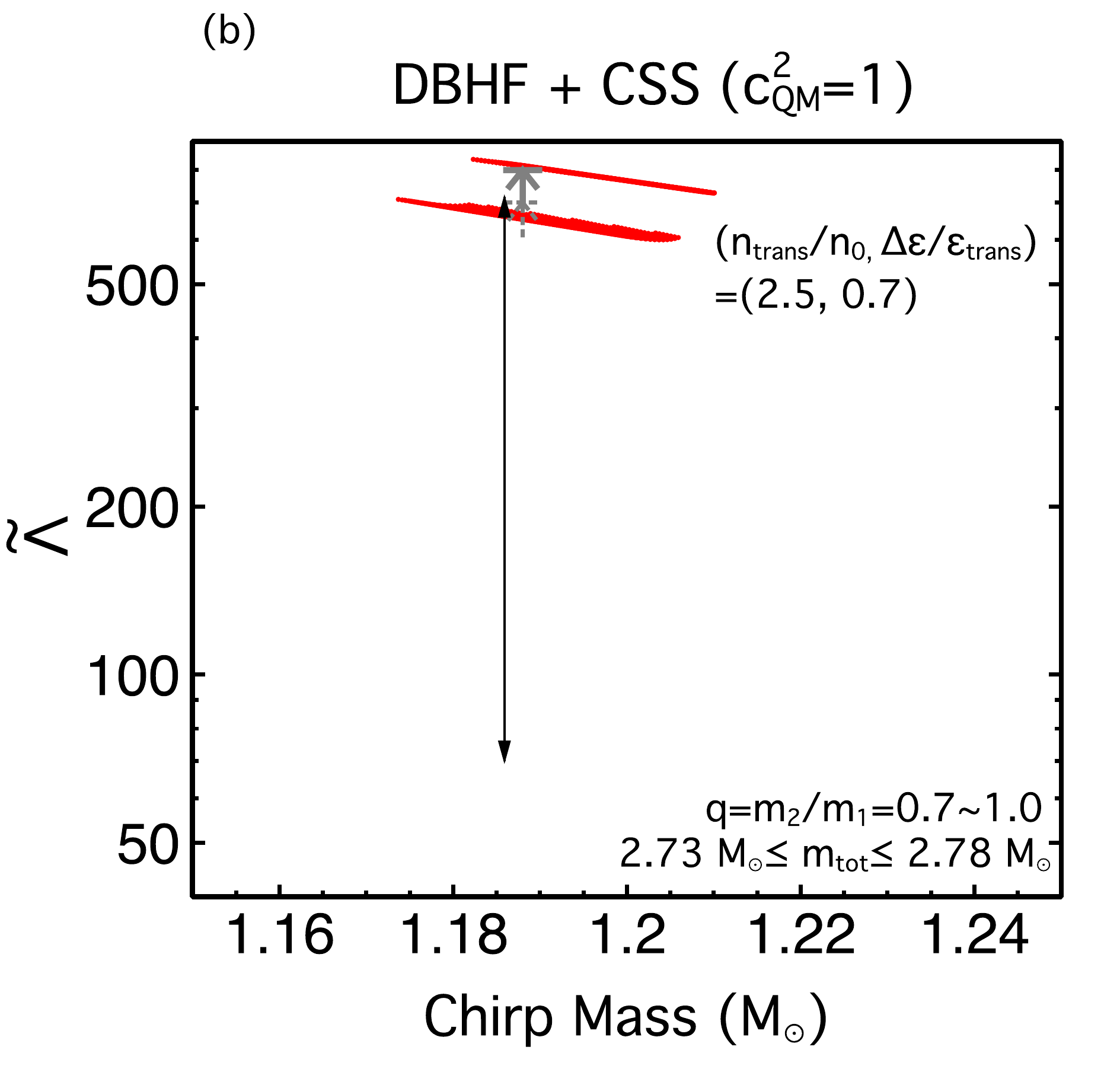}\\[-2ex]
}\\[3ex]
\parbox{0.5\hsize}{
\includegraphics[width=\hsize]{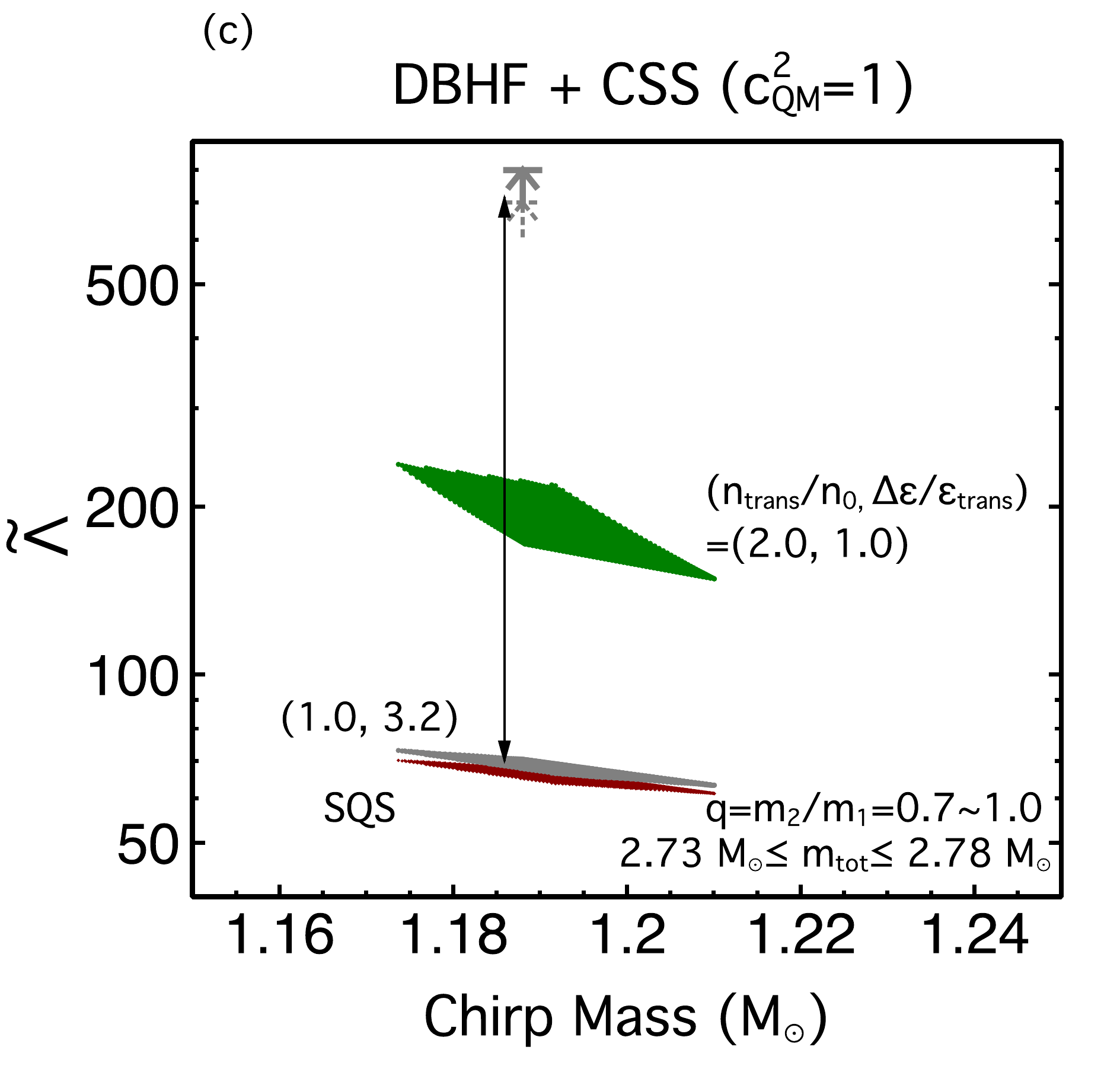}\\[-2ex]
}\parbox{0.5\hsize}{
\includegraphics[width=\hsize]{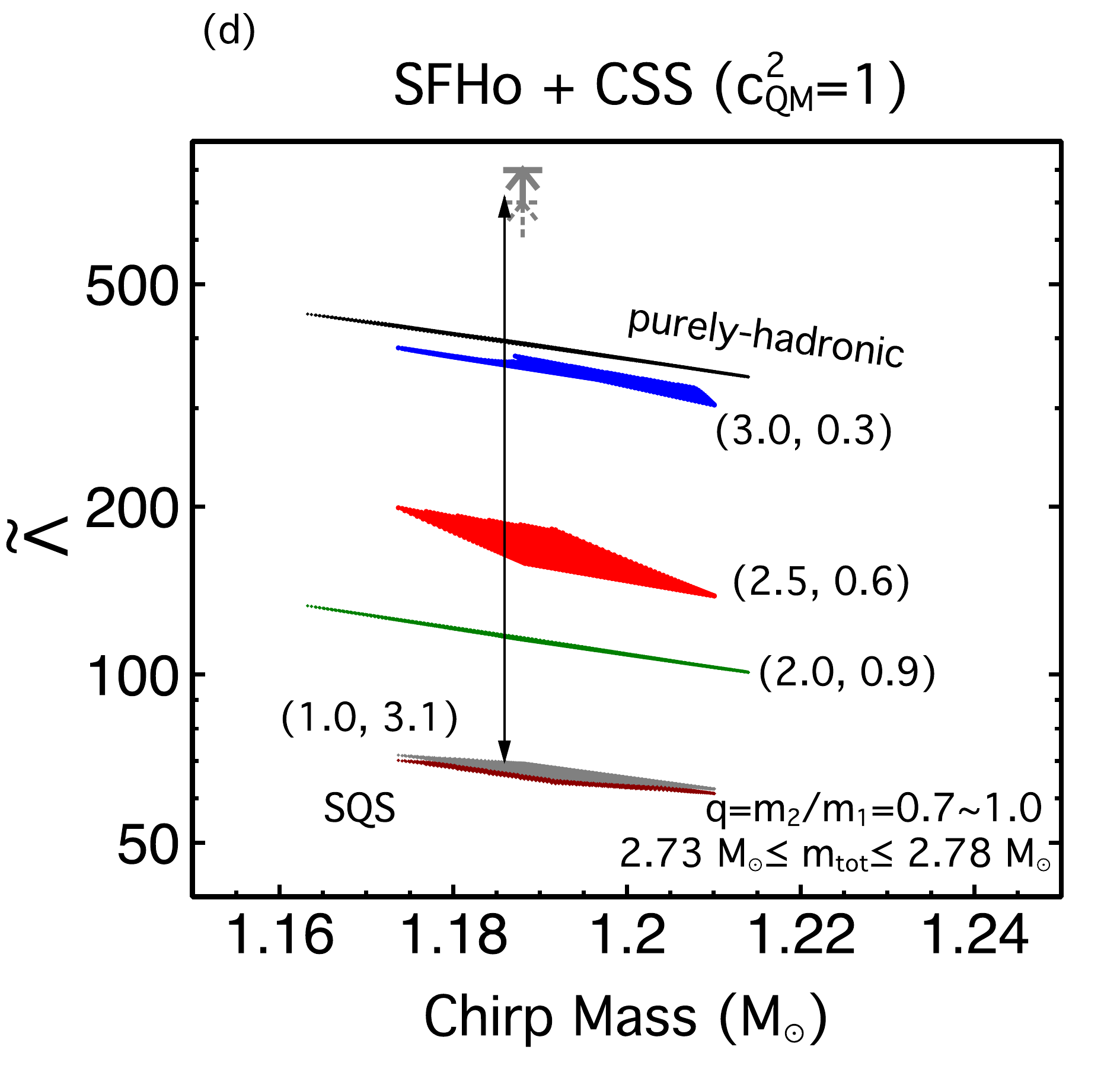}\\[-2ex]
}
\caption{(Color online) The weighted-average dimensionless tidal deformability $\tilde{\La}$ as a function of the chirp mass $\Mchirp$ for SFHo/DBHF + CSS (set I) hybrid EoSs applied in Fig.~\ref{fig:k2-Lam-set1}, within ranges of the estimated mass ratio $q$ and total mass $m_{\rm tot}$ in the binary system of GW170817 (see text). Two gray arrows $\tilde{\La}=700, 800$ correspond to upper bound on the tidal deformability for $\Mchirp=1.188_{-0.002}^{+0.004}\Msolar$ with high (dashed) or low (solid) spin priors from gravitational-wave detection~\citep{LIGO:2017qsa}, while the vertical line with double arrowheads represents updated limits $\tilde\La (\Mchirp=1.186_{-0.001}^{+0.001}\Msolar)=300_{-230}^{+420}$ for the low-spin prior (using a 90\% highest posterior density interval) \cite{LIGO:2018wiz}.The dark red band at the bottom refers to the ``maximally compact EoS'' for self-bound SQSs~\cite{Lattimer:2012nd,Lattimer:2015nhk}. If in the future more data from advanced LIGO of multiple BNS merger events were to provide a refined range estimate of $\tilde{\La}$ for given chirp mass statistically, capable of distinguishing among different scenarios, we anticipate increasingly better constraint on the phase transition parameters and the quark matter EoS.}
\label{fig:Lam-Mchirp-pt}
\end{figure*}

Assuming both neutron stars  in GW170817 obey the same EoS, we scan all possible combinations of the primary mass $m_1 \in [1.36, 1.60]\, \Msolar$ and secondary mass $m_2 \in [1.17, 1.36]\, \Msolar$, imposing bounds on the total mass $m_{\rm tot}=2.74_{-0.01}^{+0.04}\, \Msolar$ and mass ratio $q\in[0.7,1.0]$ for low-spin priors, and compute $\La_1$, $\La_2$ and eventually $\tilde{\La}$. The results are shown in Fig.~\ref{fig:Lam-Mchirp-pt}; for all hybrid star configurations we fix $\cQMsq=1$.

\noindent $\bullet$ Panel (a): 
purely-hadronic DBHF EoS, and DBHF + CSS with 

$\ntrans/n_0=3.0$ $(\Mtrans=1.78\,\Msolar)$, $\De\ep/\etrans=0.4$.
Since the mass threshold $\Mtrans$ lies above the upper limit on the primary mass $m_1 \in [1.36, 1.60]\, \Msolar$, phase transition is not realized for both stars which constitute a normal ``neutron star-neutron star'' (NS-NS) binary. Despite bearing separate $\La(M)$ relation on the massive hybrid branch (Fig.~\ref{fig:k2-Lam-set1}) that might be probed in post-merger observables, this hybrid EoS yields $\tilde{\La} (\Mchirp)$ behavior indistinguishable from that based on purely-hadronic EoS. Note that DBHF predicting a narrow range of $\tilde{\La} (\Mchirp=1.188\, \Msolar) \in [809.8, 816.3]$ is too stiff to be compatible with the LIGO constraint $\leq 800$ ($\tilde{\La}=300_{-190}^{+500}$)~\cite{LIGO:2018wiz}, which remains unaltered if there is a late onset of phase transition.

\noindent $\bullet$ Panel (b):

$\ntrans/n_0=2.5$ $(\Mtrans=1.49\,\Msolar)$, $\De\ep/\etrans=0.7$.

In this case the onset for phase transition is within the primary mass range $m_1 \in [1.36, 1.60]\, \Msolar$, thus depending on the mass ratio there can be ``neutron star-neutron star'' (NS-NS) or ``neutron star-hybrid star'' (NS-HS) binaries. This is manifested in two separate bands of $\tilde{\La} (\Mchirp)$, of which the upper one representing NS-NS binaries exhibits $\tilde{\La} (\Mchirp=1.188\, \Msolar)$ identical to that in panel (a), whereas the lower one refers to NS-HS binaries of which the heavier star harbors a quark core ($m_1 > \Mtrans$) inducing smaller $\La_{1}$, bringing down the weighted-average $\tilde{\La}$ at a given chirp mass. For this slightly broader band, $\tilde{\La} (\Mchirp=1.188\, \Msolar) \in [652.3, 668.1]$.

\begin{table}[htb]
\begin{center}
\begin{tabular}{c|cccc}
\hline \\[-2ex]
 & $\left(\frac{\ntrans}{n_0}, \frac{\De\ep}{\etrans} \right)$ & $\Mtrans$  & $R_{1.4}$/km & $\tilde{\La}_{1.188}$\\[0.5ex]
\hline \\[-2ex]
SFHo & (3.0, 0.3) & $1.31\,\Msolar$ & 11.73 & [354.1, 369.7] \\
+ CSS & (2.5, 0.6) & $1.01\,\Msolar$ & 10.67 & [158.7, 185.4]\\
 ($\cQMsq$ & (2.0, 0.9) & $0.68\,\Msolar$ & 10.09 & [115.3, 116.5]\\
 $=1$) & (1.0, 3.1) &$0.20\,\Msolar$& 8.78& [66.51, 69.45] \\[0.5ex]
\hline \\[-2ex]
DBHF & (3.0, 0.4) & $1.78\,\Msolar$ & 13.41 & [809.8, 816.3] \\
+ CSS & (2.5, 0.7) & $1.49\,\Msolar$ & 13.41 & [809.8, 816.3](N-N)\\
  & &  &  & [652.3, 668.1](N-H)\\
 ($\cQMsq$ & (2.0, 1.0) & $1.13\,\Msolar$ & 11.02 & [171.2, 220.2]\\
 $=1$) & (1.0, 3.2) &$0.34\,\Msolar$& 8.96& [67.71, 70.54] \\[0.5ex]
\hline \\[-2ex]
 & $\left(B, \cQMsq \right)$ &  & $R_{1.4}$/km & $\tilde{\La}_{1.188}$\\[0.5ex]
 \hline \\[-2ex]
SQS & (186 MeV, 1)&& 8.09 & [64.97, 68.08] \\[0.5ex]
\hline \\[-2ex]
SFHo & & & 11.97 & [388.8, 392.2]\\
\hline \\[-2ex]
DBHF & & & 13.41 & [809.8, 816.3]\\
\hline
\end{tabular}
\end{center}
\caption{Tidal properties of EoSs used in Fig.~\ref{fig:Lam-Mchirp-pt}, panels (a)--(d); (N-N) and (N-H) represent the two separate bands in panel (b), ``neutron star-neutron star'' and ``neutron star-hybrid star'' binaries, respectively (see text for discussion). Related $M(R)$ and $\La(M)$ relations can be found in Figs.~\ref{fig:MR-curves-1} and \ref{fig:k2-Lam-set1}.
}
\label{tab:Lamwt_panel_d}
\end{table}

\noindent $\bullet$ Panel (c):

\begin{figure*}[htb]
\parbox{0.5\hsize}{
\includegraphics[width=\hsize]{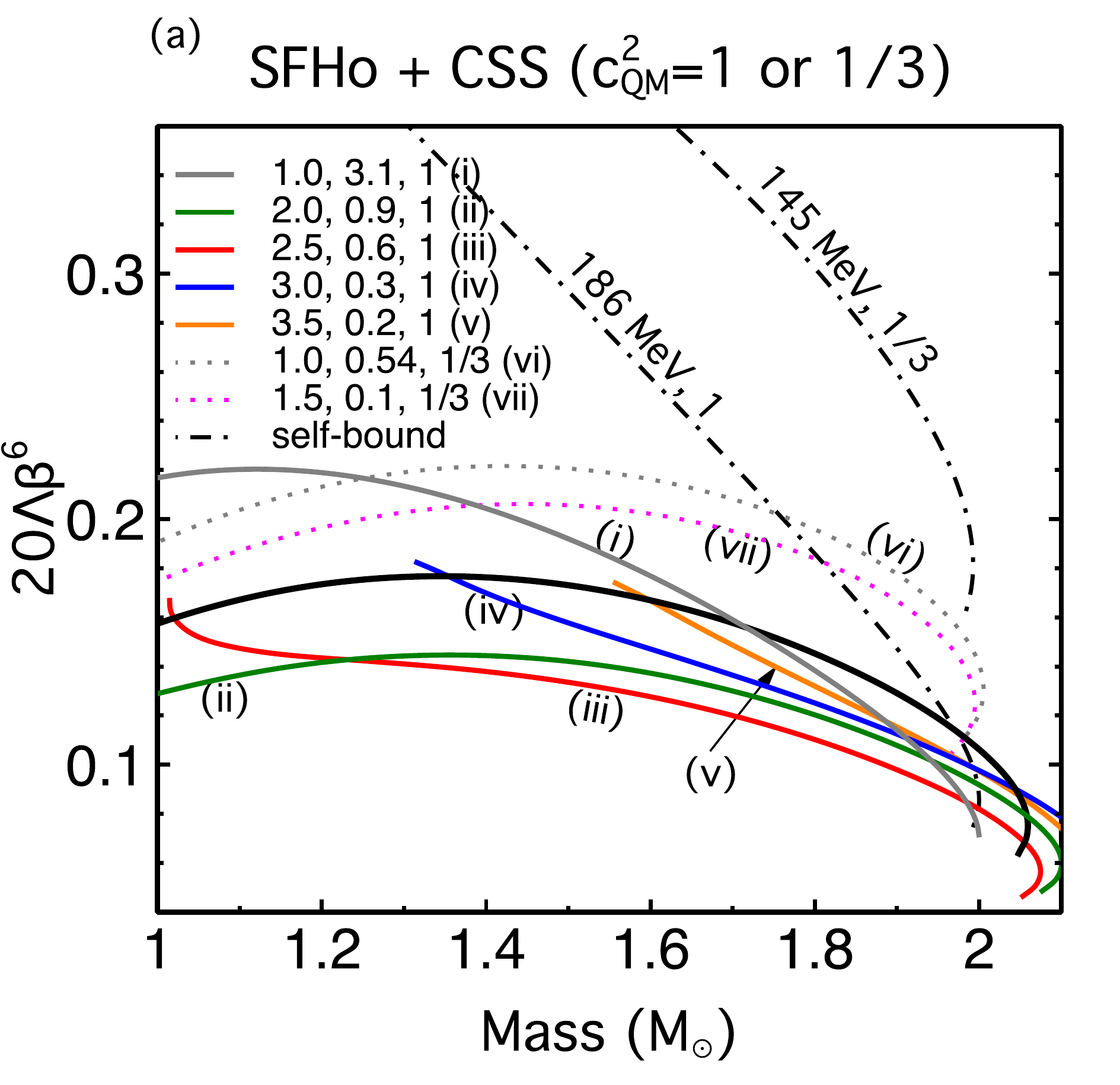}\\[-2ex]
}\parbox{0.5\hsize}{
\includegraphics[width=\hsize]{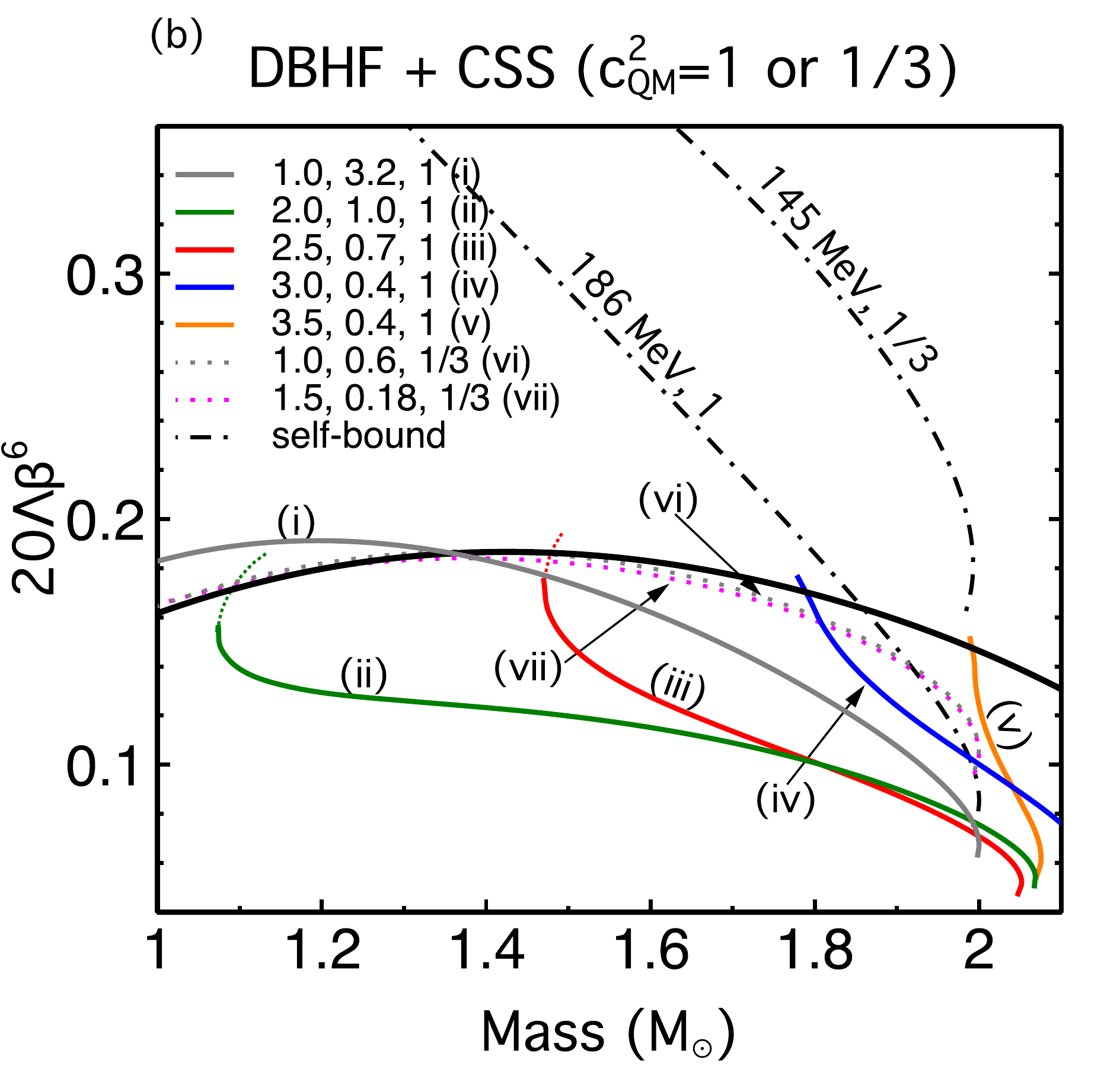}\\[-2ex]
}
\caption{(Color online) For SFHo/DBHF + CSS EoSs with selected phase transition parameters ($\ntrans$, $\De\ep/\etrans$, $\cQMsq$), variation in the quantity $k_2 \beta\propto \La\be^6$ is shown (we choose the renormalizing factor 20 following Ref.~\cite{Steiner:2015aea}). Taking into account hybrid stars with sharp phase transition, the assumption $\La\beta^6\simeq\text{constant}$ in the mass range relevant to BNS~\cite{De:2018uhw} becomes inapplicable.}
\label{fig:k2beta-M}
\end{figure*}

$\ntrans/n_0=2.0$ $(\Mtrans=1.13\,\Msolar)$, $\De\ep/\etrans=1.0$,

represented by the broader (green) band with $\tilde{\La} (\Mchirp=1.188\, \Msolar) \in [171.2, 220.2]$;

$\ntrans/n_0=1.0$ $(\Mtrans=0.34\,\Msolar)$, $\De\ep/\etrans=3.2$,

represented by the narrower (gray) band with $\tilde{\La} (\Mchirp=1.188\, \Msolar) \in [67.71,70.54]$; it corresponds to the minimal $R_{1.4}= 8.96\, \rm{km}$ (see Fig.~\ref{fig:MR-curves-1}) and $\La_{1.4}\approx60$ (Fig.~\ref{fig:lam_bar-contour}).

Also presented on this panel is the bottom (dark red) band calculated from the ``maximally compact EoS'' for self-bound SQSs~\cite{Lattimer:2012nd,Lattimer:2015nhk}, by setting coefficients in \Eqn{eqn:eos_bare} to be $B=186 \,\rm{MeV}$, $\cQMsq=1$. It gives rise to the smallest possible $R_{1.4}= 8.09\, \rm{km}$ (see Fig.~\ref{fig:MR-curves-1}), and predicts $\tilde{\La} (\Mchirp=1.188\, \Msolar) \in [64.97,68.08]$.

\noindent $\bullet$ Panel (d): 
purely-hadronic SFHo EoS, SFHo + CSS with selected parameter values, and maximally compact SQS.
We list tidal properties in Table~\ref{tab:Lamwt_panel_d}.
Note that for $\ntrans/n_0=3.0$ $(\Mtrans=1.31\,\Msolar)$, $\De\ep/\etrans=0.3$ (the second band (blue) from the top), onset mass for phase transition is within the range of the secondary mass $m_2 \in [1.17, 1.36]\, \Msolar$, thus depending on the mass ratio there can be `neutron star-hybrid star''  (NS-HS) or ``hybrid star-hybrid star'' (HS-HS) binaries.

A key finding here is a signature of strong phase transition by tidal effects if distinct separation between allowed ranges of $\tilde\La$ were to be observed in multiple merger events. It was also suggested that for normal nuclear matter there is a quasi-universal correlation $\tilde{\La} \propto (G\Mchirp/\Rtyp)^{6}$ in the relevant mass range assuming a common radius \cite{De:2018uhw,Zhao:2018nyf}. To estimate to what extent the correlation is affected by phase transitions, we plot the quantity $20\,\La \be^6$ as a function of mass for a variety of hybrid EoSs in Fig.~\ref{fig:k2beta-M}, and find that its value can vary by as large as $\sim20\%$. This variation combined with much wider spread in the hybrid star radii set bounds on the $(\La_2 \,q^{6}/\La_1)$ parameter~\cite{De:2018uhw,Zhao:2018nyf} in a binary system.

\begin{figure*}[htb]
\parbox{0.34\hsize}{
\includegraphics[width=\hsize]{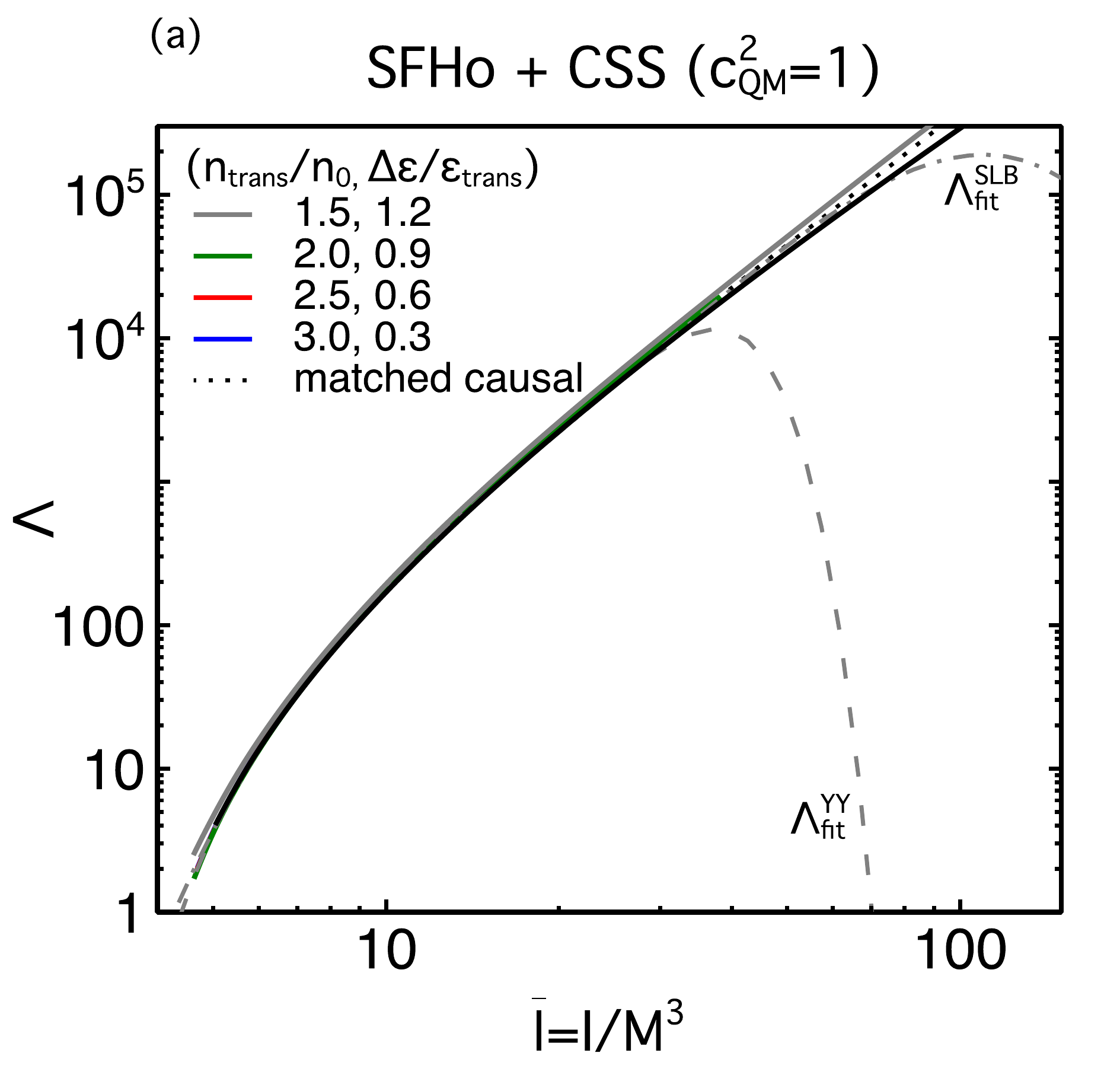}\\[-2ex]
}\parbox{0.34\hsize}{
\includegraphics[width=\hsize]{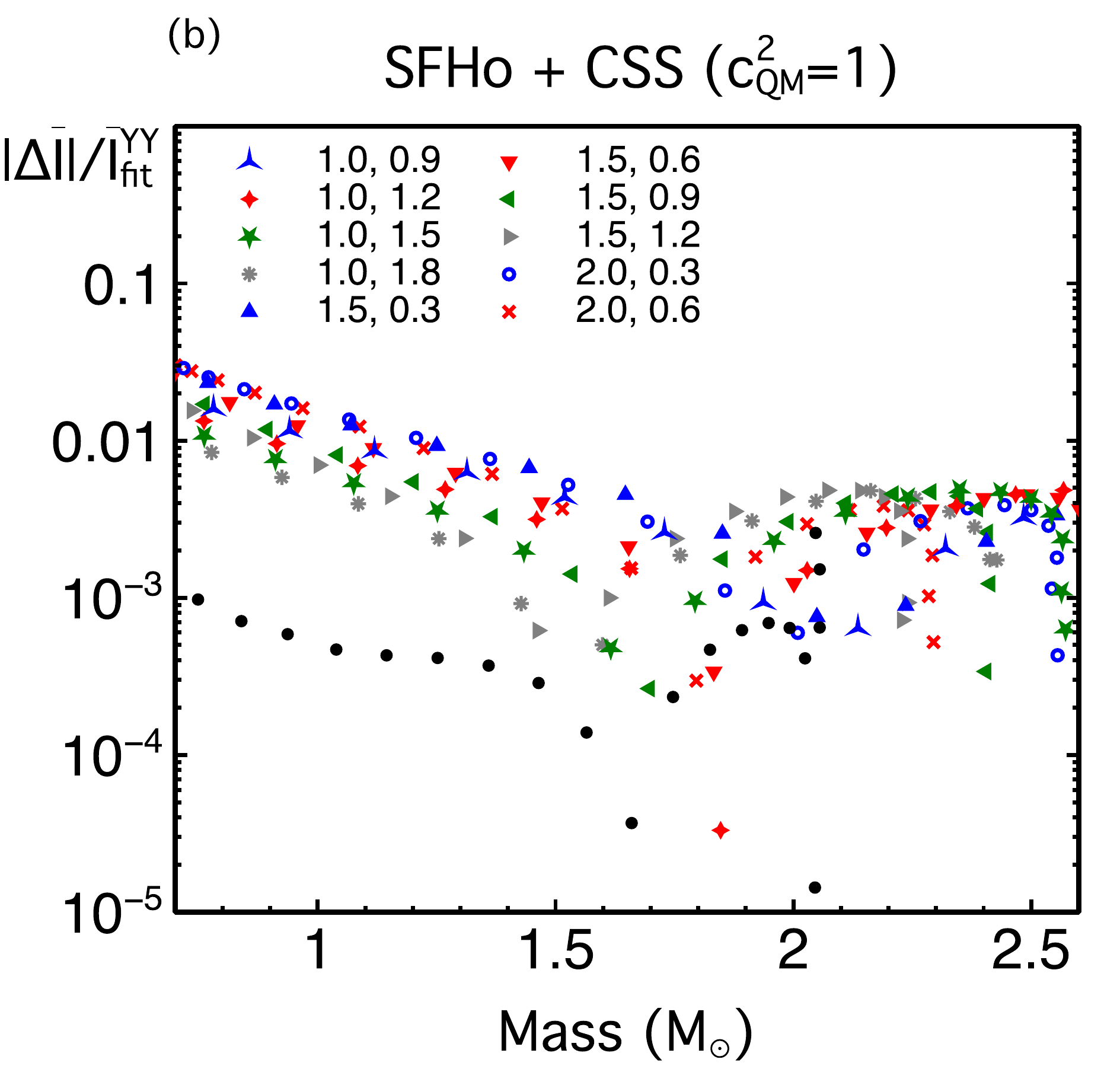}\\[-2ex]
}\parbox{0.34\hsize}{
\includegraphics[width=\hsize]{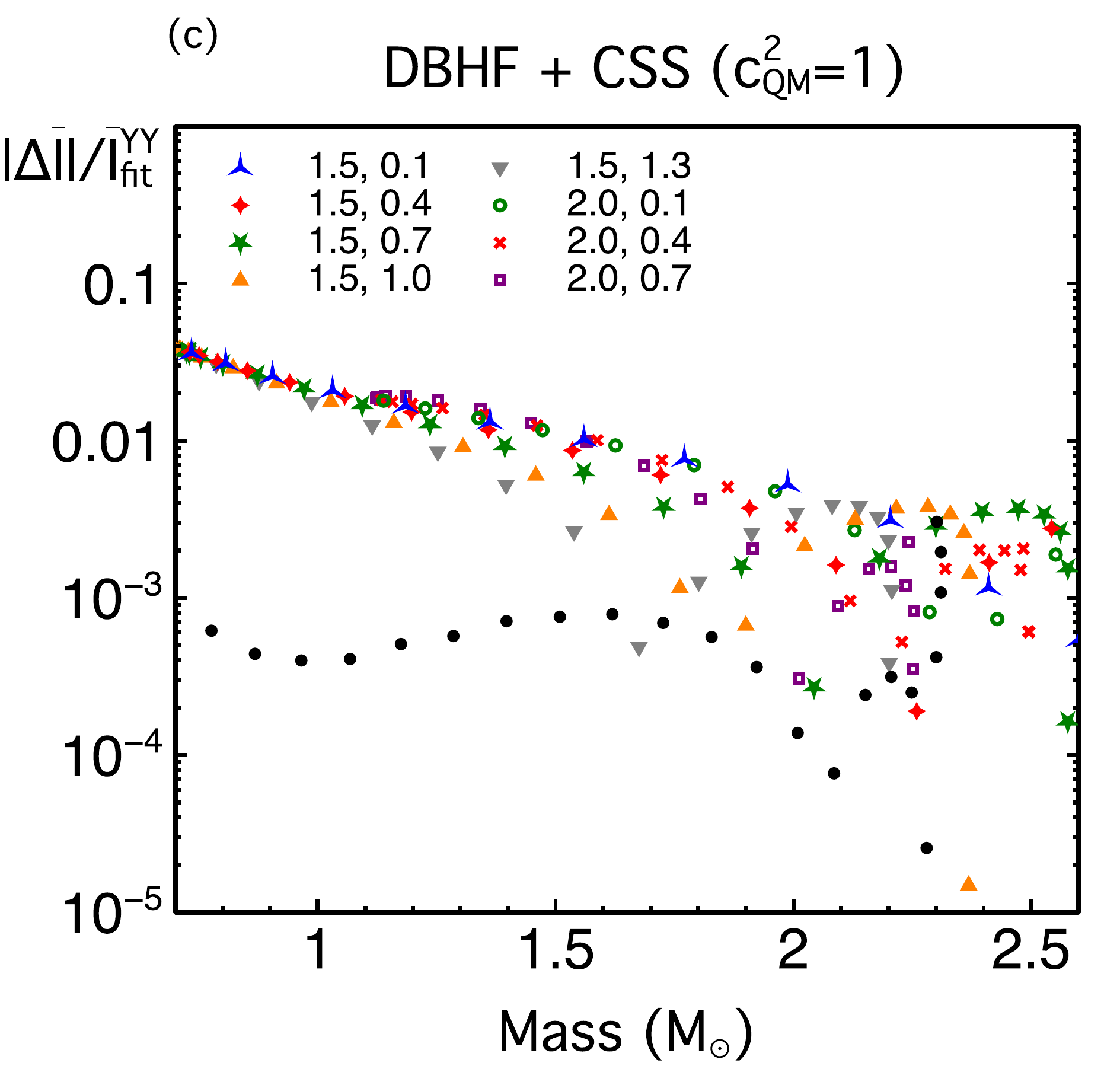}\\[-2ex]
}
\caption{(Color online) Panel (a): I-Love correlation $\Lambda (\bar{I})$ for selected SFHo + CSS EoSs with $\ntrans$ ranging from 1.5 to 3.0 $n_0$; black solid curve refers to SFHo only. Two fit functions are adapted from~\cite{Yagi:2016bkt} (gray dashed) and \cite{Steiner:2015aea} (gray dash-dotted); ``matched causal'' stands for continuation to stiffest linear EoS from nuclear matter at saturation density without any energy discontinuity ($\ntrans=n_0$, $\De\ep=0$, $\cQMsq=1$)~\cite{VanOeveren:2017xkv}. For light neutron stars $\lesssim0.89\,\Msolar$ ($\bar{I}\gtrsim35$), deviation from the analytical fits was discussed in Ref.~\cite{Silva:2016myw}. For typical neutron stars $\gtrsim1.0\,\Msolar$, discrepancies are negligible. Panels (b) and (c): relative error $|\De \bar{I}|/\bar{I}_{\rm fit}^{\rm YY}$ plotted against the neutron star mass for multiple SFHo/DBHF + CSS EoSs with $\ntrans\lesssim 2 n_0$; black dots refer to purely-hadronic stars.}
\label{fig:lam-I-pt}
\end{figure*}

\subsection{I-Love universal relation and moment of inertia}
\label{sec:correlation}

In the slow-rotation approximation, the dimensionless moment of inertia $\bar{I}= I/M^3$ and the dimensionless tidal deformability $\La=\la/M^5$ of neutron stars without phase transitions are related by EoS-independent universal relation to within $1\%$~\cite{Yagi:2013awa,Yagi:2014qua}, for which we apply analytical fit functions in Ref.~\cite{Yagi:2016bkt} (labelled ``YY'') and Ref.~\cite{Steiner:2015aea} (labelled ``SLB''); see \Eqn{eqn:I_Lam_fit}.

\bea
&\ln \bar{I}_{\rm fit}^{\rm YY}&=1.496+0.05951\ln \La+0.02238 (\ln \La)^2 \nn
&&-0.0006953(\ln \La)^3+8.345\times10^{-6}(\ln \La)^4, \nn
&\ln \bar{I}_{\rm fit}^{\rm SLB}&=1.417+0.0817\ln \La+0.0149(\ln \La)^2 \nn
&&+0.000287(\ln \La)^3-3.64\times10^{-5}(\ln \La)^4.
\label{eqn:I_Lam_fit}
\eea

In panel (a) of Fig.~\ref{fig:lam-I-pt}, we plot $\La$ and $\bar{I}$ for selected SFHo + CSS hybrid EoSs with high transition densities ($\ntrans\gtrsim2.0\, n_0$), and compare them to the two fit functions. As can be seen from the plot, the relative error in $\La (\bar{I})$ remains small. In fact, the largest error bands $|\De \bar{I}|/\bar{I}_{\rm fit}^{\rm YY}$ for typical masses ($1.1 - 1.6 \,\Msolar$) are acquired when transition densities are low ($\ntrans\lesssim 2.0 \,n_0$), as depicted in panels (b) and (c). For both SFHo and DBHF nuclear matter that are applied in the parametrization, the maximal deviation is around $|\De \bar{I}|/\bar{I}_{\rm fit}^{\rm YY}\approx2\%$, which doubles the value for all EoSs without sharp transitions. 

\begin{table}[htb]
\begin{center}
\begin{tabular}{c c@{\quad} c}
\hline \\[-2ex]
Parameter & Limits \\[0.5ex]
\hline \\[-2ex]
$\nb^{\rm trans,1} (\rm{fm}^{-3})$          & $[0.16,\,0.80]$ \\[0.5ex]
$\De \ep_1/\ep_1$          & $[0,\, 1.5]$  \\[0.5ex]
$\De \ep_{\rm 2SC}/\ep_1$          & $[0,\, 2.0]$ \\[0.5ex]
$\De \ep_2/\De \ep_1$ & $[0,\, 1.0]$ \\[0.5ex]
$s_1=c_{\rm s,2SC}^2$    & $[0,\,1.0]$  \\[0.5ex]
$s_2=c_{\rm s,CFL}^2$ & $[0,\,1.0]$ \\[0.5ex]
\hline
\end{tabular}
\end{center}
\caption{Ranges of the input parameters for quark matter EoS with nuclear $\to$ 2SC and 2SC $\to$ CFL phase transitions (\Eqn{eqn:seq_EoS}) that are applied in the MC calculation.}
\label{tab:mc-paras}
\end{table} 

\subsection{Monte Carlo analysis}
 
In addition to analyzing how phase transitions can modify the tidal deformability and the moment of inertia, it is helpful to assess how likely these modifications are using a Monte Carlo (MC) simulation. Following Ref.~\cite{Alford:2017qgh}, we then parametrize sequential nuclear $\to$ 2SC $\to$ CFL  sharp transitions to two quark-matter phases (see \Eqn{eqn:seq_EoS}). There are in total six independent parameters as listed in Table~\ref{tab:mc-paras}: i) $\nb^{\rm trans,1}$ is the baryon density at the transition pressure $p_1=\ptrans$ for nuclear $\to$ 2SC; ii) $\De\ep_1$ and $\De\ep_2$ are the two energy density jumps at two phase transitions; iii) $\De\ep_{\rm 2SC}$ is the width of the 2SC phase; and iv) $s_1$ and $s_2$ are sound-speed squared for the 2SC and CFL phases respectively. We randomly select from these six parameters using the ranges outlined in Table~\ref{tab:mc-paras}, and remove all configurations where the maximum mass is below $2\,\Msolar$. For each point in the six-dimensional parameter space, we compute the gravitational mass, the radius, the moment of inertia, and the tidal deformability as a function of the central pressure in the neutron star. We also compute the moment of inertia using the Yagi-Yunes I-Love correlation $\bar{I}_{\rm fit}^{\rm YY}$ in \Eqn{eqn:I_Lam_fit} for comparison with the exact results.

\begin{figure*}[htb]
\parbox{0.5\hsize}{
\includegraphics[width=\hsize]{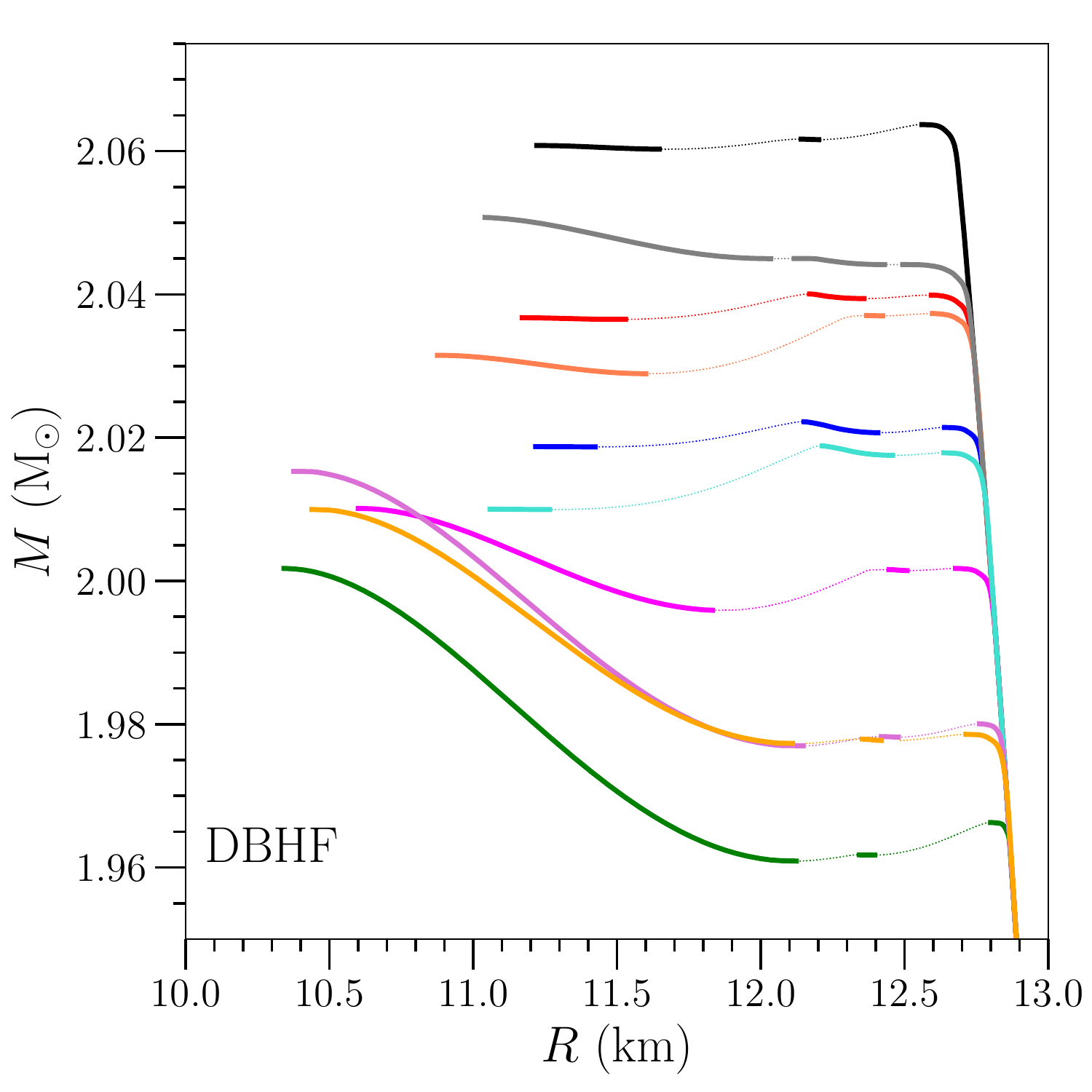}\\[-2ex]
}\parbox{0.5\hsize}{
\includegraphics[width=\hsize]{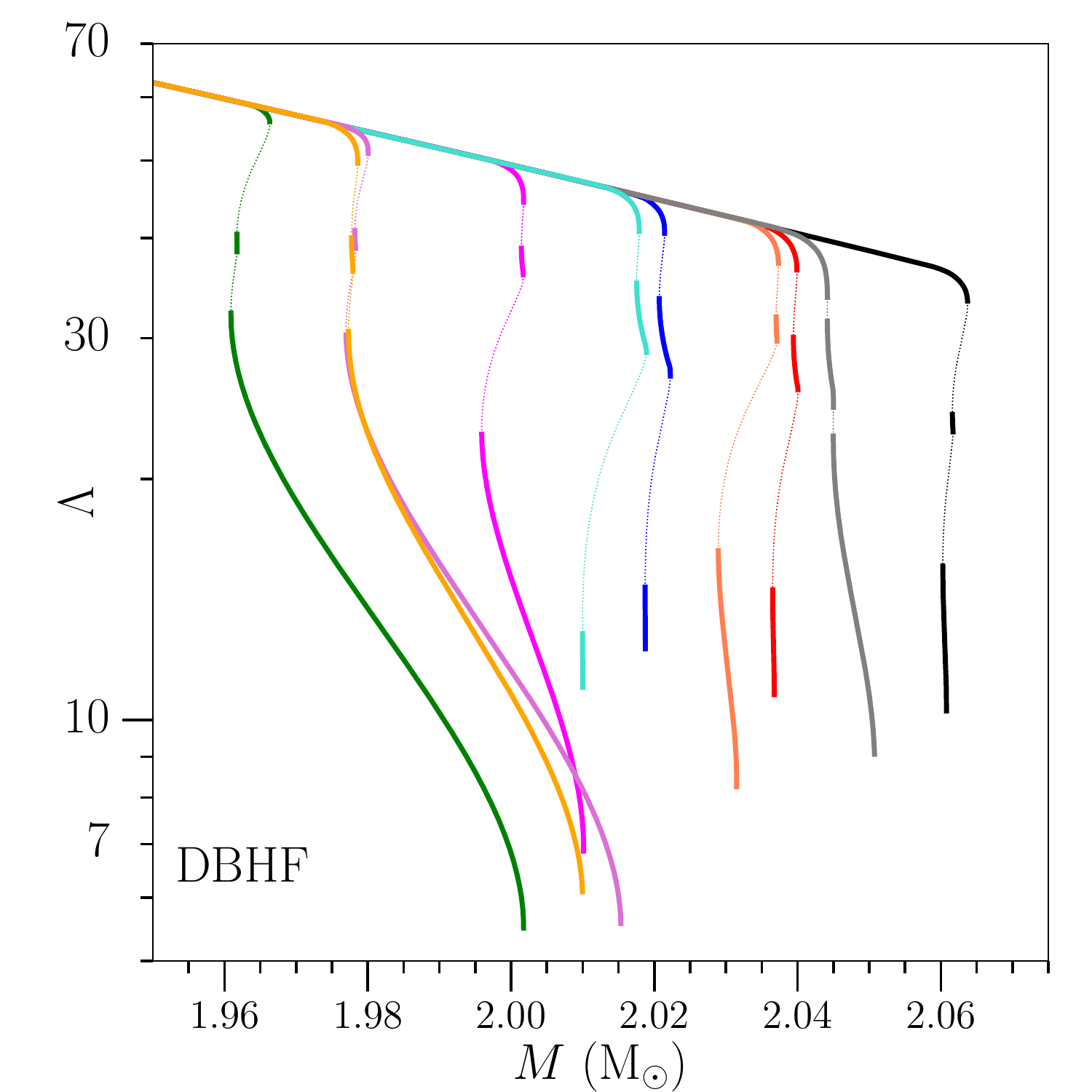}\\[-2ex]
}
\caption{(Color online)$M(R)$ and $\La(M)$ relation for triplet configurations with three maxima of mass on the curves; DBHF EoS is applied for the hadronic part. Corresponding EoS parameter values are listed in Table~\ref{tab:triplets-paras} (from low to high onset mass $\Mtrans$).}
\label{fig:mc-mr-mlam}
\end{figure*}

\begin{table}[htb]
\begin{center}
\begin{tabular}{c|c|c|c|@{\quad}c@{\quad}|@{\quad}c}
\hline 
&&&&&\\[-2ex]
$\nb^{\rm trans,1} (\rm{fm}^{-3})$ & $\De \ep_1/\ep_1$  & $\De \ep_{\rm 2SC}/\ep_1$  & $\De \ep_2/\De \ep_1$ & $s_1$ & $s_2$ \\[0.5ex]
\hline  
&&&&&\\[-2ex]
 0.549 & 0.445 & 0.383 & 0.061& 0.868 &0.992 \\[0.5ex]
 0.554 & 0.409 & 0.472 & 0.055 & 0.754 &0.974 \\[0.5ex]
 0.555 & 0.430 & 0.308 & 0.070& 0.962 &0.999 \\[0.5ex]
 0.564 & 0.408 & 0.296 & 0.098& 0.981 &0.969 \\[0.5ex]
 0.572 & 0.398 & 0.425 & 0.128& 0.963 &0.971 \\[0.5ex]
 0.574 & 0.399 & 0.462 & 0.098& 0.961 &0.914 \\[0.5ex]
 0.581 & 0.387 & 0.333 & 0.092& 0.982 &0.986 \\[0.5ex]
 0.582 & 0.381 & 0.493 & 0.077& 0.909 &0.948 \\[0.5ex]
 0.584 & 0.372 & 0.492 & 0.042& 0.889 &0.921 \\[0.5ex]
 0.596 & 0.38 & 0.465 & 0.047& 0.995 &0.945 \\[0.5ex]
\hline
\end{tabular}
\end{center}
\caption{Values of sequential phase transition parameters (Table~\ref{tab:mc-paras}) that give rise to the triplet configurations in Fig.~\ref{fig:mc-mr-mlam}.}
\label{tab:triplets-paras}
\end{table} 

\begin{figure*}[htb]
\parbox{0.5\hsize}{
\includegraphics[width=\hsize]{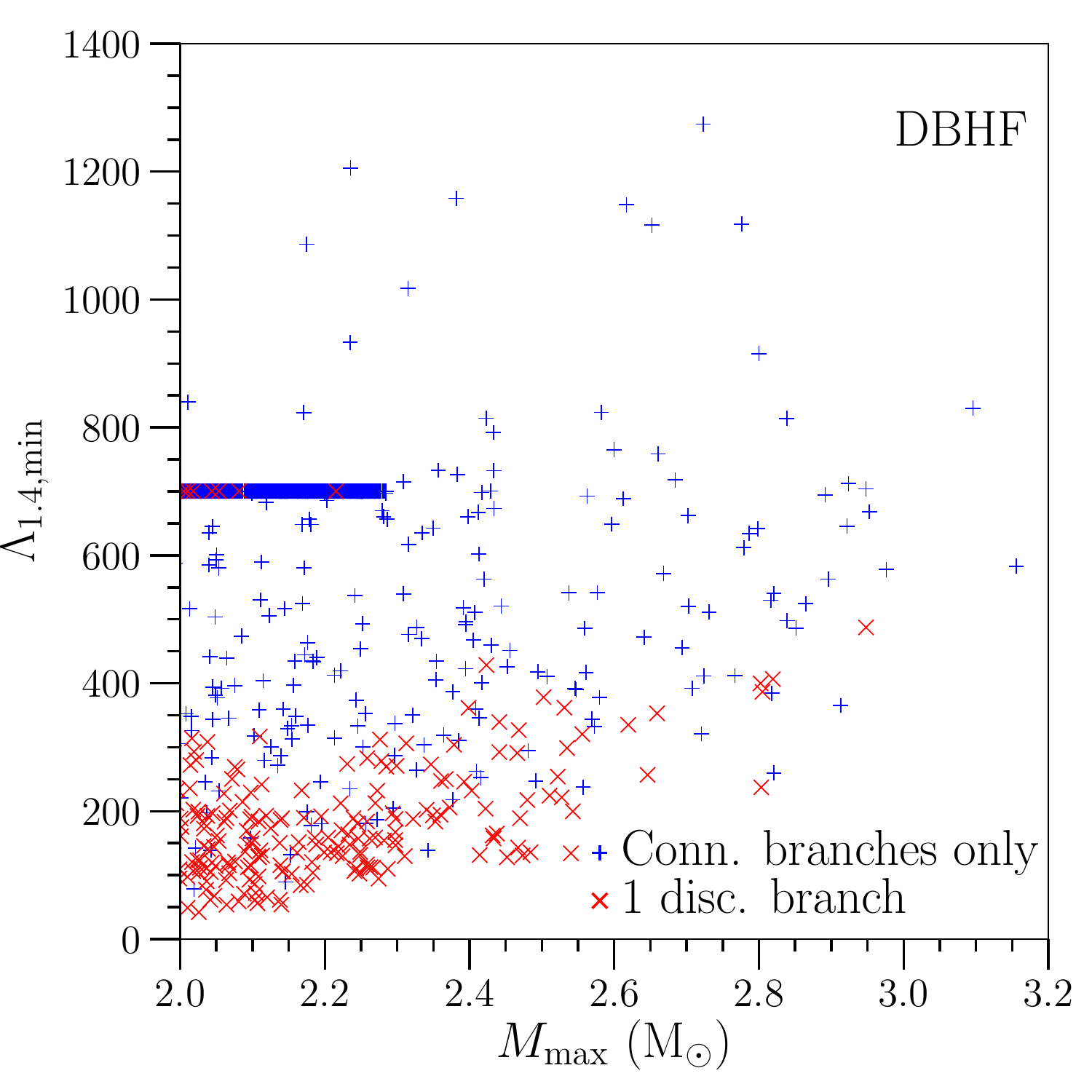}\\[-2ex]
}\parbox{0.5\hsize}{
\includegraphics[width=\hsize]{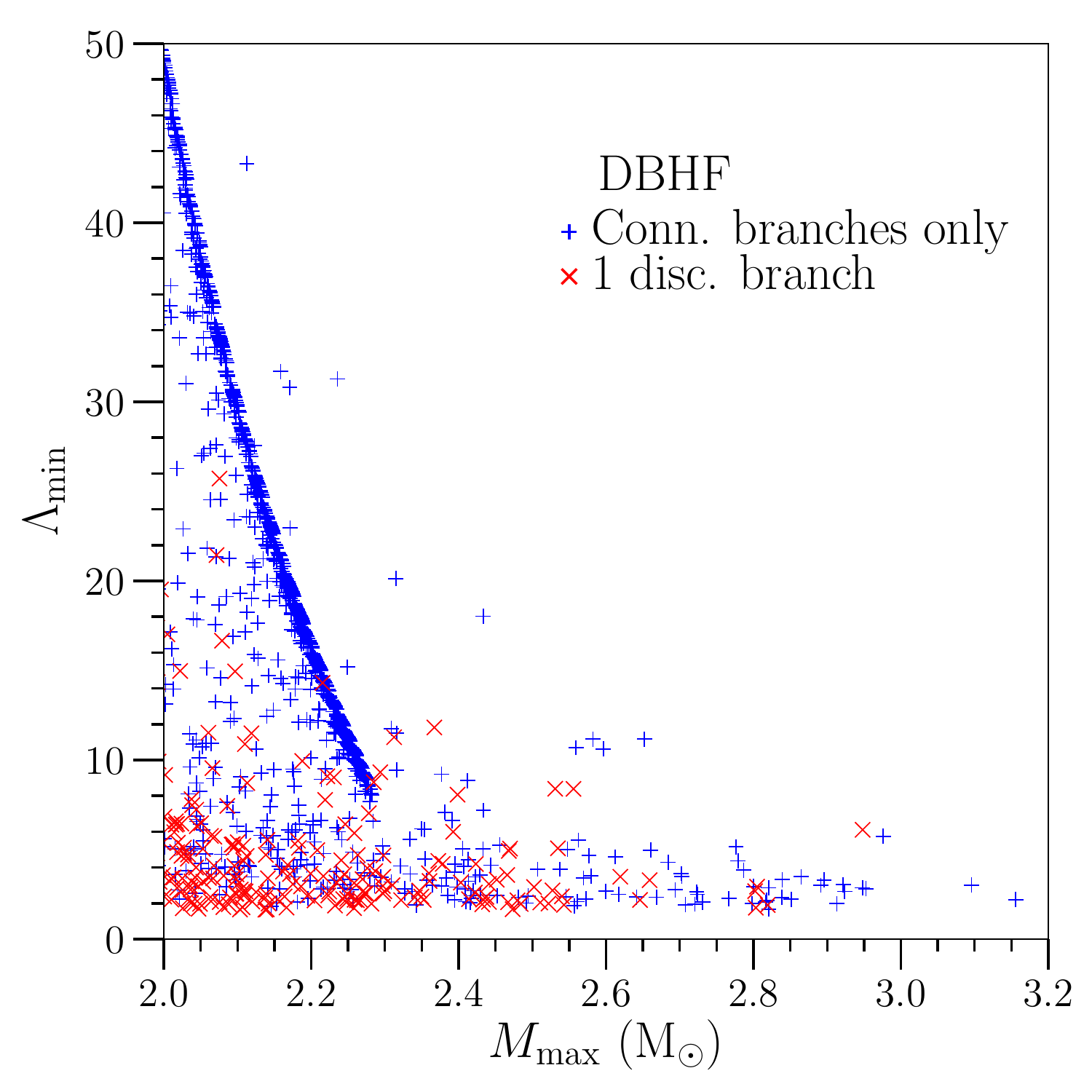}\\[-2ex]
}
\caption{(Color online) $\La_{1.4}^{\rm min} (\Mmax)$ and $\La_{\rm min} (\Mmax)$ relation for EoS parameters reported in Table~\ref{tab:mc-paras}; DBHF EoS is applied for the hadronic part. Blue pluses correspond to two connected branches (weak nuclear $\to$ 2SC $\to$ CFL transitions), and red crosses represent configurations with a disconnected branch (either nuclear $\to$ 2SC or 2SC $\to$ CFL transition being strong). Note that triplet configurations in Fig.~\ref{fig:mc-mr-mlam} barely affect $\La_{1.4}$ due to their high transition densities ($\Mtrans\gtrsim 1.8\,\Msolar>1.4\,\Msolar$), and they are from a much smaller parameter set variation that we do not show here.
}
\label{fig:mc-mmax-lam14}
\end{figure*}

\begin{figure*}[htb]
\parbox{0.5\hsize}{
\includegraphics[width=\hsize]{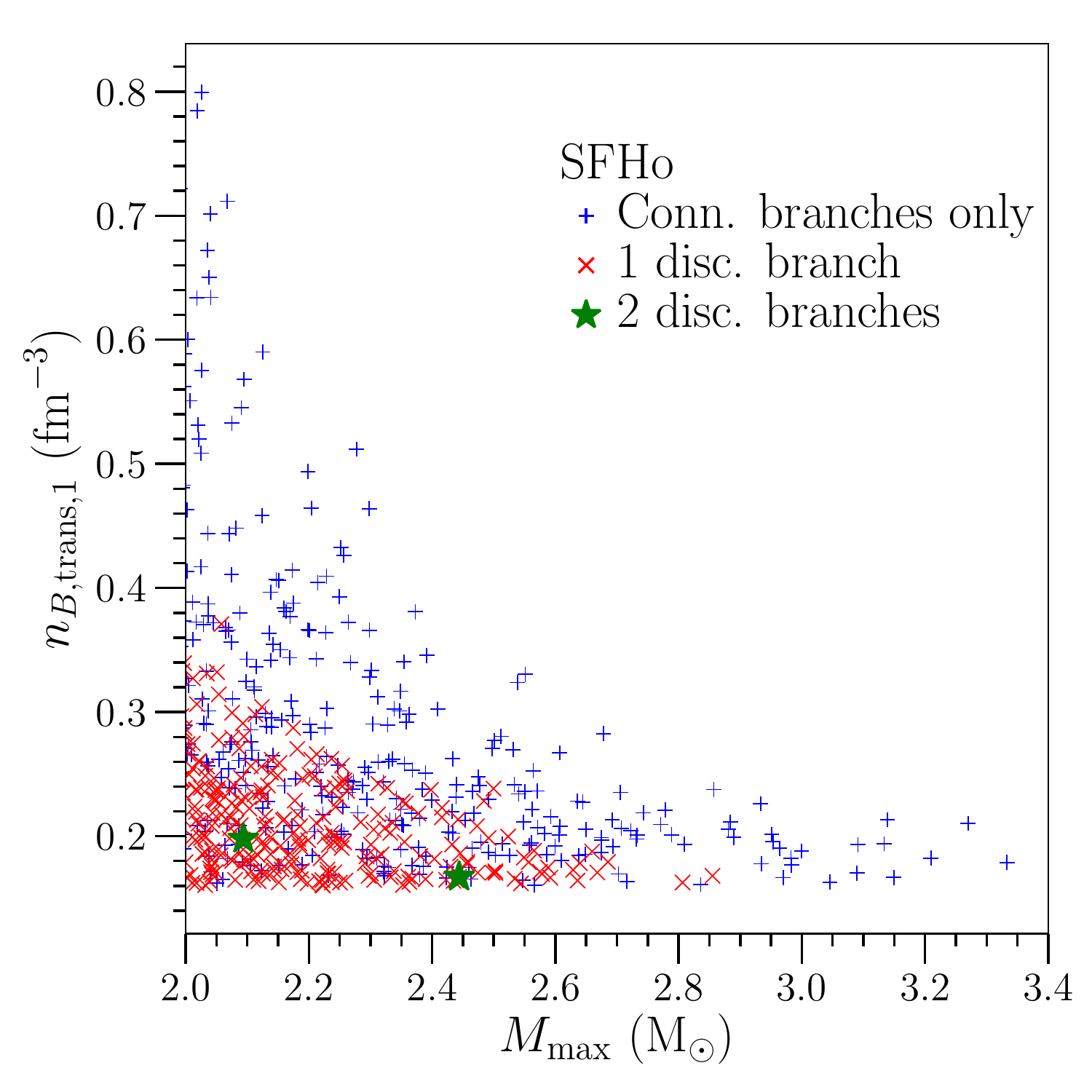}\\[-2ex]
}\parbox{0.5\hsize}{
\includegraphics[width=\hsize]{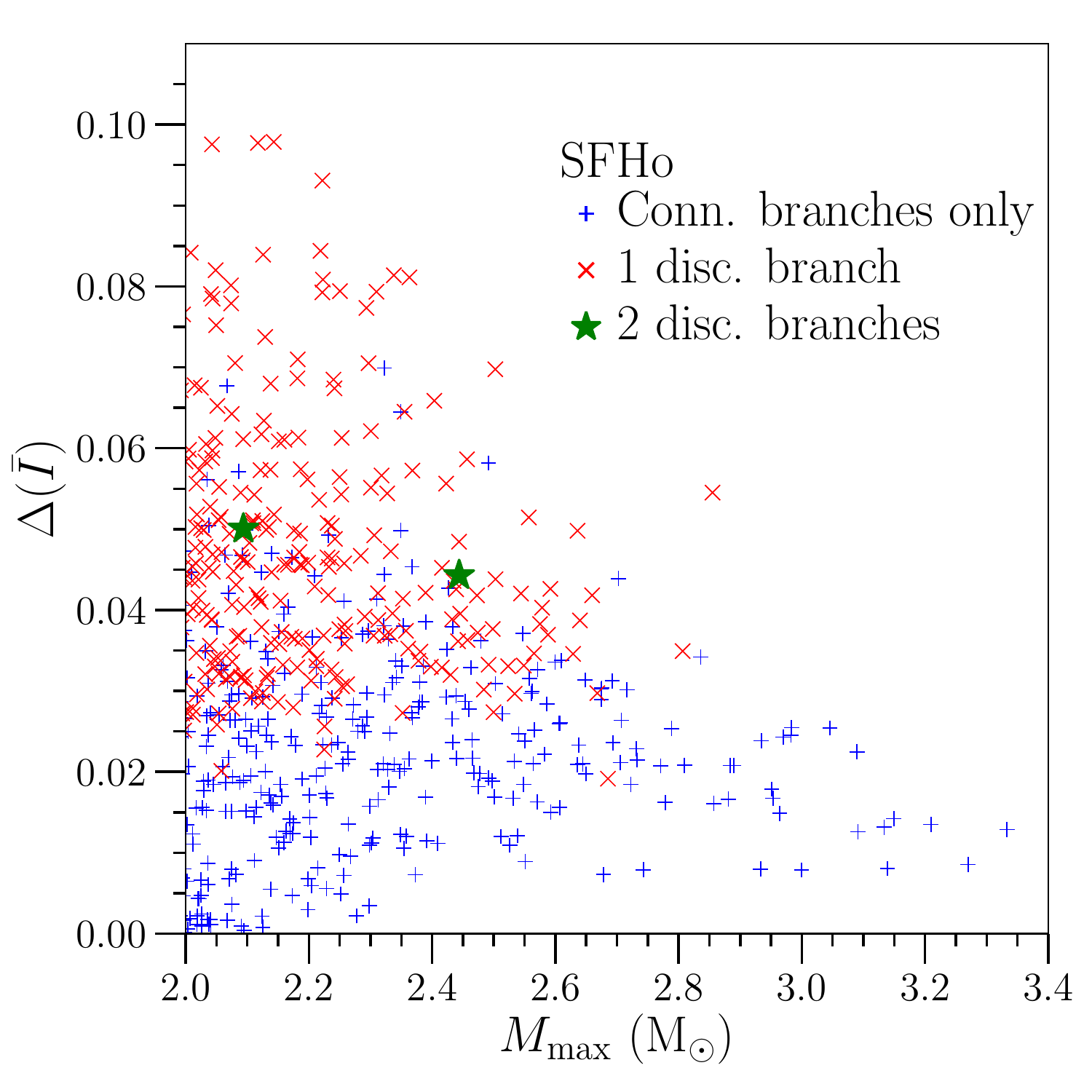}\\[-2ex]
}\\[2ex]
\parbox{0.5\hsize}{
\includegraphics[width=\hsize]{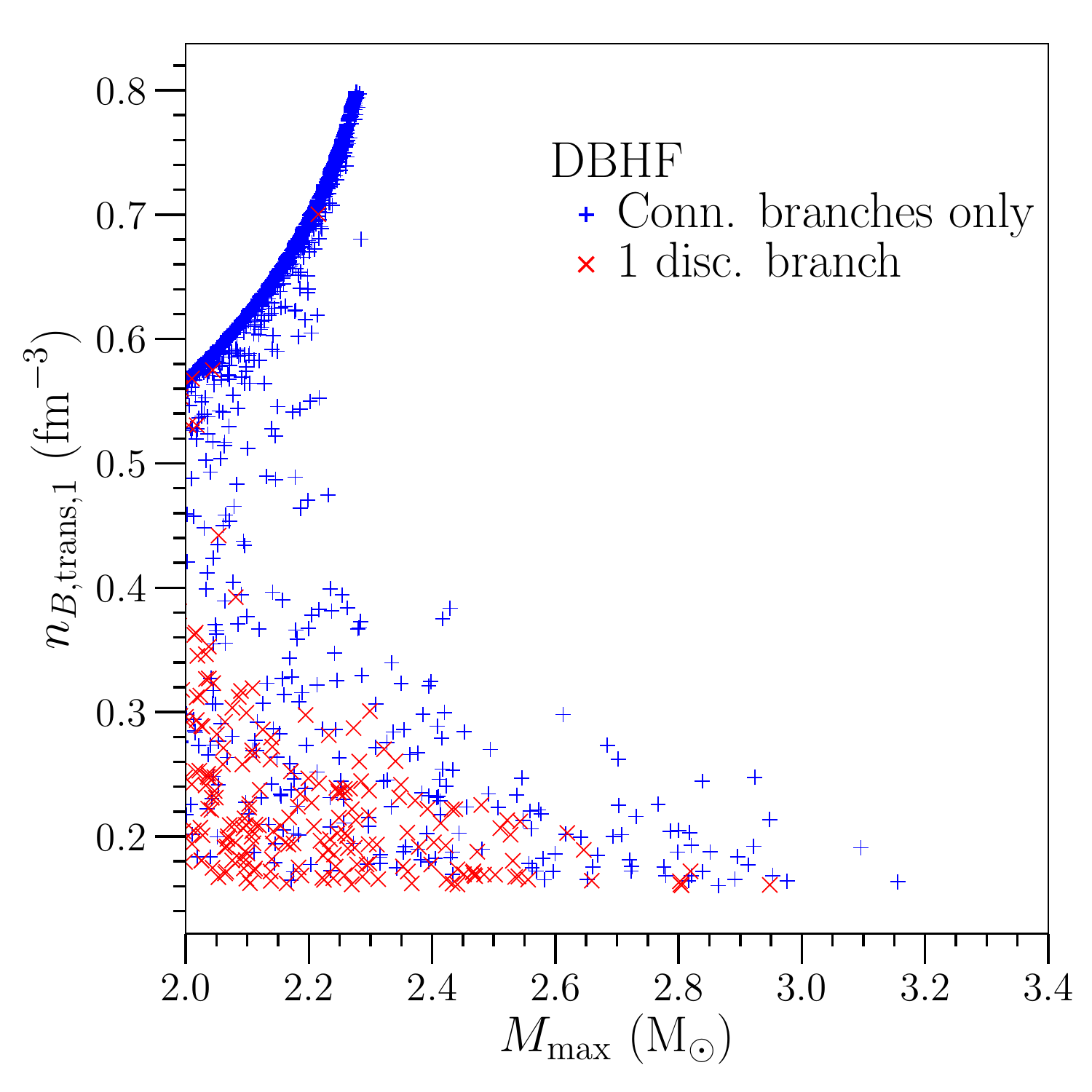}\\[-2ex]
}\parbox{0.5\hsize}{
\includegraphics[width=\hsize]{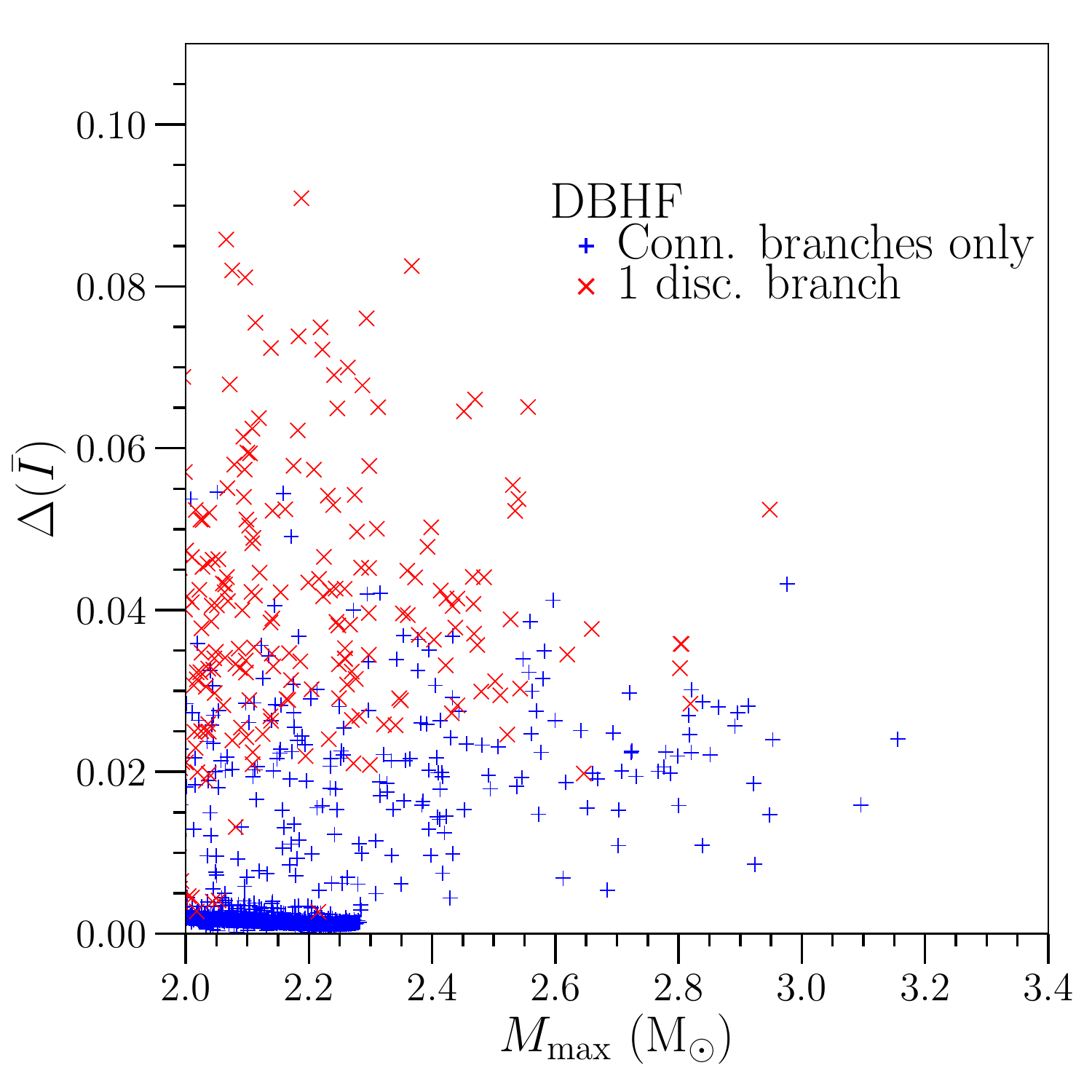}\\[-2ex]
}
\caption{(Color online) Data of transition threshold density and the maximum mass $\ntrans(\Mmax)$ and the deviation from Yagi-Yunes I-Love correlation $\De\bar{I} (\Mmax)$ for two sequential phase transition configurations. Blue pluses correspond to two connected branches (weak nuclear $\to$ 2SC $\to$ CFL transitions), and red crosses represent a disconnected branch (either nuclear $\to$ 2SC or 2SC $\to$ CFL transition being strong). For SFHo we also show examples with both transitions being strong (two disconnected branches) for which the parameter space is rare.
}
\label{fig:mc-mmax-nb-delta} 
\end{figure*}

Our results are summarized in Figs.~\ref{fig:mc-mr-mlam}, \ref{fig:mc-mmax-lam14} and \ref{fig:mc-mmax-nb-delta}. ``Triplet configurations'' that are able to support $\Mmax\gtrsim 2\,\Msolar$ necessitate the hadronic EoS being rather stiff and the first phase transition onset density $\nb^{\rm trans,1}$ very close to the central density of a $2\,\Msolar$ star. Fig.~\ref{fig:mc-mr-mlam} displays the mass-radius relation and the tidal deformability as a function of mass for such configurations, where the stiffer DBHF EoS is applied. The high transition density $\nb^{\rm trans,1}$ from hadrons to quarks implies that for typical component masses ($1.1 - 1.6 \,\Msolar$) observed in a binary, quark matter is nonexistent even in the densest cores. As a result, during the pre-merger stage tidal deformabilities $\La_1$ and $\La_2$ (and other observables) are entirely determined by the nuclear matter EoS. By contrast, the post-merger remnant (if it survives as a supramassive or hypermassive neutron star) might attain densities above the phase transition threshold, and therefore gravitational-wave signatures during the post-merger stage potentially probe the densest quark matter \cite{Most:2018eaw,Bauswein:2018bma}.

If the onset mass for phase transition lies within the range of component masses, interpretation of the pre-merger detection can differ significantly (see Sec.~\ref{sec:bns-obs}). In Fig.~\ref{fig:mc-mmax-lam14} we show $\La_{1.4}^{\rm min} := \bm{\min} \{ \La_{1.4}^{\rm hadronic}, \La_{1.4}^{\rm hybrid} \}$ obtained for EoSs with sequential phase transitions (excluding triplet configurations), and also tidal deformabilities of the maximum-mass stars $\La_{\rm min}$. We find that hybrid EoSs with a strong nuclear $\to$ 2SC transition followed by a weak 2SC $\to$ CFL transition can lead to $\La_{1.4}^{\rm min}\approx 40$. This is a possible scenario, given the smallness of the baryon number density discontinuities between different phases of quark matter compared to a larger jump expected at the hadron/quark interface. It in addition requires $\nb^{\rm trans,1} =1\sim2\, n_0$, and the tidal Love numbers $k_2$ and radii for corresponding $1.4\,\Msolar$ stars are displayed in upper panels of Fig.~\ref{fig:lam_bar-contour}. In general, there is a cut-off on the threshold $\nb^{\rm trans,1}$ to ensure $\Mmax\geq2\,\Msolar$ if disconnected branches are present; see (red) crosses in Fig.~\ref{fig:mc-mmax-nb-delta}. We find that sequential phase transitions can augment the deviation from I-Love universal relation (Eq.~\eqn{eqn:I_Lam_fit}) up to $|\De \bar{I}|/\bar{I}_{\rm fit}^{\rm YY}\approx9\%$, in excess of those based on purely-hadronic EoSs and self-bound quark EoSs ($\lesssim 1\%$) \cite{Steiner:2015aea,Yagi:2016bkt}, and hybrid EoSs with a single phase transition ($\lesssim 3\%$) \cite{Paschalidis:2017qmb,Wei:2018dyy}.

\section{Comparison with other works}
\label{sec:cmp}

In the normal hadronic matter scenario, EoSs with very steep slope in the symmetry
energy that relate to large stellar radii and large tidal deformabilities are ruled out by observations, and predictions on the range of $\La$ for typical masses are model-dependent. Examples include relativistic mean-field (RMF) models~\cite{Fattoyev:2017jql}, models based on chiral effective field theory (EFT) \cite{Lim:2018bkq}, piecewise polytropic EoSs~\cite{Annala:2017llu}, and other parameterizations of the specific energy and symmetry energy \cite{Tews:2018iwm,Zhang:2018vrx} et cetera. In this work, we choose two representative (soft/stiff) hadronic EoSs that are in fairly good agreement with currently available constraints.

Tidal deformation of compact stars with strange quark matter has been studied in various models \cite{Lau:2017qtz, Annala:2017tqz,Paschalidis:2017qmb,Nandi:2017rhy,Alvarez-Castillo:2018pve,Burgio:2018yix,Gomes:2018eiv}. Refs. \cite{Most:2018hfd,Tews:2018iwm,Sieniawska:2018zzj,Zhao:2018nyf,Christian:2018jyd} utilized phenomenological parameterization for a first-order phase transition similar to the CSS framework in the present paper; the recent work by Montana et al.~\cite{Montana:2018bkb} focused on constraining the ``twin-star'' configurations, of which the Maxwell construction followed the same procedure. Our results are consistent with most of the previous studies on a single phase transition (except for the lower limit $\La_{1.4}>35.5$ at 2-$\sigma$ level reported in Ref.~\cite{Most:2018hfd}), giving the conservative lower bound $\La_{1.4}\simeq60$ which is much smaller than that in the purely-hadronic case. Furthermore, we performed calculations for sequential nuclear $\to$ 2SC $\to$ CFL phase transitions with reasonable range of parameter values, which push down the minimum value of $\La_{1.4}$ to around 40, and we for the first time clearly identify these minima are associated with a strong (large energy density discontinuity) phase transition around $1\sim2 \,n_0$ followed by a weak (small energy density discontinuity) phase transition  at higher densities, 
not with the triplet configurations that require the first transition to set in fairly late ($\Mtrans \gtrsim 1.8\,\Msolar$; see Fig.~\ref{fig:mc-mr-mlam}). Other exotic sources in the binary system such as boson stars \cite{Sennett17}, vacuum energy~\cite{Csaki:2018fls} and dark matter \cite{Nelson:2018xtr, Ellis:2018bkr} have also been studied with regard to their tidal effects in the gravitational-wave signal, which are beyond the scope of this paper. We do not incorporate potential corrections due to the superfluid component in neutron stars as examined in Ref.~\cite{Char:2018grw}, where no crust model was included.

Apart from gravitational-wave detections, there are a plenty of analyses extracting constraints on EoS properties from electromagnetic post-merger data with the assistance of numeric relativity simulations. By relating the black hole formation time in the merger and remnant disk plus dynamic ejecta masses to the tidal parameter, a lower bound on the combined tidal deformability $\tilde{\La}\gtrsim 400$ was derived~\cite{Radice:2017lry}. To explain the kilonova, modeling the light curves and spectra of AT2017GFO with fits between the ejecta mass and source properties suggests a more conservative bound $\tilde{\La}\gtrsim197$ \cite{Coughlin:2018miv}. Ref.~\cite{Margalit:2017dij} argued that the BNS merger in GW170817 formed a short-lived hypermassive neutron star before collapsing to a black hole (which was confirmed later by separate work e.g. in Ref.~\cite{Rezzolla:2017aly}), and to be consistent with the ejected rotational energy inferred from the electromagnetic emission, an upper limit was placed on the neutron star maximum mass $\Mmax\lesssim2.17\,\Msolar$. We anticipate these bounds on $\tilde{\La}$ and $\Mmax$ from modeling to be updated for the separate class of hybrid EoSs with phase transitions once finite-temperature effects are appropriately incorporated. Our predictions based on zero-temperature EoSs already provide robust constraints on the high-density quark matter parameter space (see e.g. Fig.~\ref{fig:lam_bar-contour} and Fig.~\ref{fig:Lam-Mchirp-pt}). 

\section{Conclusions and outlook}
\label{sec:con} 

In this work we calculated the dimensionless Love number $k_2$ and dimensionless tidal deformability $\La$ for normal neutron stars, strange quark stars and hybrid stars with sharp quark/hadron phase transitions using the model-independent ``constant-sound-speed (CSS)'' parametrization. We evaluated the weighted-average tidal deformability $\tilde{\La}$ at given chirp mass $\Mchirp$ in a binary system, assuming that both stars are governed by the same equation of state. In particular, we investigated its sensitivity to phase transition parameters, hoping that future gravitational-wave detections with improved resolution could rule out certain parameter space for quark matter.

The role of speed of sound in dense matter and neutron star properties were stressed recently~\cite{Bedaque:2014sqa,Moustakidis:2016sab,Tews:2018kmu}, and we apply a generic classification of EoSs with regard to their sound-speed behavior as following:

i) $c_s^2 (p)$ is monotonically increasing and smooth, which is the case for numerous hadronic EoS models with the common feature that, tidal properties are in general mostly sensitive to the slope of symmetry energy $S_{0}$ at saturation density, the quantity $L$, or equivalently the radius of a $1.4\,\Msolar$ star $R_{1.4}$.

ii) $c_s^2 (p)$ undergoes some abrupt discontinuity which induces a singularity in its inverse. For self-bound quark stars with a bare surface, $k_2$ generally has much larger values but $\La$ is reduced due to the very small radius attained ($\La\propto R^5$). For hybrid stars with a sharp hadron-quark interface, strong discontinuities in both $k_2$ and $\La$ occur only over a small range in masses, and at high density $k_2(M)$ and $\La(M)$ notably deviate from the smooth curves in the purely-hadronic case. 

iii) $c_s^2 (p)$ is smooth but varies rapidly in a short range of pressures, which typically indicates special forms of phase transition in realistic models. We show that smoothing of the sharp first-order phase transition to a rapid crossover does not affect tidal properties significantly.

We find that the smallest value of the dimensionless tidal deformability for a neutron or hybrid star (if not a bare strange quark star) with typical mass is given by a sharp phase transition immediately above nuclear saturation density ($\ntrans =n_0$), transforming from soft nuclear matter to stiff quark matter. Sequential phase transitions could further lower this limit. The new minimum of $\La_{1.4}$ we obtained is $\approx 60$ for EoSs with one phase transition, and $\approx 40$ for those with two sequential phase transitions. We followed by demonstrating the prospects for distinguishing NS-NS, NS-HS and HS-HS binary mergers, and it turns out that $\ntrans$ is the primary parameter. Since increasing the phase transition threshold $\ntrans$ to densities above $n_0$ would merely result in larger values of the lower bound on $\La_{1.4}$, refined constraints on nuclear EoS uncertainties at $1\sim 2\, n_0$ from either experiment or theory is crucial for eliminating certain quark EoS parameter space. It is worth mentioning that in order to rescue the possible contradiction with low-density neutron skin measurement, a not-too-high $\ntrans$ is inevitable~\cite{Fattoyev:2017jql}.

We test two universal relations, i.e. $\La\propto\be^6$ and the Yagi-Yunes I-Love $\bar{I}(\La)$ correlation, and show that their accuracy can fail more than has been anticipated previously because of the sharp phase transitions. Possible sequential phase transitions e.g. nuclear $\to$ 2SC $\to$ CFL can aggravate the deviation from universality, which is contingent on the choice of parameter values. We confirm that in this case triplet configurations are relatively rare even with a parameterization optimized to produce them, and that they depend on a stiff hadronic EoS with $\Mtrans$ close to $2\,\Msolar$. The underlying physics sensitive to other observables such as dynamic twin-star oscillation/mini-collapse that obeys baryon number conservation or the rapid rotation effects~\cite{Bejger:2016emu} is not covered in this work, and we believe it warrants further investigation in the future.

\section*{Acknowledgements} 
This work was supported by Chandra Award TM8-19002X (S.H.), NSF grant PHY-1554876 (S.H. and A.W.S.), and the U.S. DOE Office of Nuclear Physics. S.H. is also supported by National Science Foundation, Grant PHY-1630782, and the Heising-Simons Foundation, Grant 2017-228. The authors greatly acknowledge the hospitality of Aspen Center for Physics during the workshop ``Neutron Stars: Linking Nuclear Physics of the Interior to Electromagnetic Observations and Gravitational Radiation'' supported by NSF grant PHY-1607611, where part of this work was performed.  S.H. would like to thank Jim Lattimer, Mark Alford, Katerina Chatziioannou, and Ingo Tews for fruitful discussions. This project used computational resources from the University of Tennessee and Oak Ridge 
National Laboratory's Joint Institute for
Computational Sciences.

\newcommand{\apjl}{Astrophys. J. Lett.\ }
\newcommand{\mnras}{Mon. Not. R. Astron. Soc.\ }
\newcommand{\aap}{Astron. Astrophys.\ }

\bibliographystyle{apsrev}
\bibliography{tidal_pt_arxiv}

\appendix
\section{Behavior of tidal parameters in close proximity to the phase transition}
\label{app:tidal_k2_pts}

The generic feature of both $k_2(M)$ and $\La(M)$ in the presence of phase transition is their abrupt changes above the transition pressure $\ptrans$, when the dense quark core emerges (if there is a stable branch). In Fig.~\ref{fig:k2-Lam-set1}, increasing the central pressure one follows the $k_2(M)$ or $\La(M)$ trajectory on the hadronic branch (black solid curve) as the neutron star mass grows. Arriving at the critical pressure for phase transition ($\pcent=\ptrans$), the hadronic branch terminates at $\Mtrans$, after which $k_2$ and $\La$ switch to one of the hybrid branches (colored curves) depending on parameter values, and the first stable hybrid configuration starts whenever $dM/d\pcent>0$ (solid). 

\begin{figure*}[htb]
\parbox{0.26\hsize}{
\includegraphics[width=\hsize]{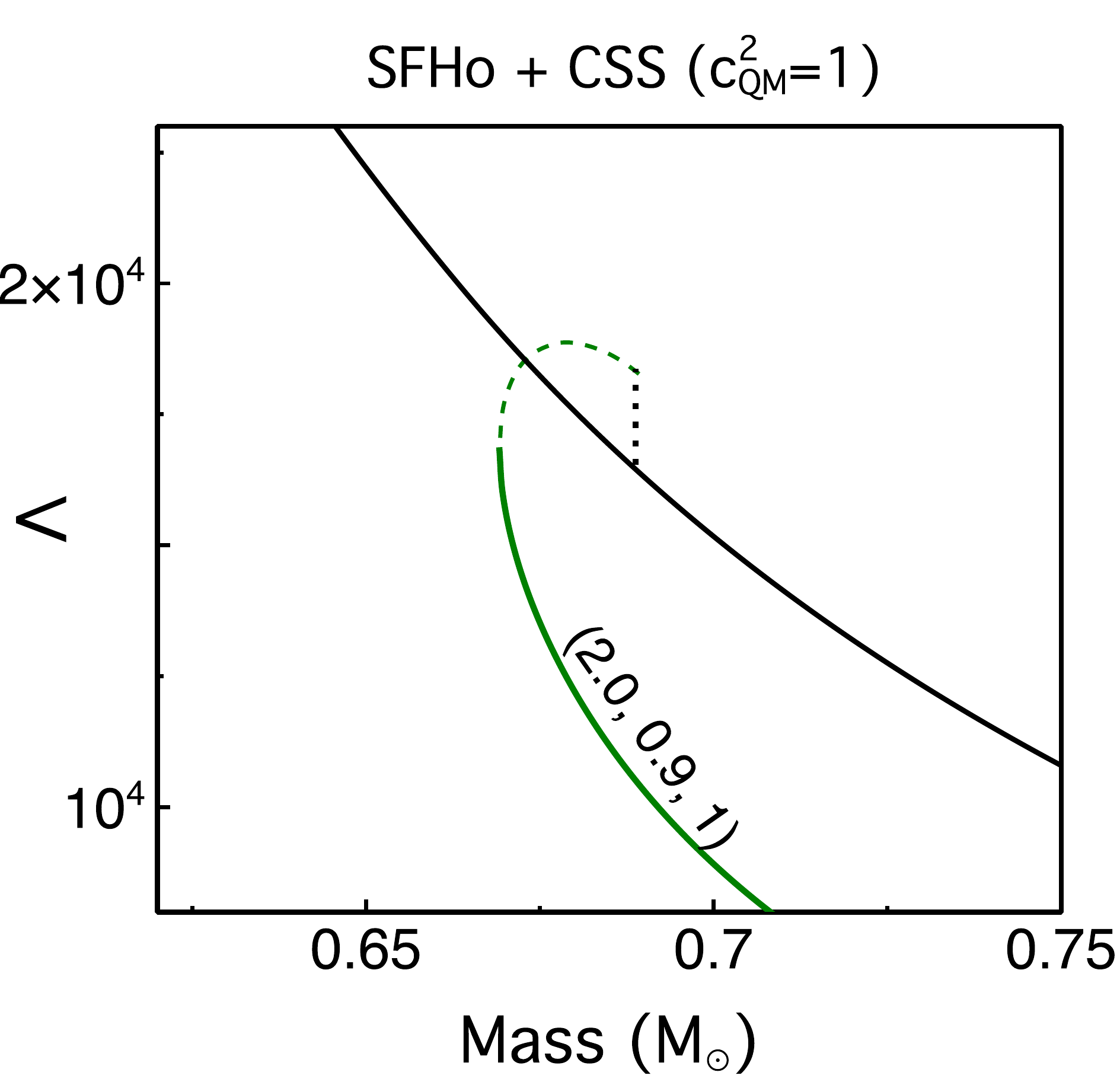}\\[-2ex]

}\parbox{0.26\hsize}{
\includegraphics[width=\hsize]{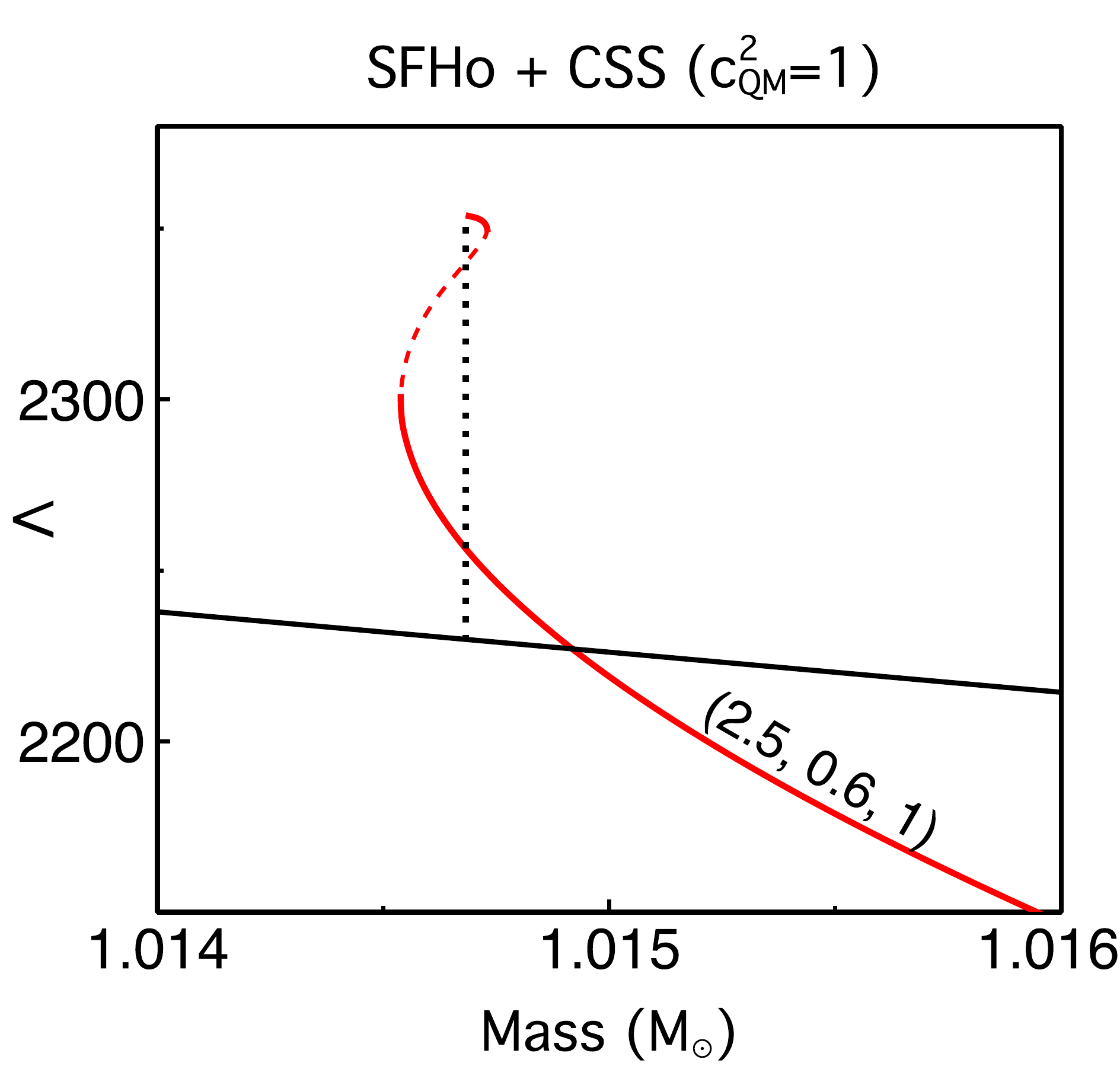}\\[-2ex]
}\parbox{0.26\hsize}{
\includegraphics[width=\hsize]{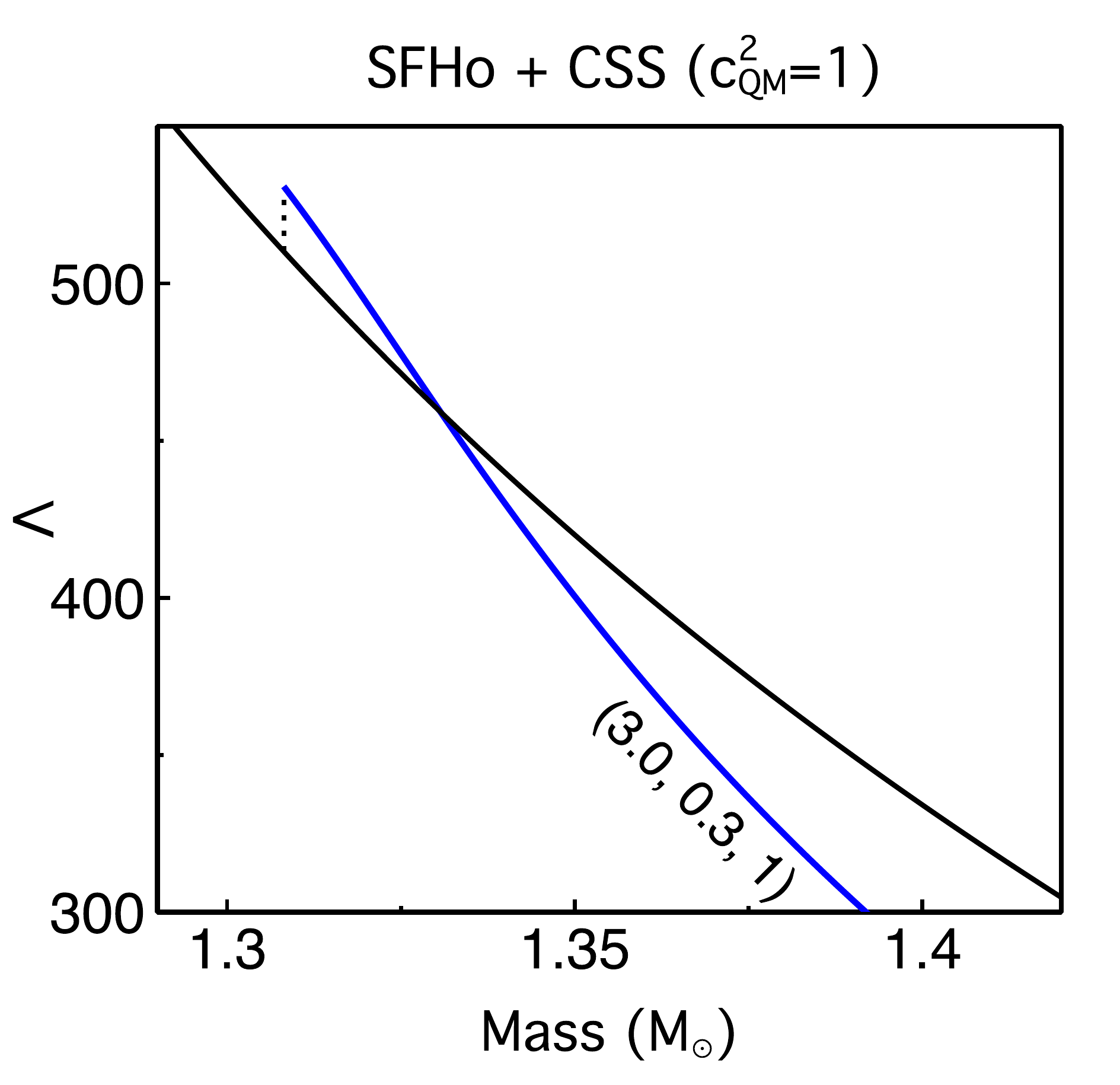}\\[-2ex]
}\parbox{0.26\hsize}{
\includegraphics[width=\hsize]{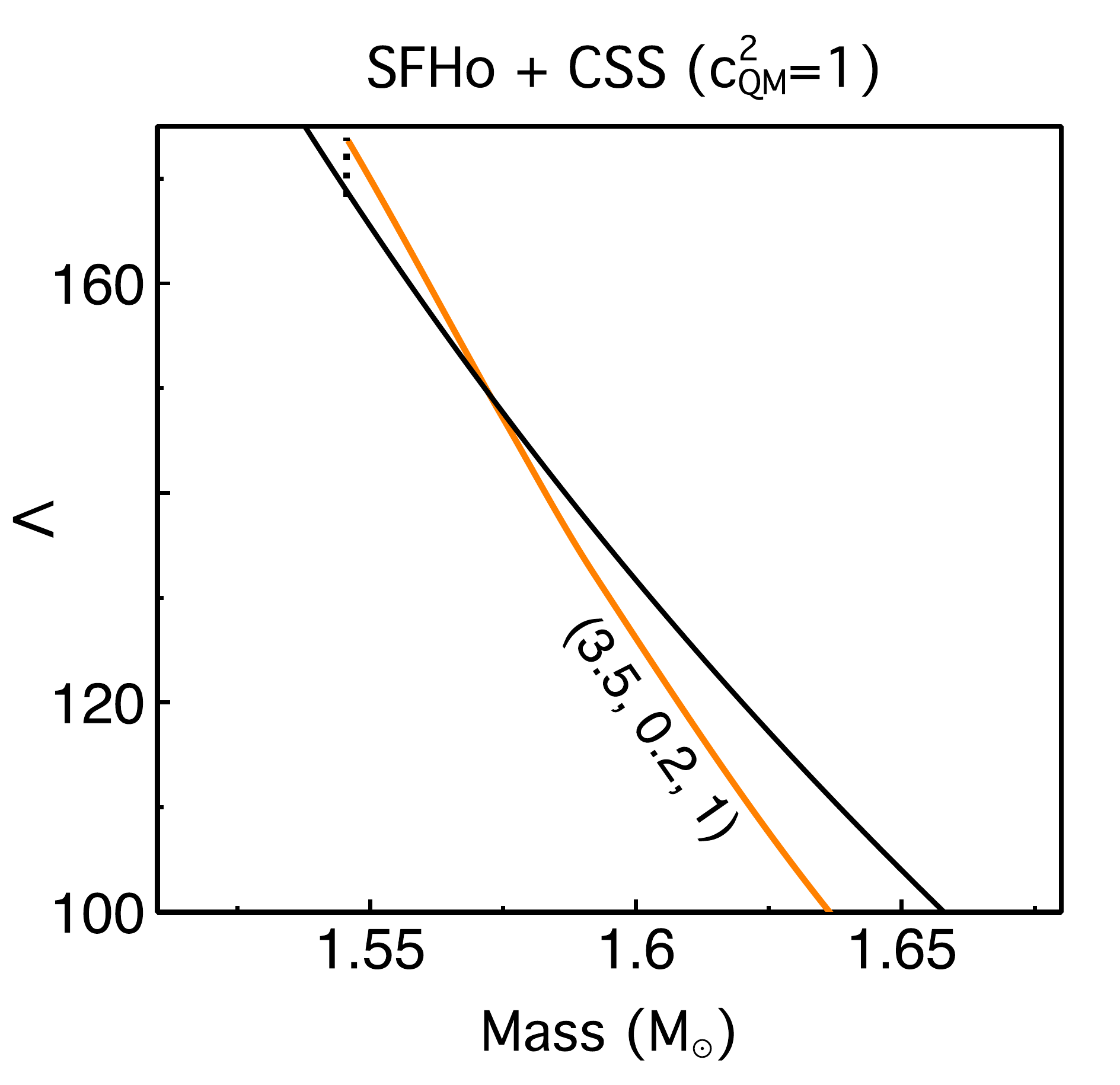}\\[-2ex]

}\\[1ex]

\parbox{0.26\hsize}{
\includegraphics[width=\hsize]{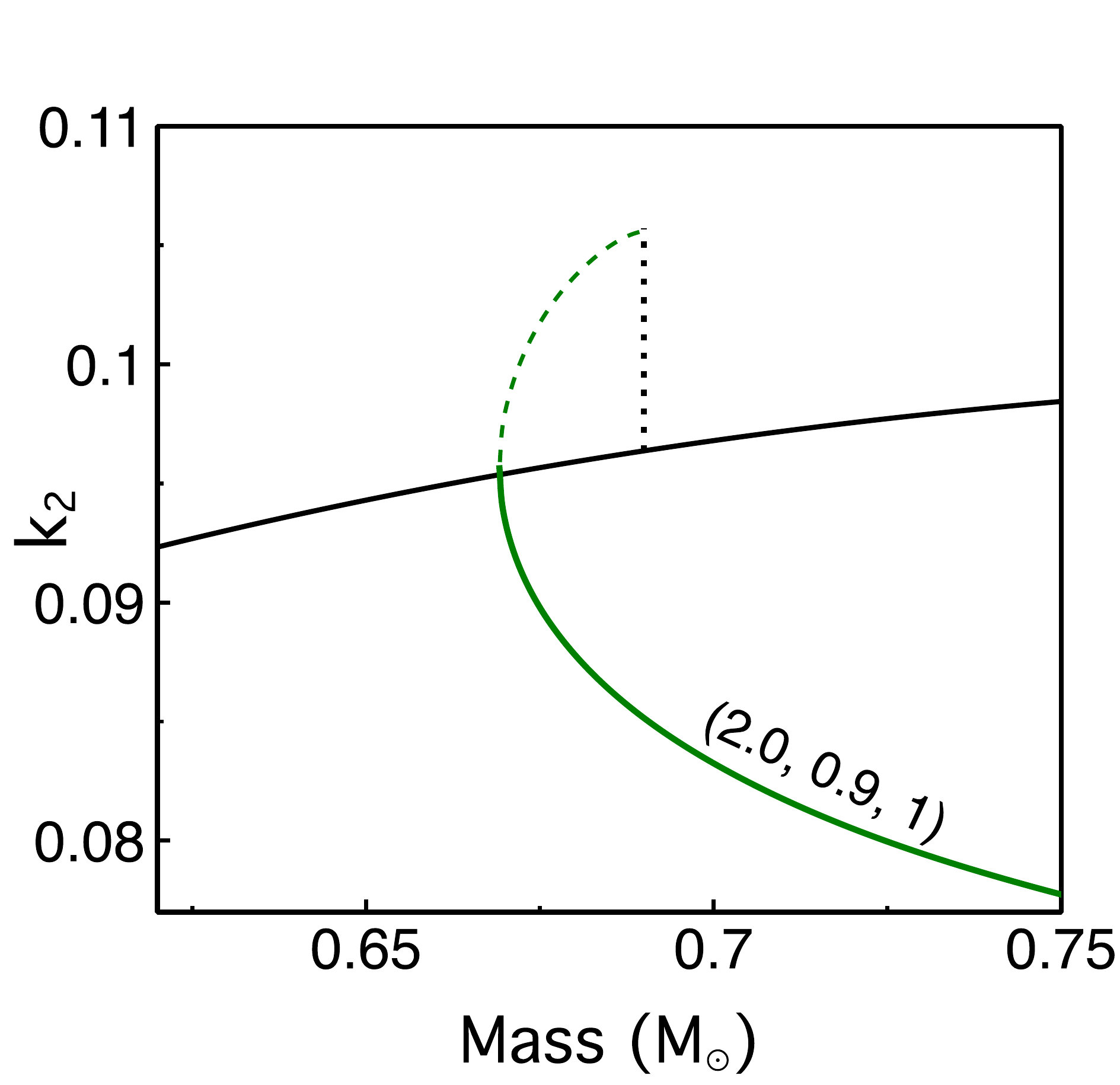}\\[-2ex]

}\parbox{0.26\hsize}{
\includegraphics[width=\hsize]{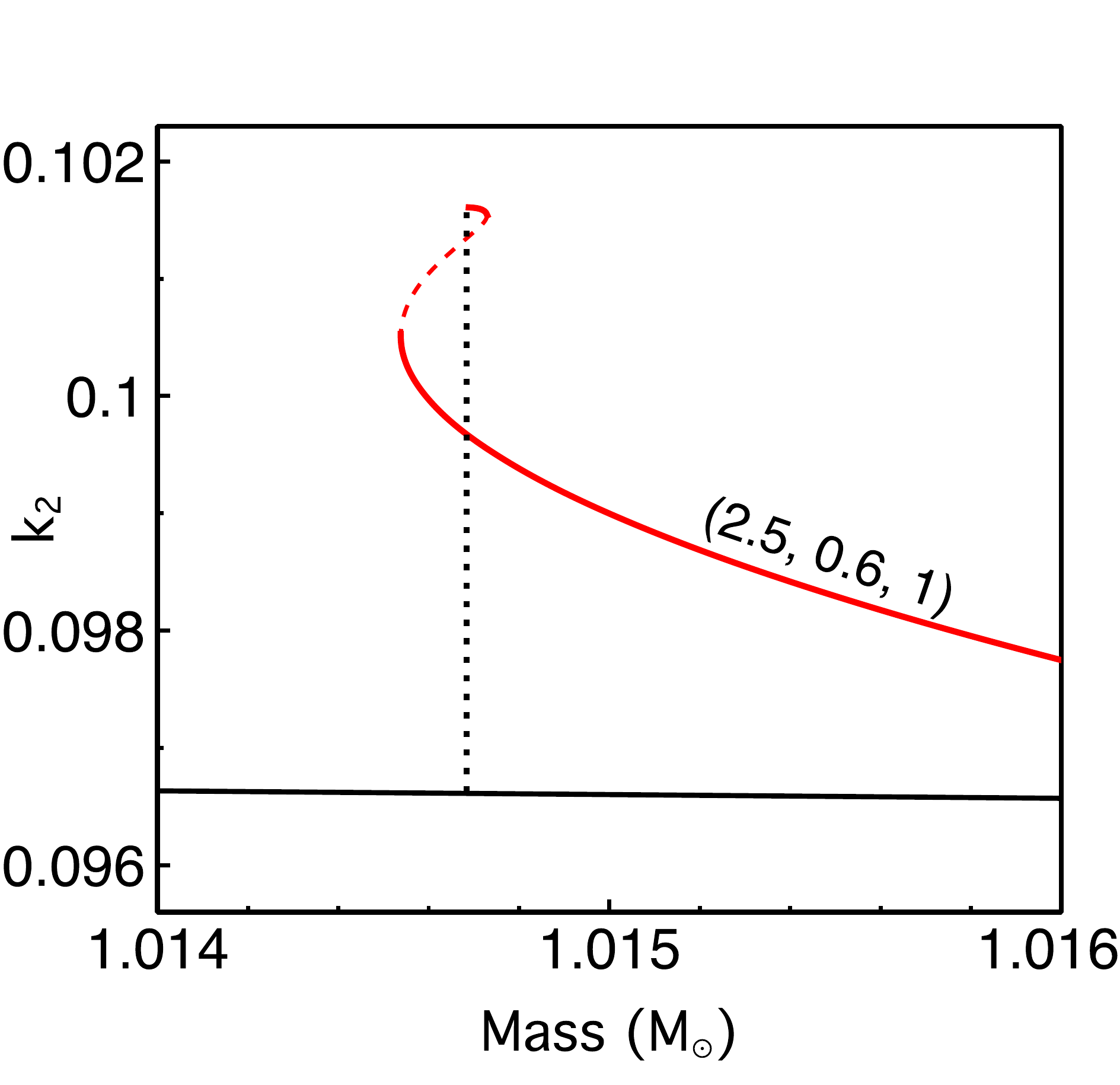}\\[-2ex]
}\parbox{0.26\hsize}{
\includegraphics[width=\hsize]{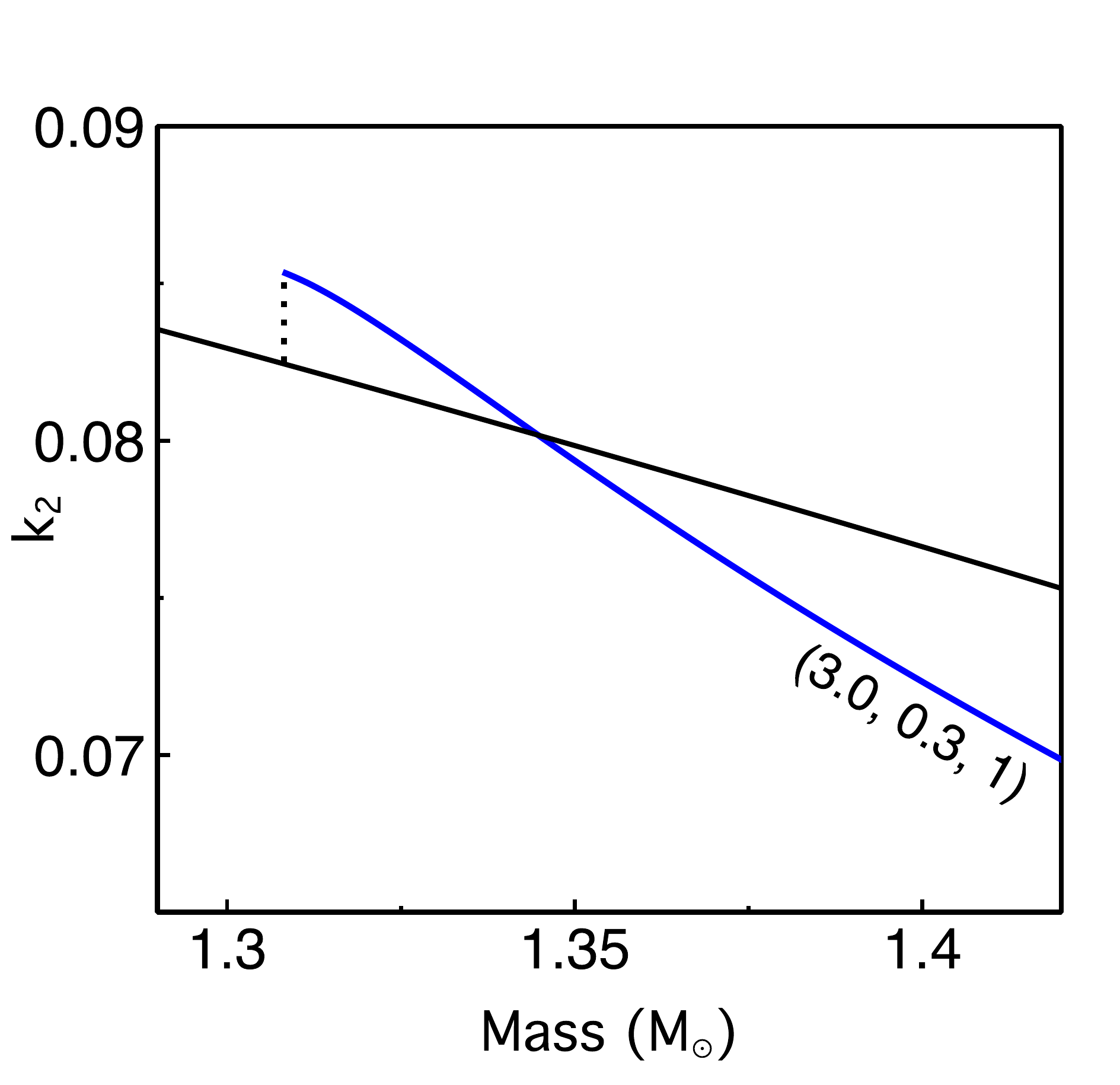}\\[-2ex]
}\parbox{0.26\hsize}{
\includegraphics[width=\hsize]{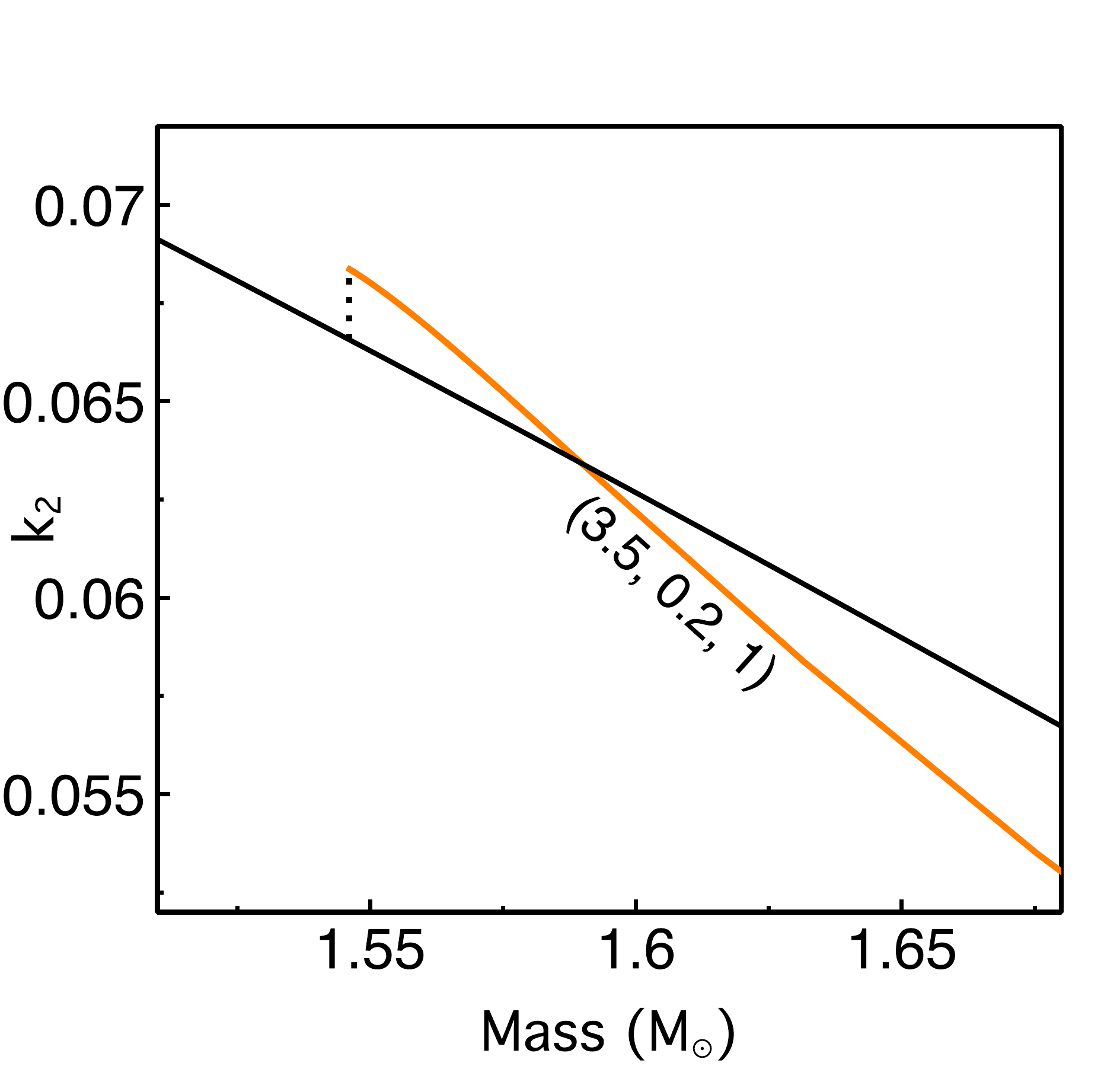}\\[-2ex]
}
\\[1ex]
\caption{(Color online) Magnified plots of $\La (M)$ (upper) and $k_{2} (M)$ (lower) near the phase transition point $\Mtrans$ when $\pcent=\ptrans$, indicated by the vertical dot line. When the central pressure is below the transition pressure $\pcent<\ptrans$, purely-hadronic stellar configurations are determined by the nuclear matter EoS SFHo (black solid curve); at $\pcent=\ptrans$ ($M=\Mtrans$) depending on CSS phase transition parameters (taken from set I in Table~\ref{tab:hyb_EoS}; see also Fig.~\ref{fig:diag-dbhf-sfho-c2-1} left panel), due to the singularity in $1/c_s^2=d\ep/dp$, both $k_2$ and $\La$ are discontinuous; above $\ptrans$, solid (dashed) curves represent stable (unstable) hybrid configurations. From left to right, four EoSs give rise to ``Disconnected'', ``Both'', ``Connected'', and ``Connected'' scenarios of mass-radius in Fig.~\ref{fig:MR-De}.
}
\label{fig:lam-k2-M-sfho-set1}
\end{figure*}

\begin{figure*}[htb]
\parbox{0.26\hsize}{
\includegraphics[width=\hsize]{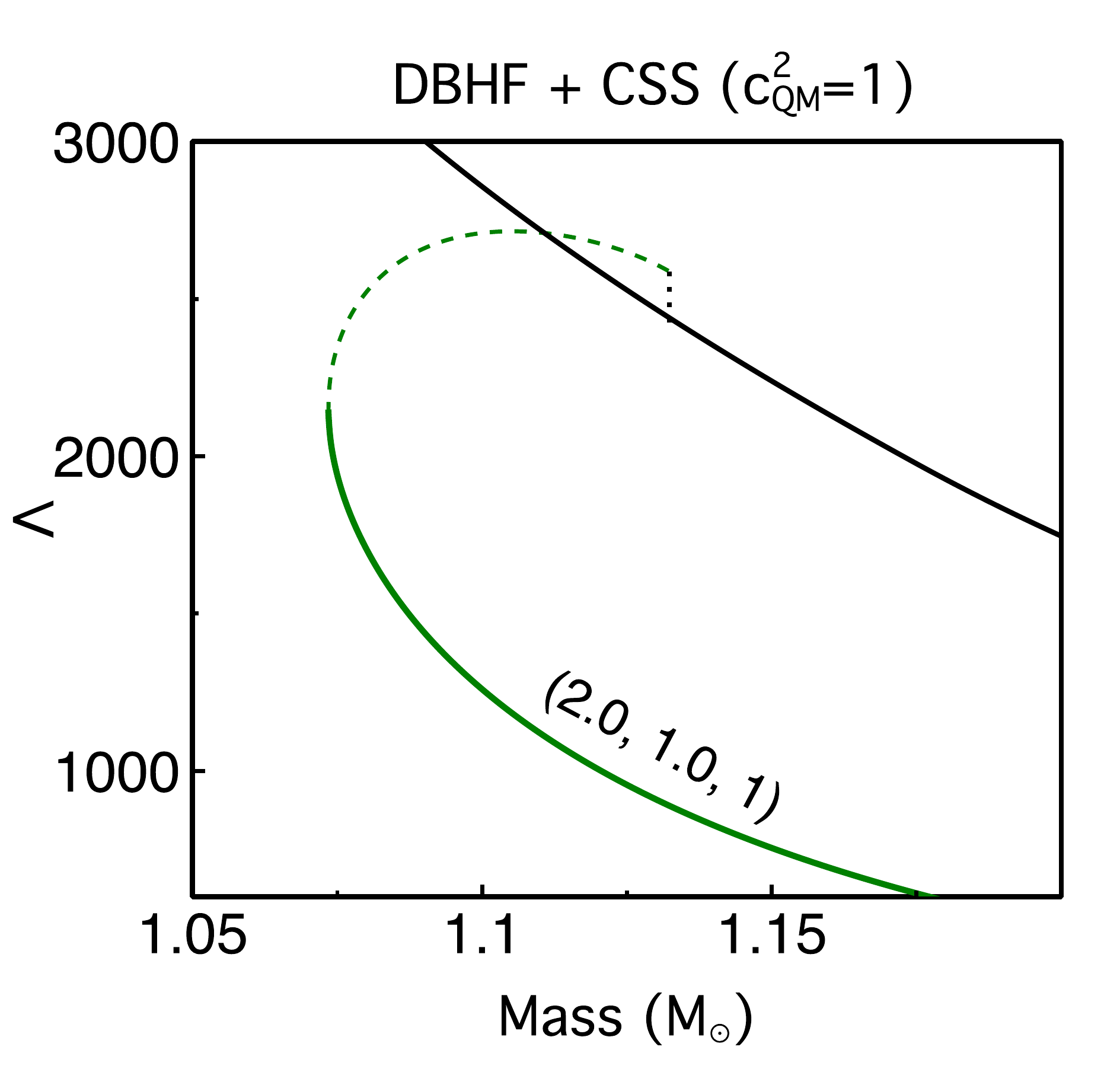}\\[-2ex]

}\parbox{0.26\hsize}{
\includegraphics[width=\hsize]{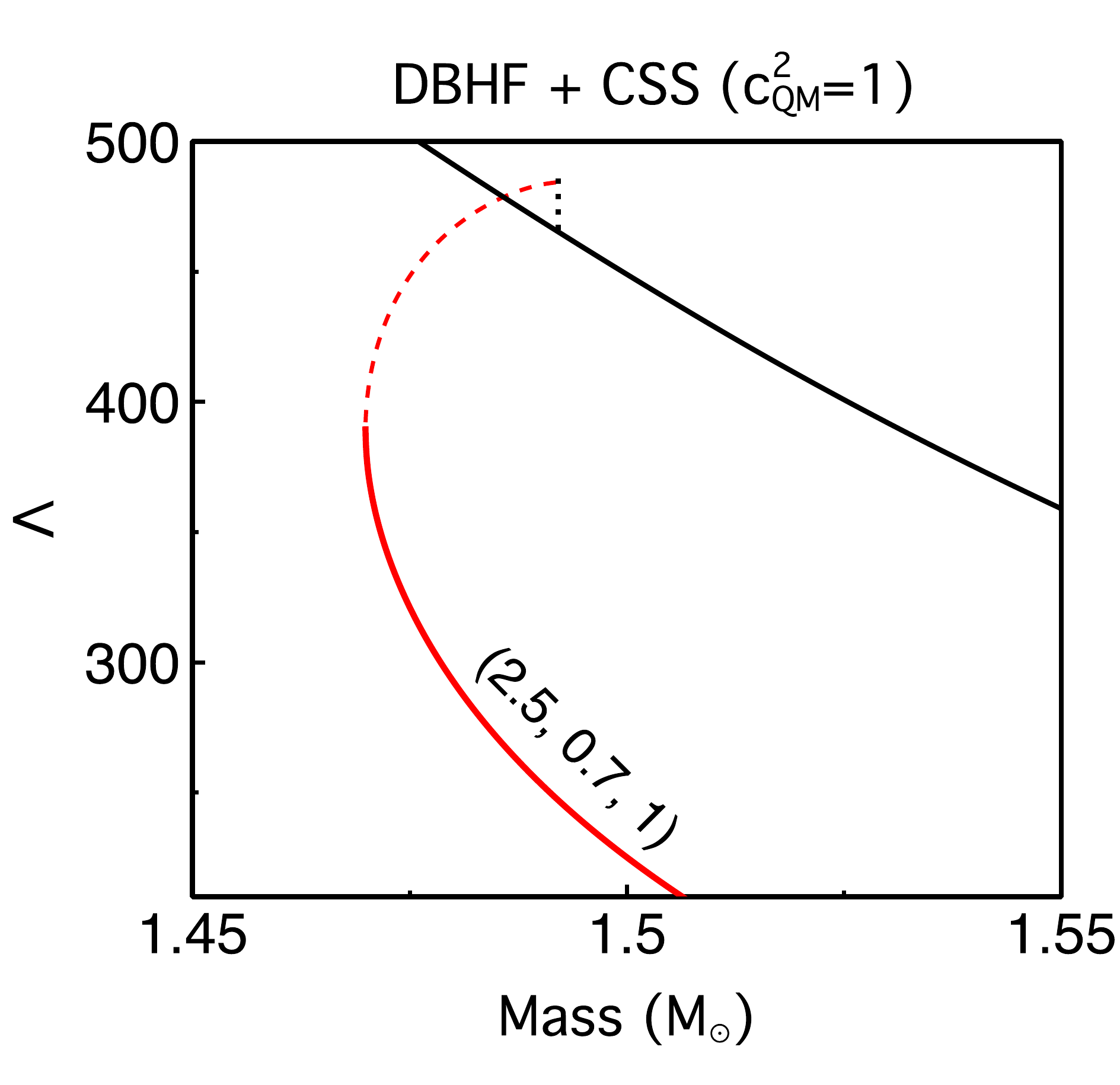}\\[-2ex]
}\parbox{0.26\hsize}{
\includegraphics[width=\hsize]{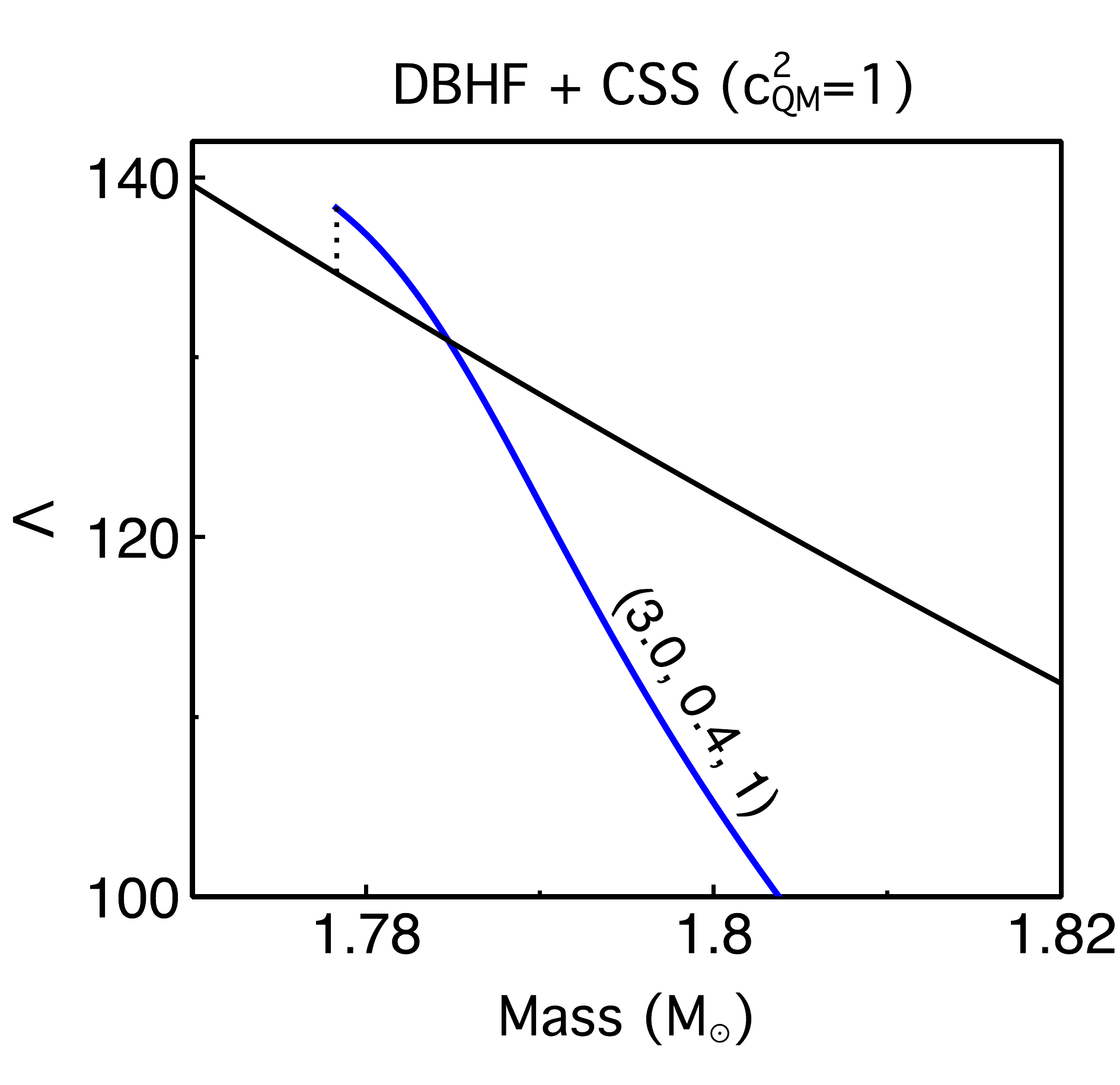}\\[-2ex]
}\parbox{0.26\hsize}{
\includegraphics[width=\hsize]{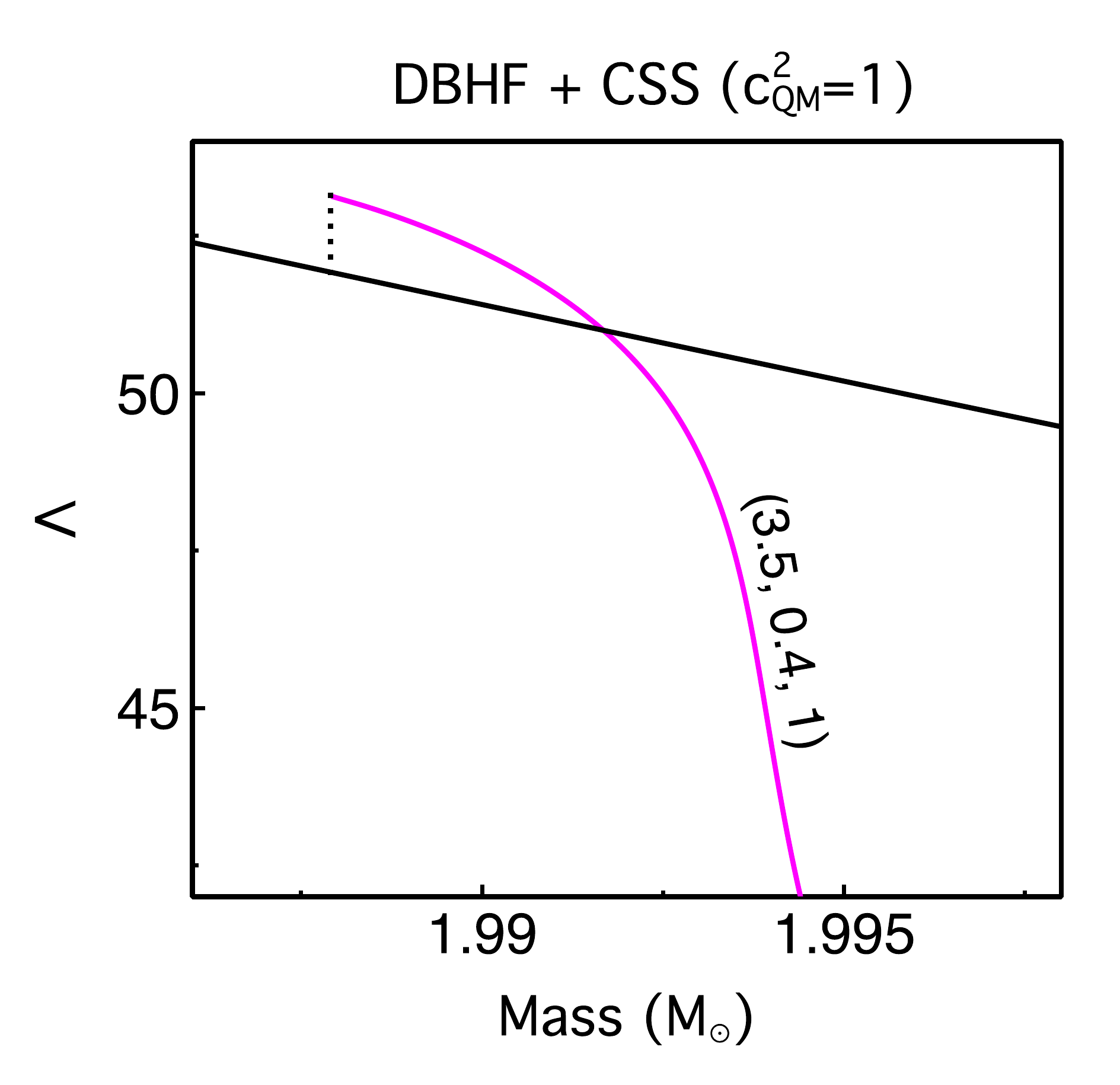}\\[-2ex]

}\\[1ex]

\parbox{0.26\hsize}{
\includegraphics[width=\hsize]{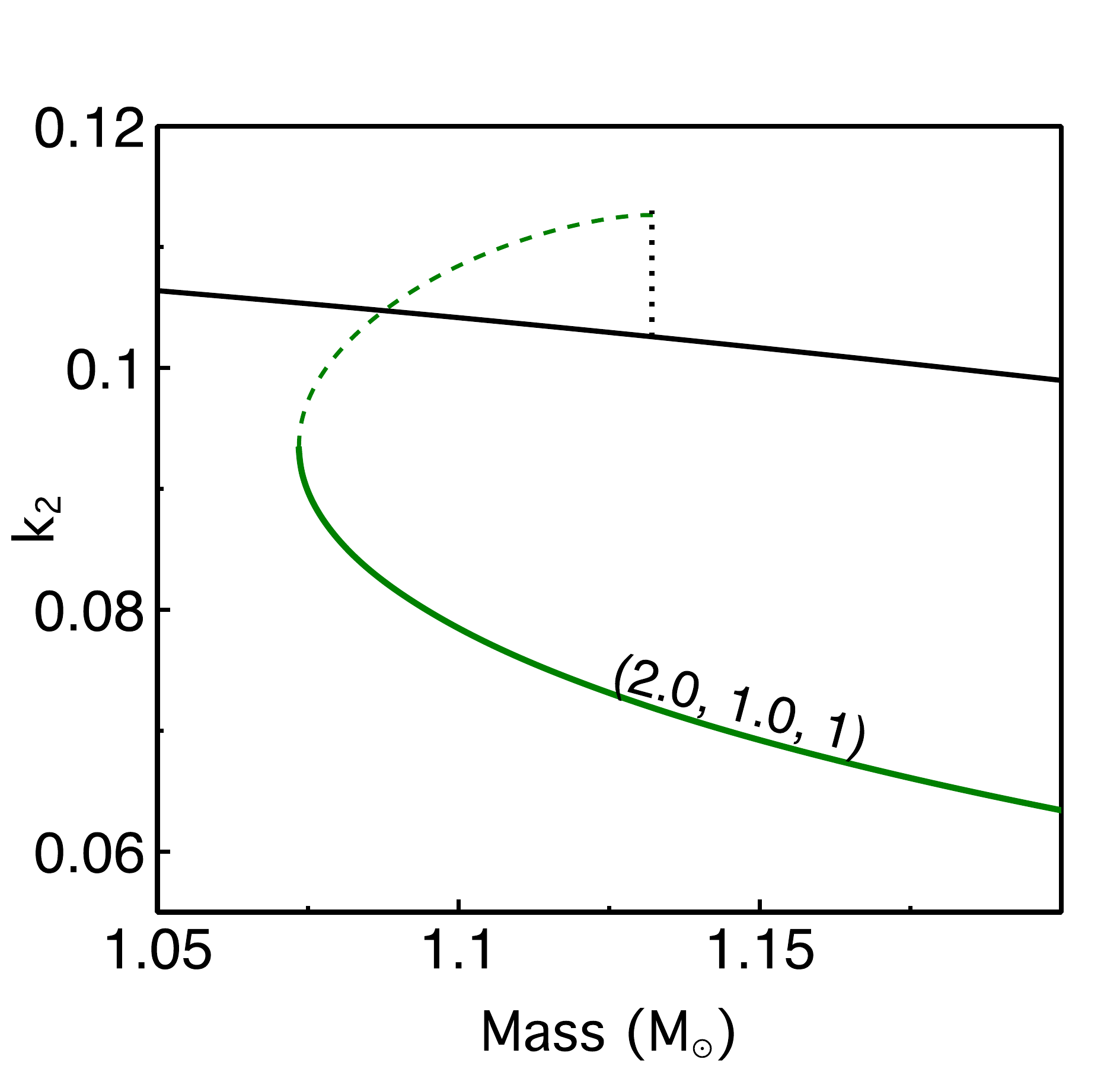}\\[-2ex]
}\parbox{0.26\hsize}{
\includegraphics[width=\hsize]{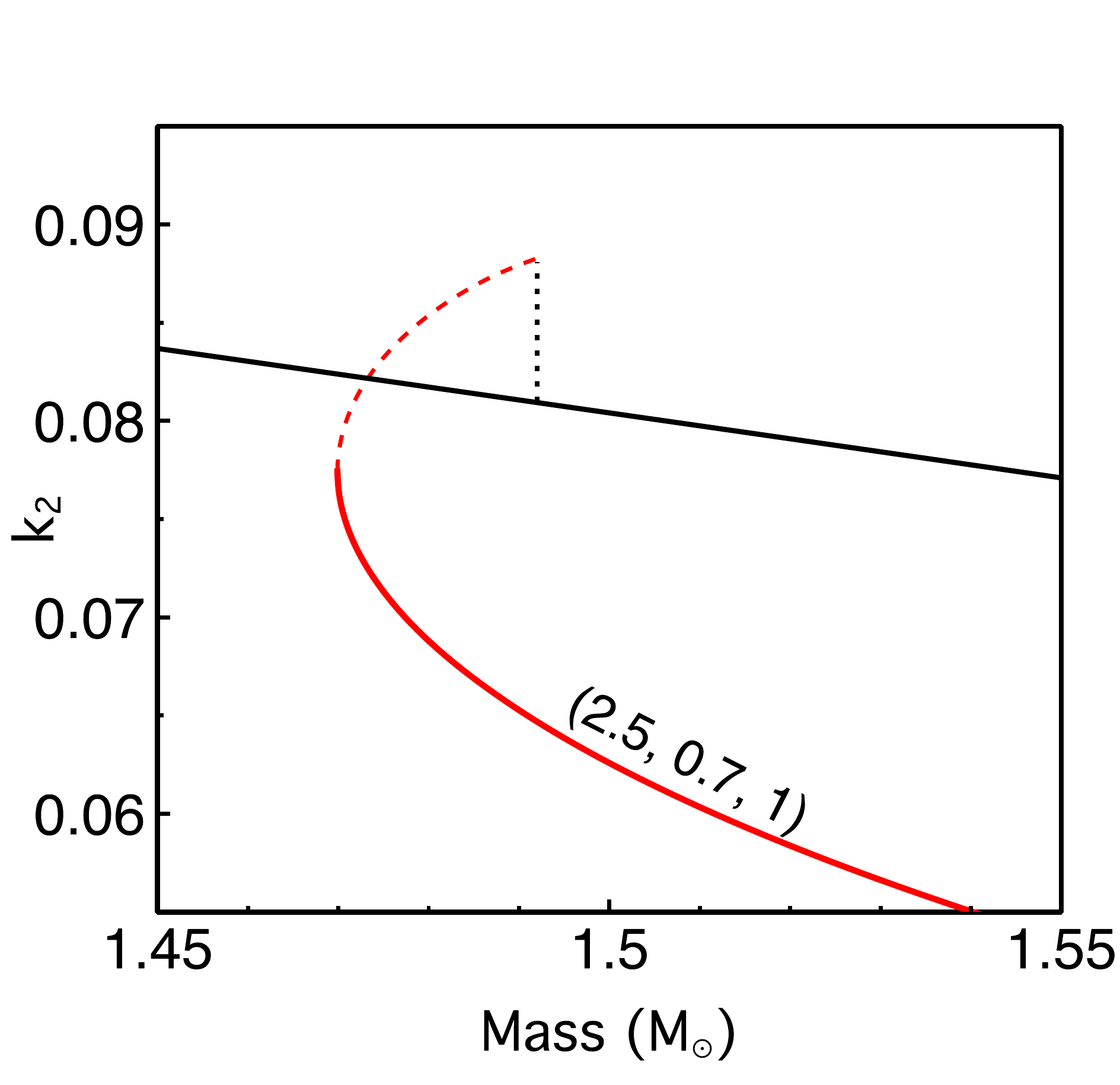}\\[-2ex]
}\parbox{0.26\hsize}{
\includegraphics[width=\hsize]{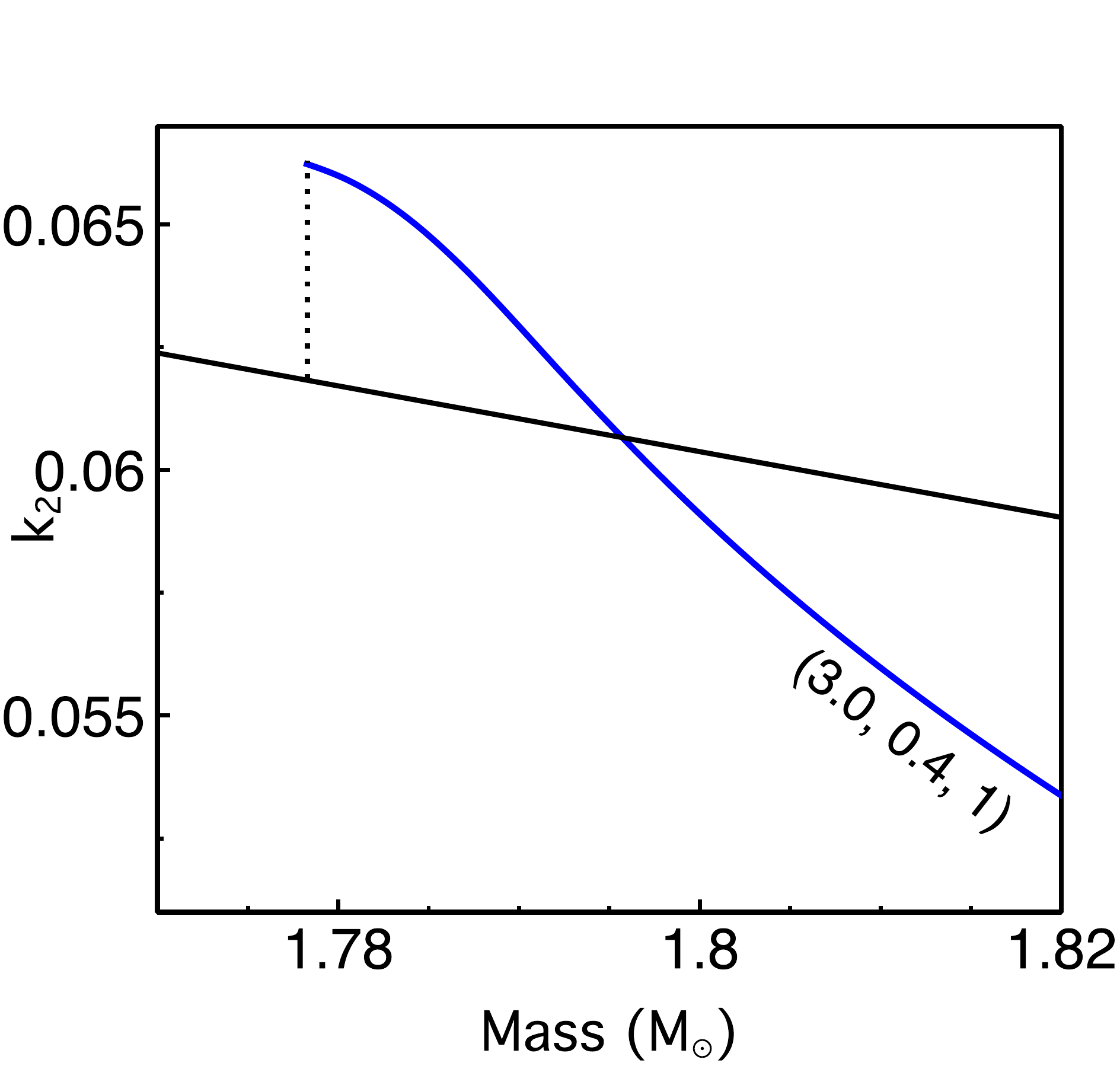}\\[-2ex]
}\parbox{0.26\hsize}{
\includegraphics[width=\hsize]{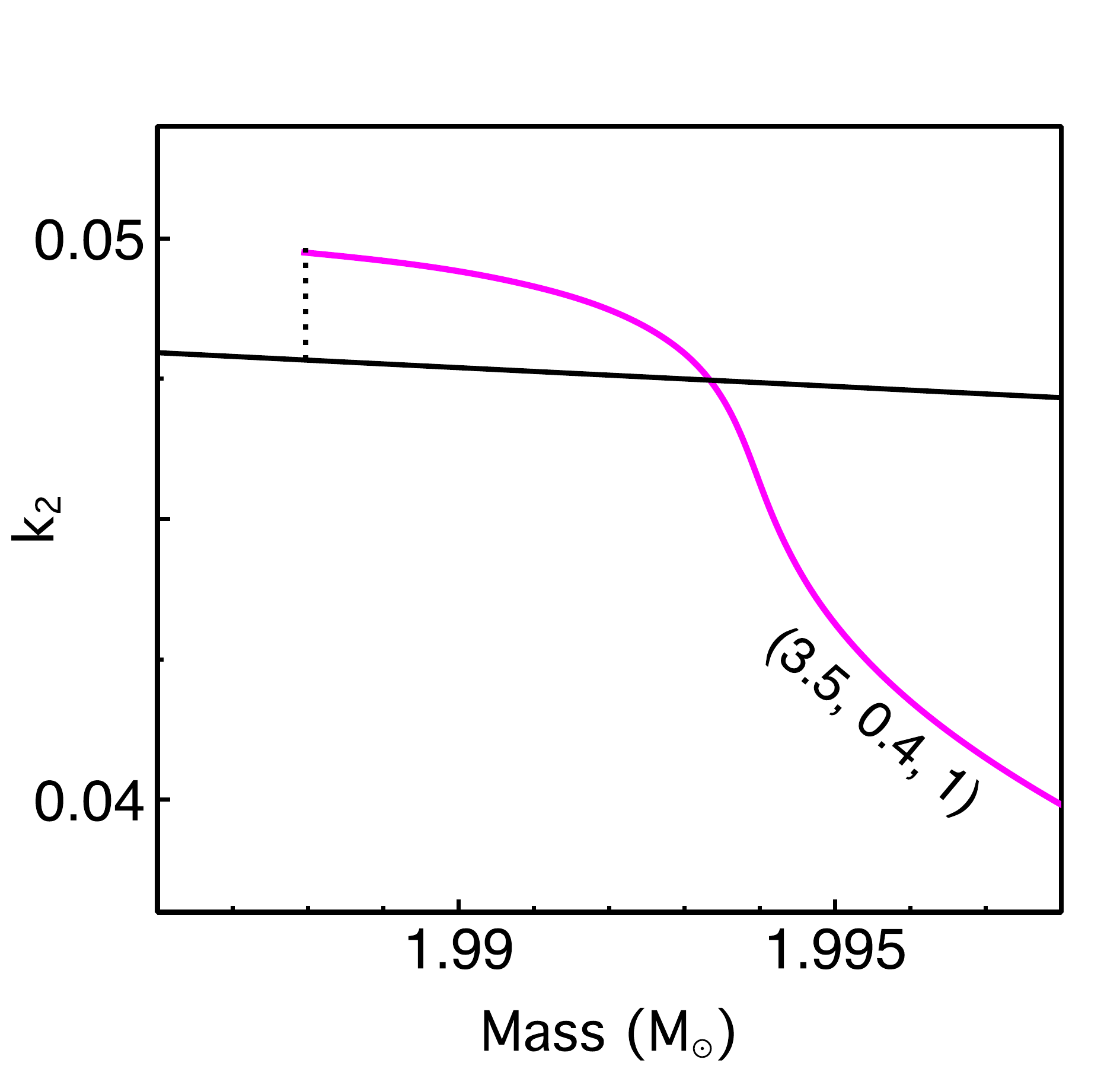}\\[-2ex]

}\\[1ex]
\caption{(Color online) Magnified plots of $\La (M)$ (upper) and $k_{2} (M)$ (lower) near the phase transition point $\Mtrans$ when $\pcent=\ptrans$, indicated by the vertical dot line. When the central pressure is below the transition pressure $\pcent<\ptrans$, purely-hadronic stellar configurations are determined by the nuclear matter EoS DBHF (black solid curve); at $\pcent=\ptrans$ ($M=\Mtrans$) depending on CSS phase transition parameters (taken from set I in Table~\ref{tab:hyb_EoS}; see also Fig.~\ref{fig:diag-dbhf-sfho-c2-1} right panel), due to the singularity in $1/c_s^2=d\ep/dp$, both $k_2$ and $\La$ are discontinuous; above $\ptrans$, solid (dashed) curves represent stable (unstable) hybrid configurations. From left to right, four EoSs give rise to ``Disconnected'', ``Disconnected'', ``Connected'', and ``Connected'' scenarios of mass-radius in Fig.~\ref{fig:MR-De}.
}
\label{fig:lam-k2-M-dbhf-set1}
\end{figure*}

Here we enlarge on hybrid stars with central pressure slightly higher than the phase transition pressure, yielding masses close to that of the heaviest purely-hadronic stars ($M=\Mtrans$). Their tidal properties are shown in Fig.~\ref{fig:lam-k2-M-sfho-set1} and Fig.~\ref{fig:lam-k2-M-dbhf-set1}, with the hadronic part being SFHo and DBHF EoS respectively. The phase transition parameters are the same as in set I of Table~\ref{tab:hyb_EoS} (excluding the cases with $\ntrans=n_0$). These hybrid EoSs are represented by asterisks, lying in regions ``D'', ``B'' and ``C'' on the CSS phase diagram Fig.~\ref{fig:diag-dbhf-sfho-c2-1}, and their mass-radius curves can be found in panels (a) and (b) of Fig.~\ref{fig:MR-curves-1}. For zoomed-out versions of $k_2(M)$ and $\La(M)$, see Fig.~\ref{fig:k2-Lam-set1}.

In Fig.~\ref{fig:lam-k2-M-sfho-set1} and Fig.~\ref{fig:lam-k2-M-dbhf-set1}, the vertical dotted lines connect the endpoint of the stable hadronic branch with $M=\Mtrans$ at $\pcent=\ptrans$, to the beginning of a possible hybrid branch being stable ($dM/d\pcent>0$, solid) or unstable ($dM/d\pcent<0$, dashed). This shows that an infinitesimal quark core invariably boosts the tidal Love number and tidal deformability, and then 

i) if the initial hybrid star is stable, $k_2$ and $\La$ continue decreasing with the mass but along the curve for hybrid stars instead of that for hadronic stars, or 

ii) if the initial hybrid star is unstable, it collapses with mass decreasing while the central pressure increases, until reaching the stable branch at $dM/d\pcent>0$ and resumes the evolution of i). 

\begin{figure*}[htb]
\parbox{0.26\hsize}{
\includegraphics[width=\hsize]{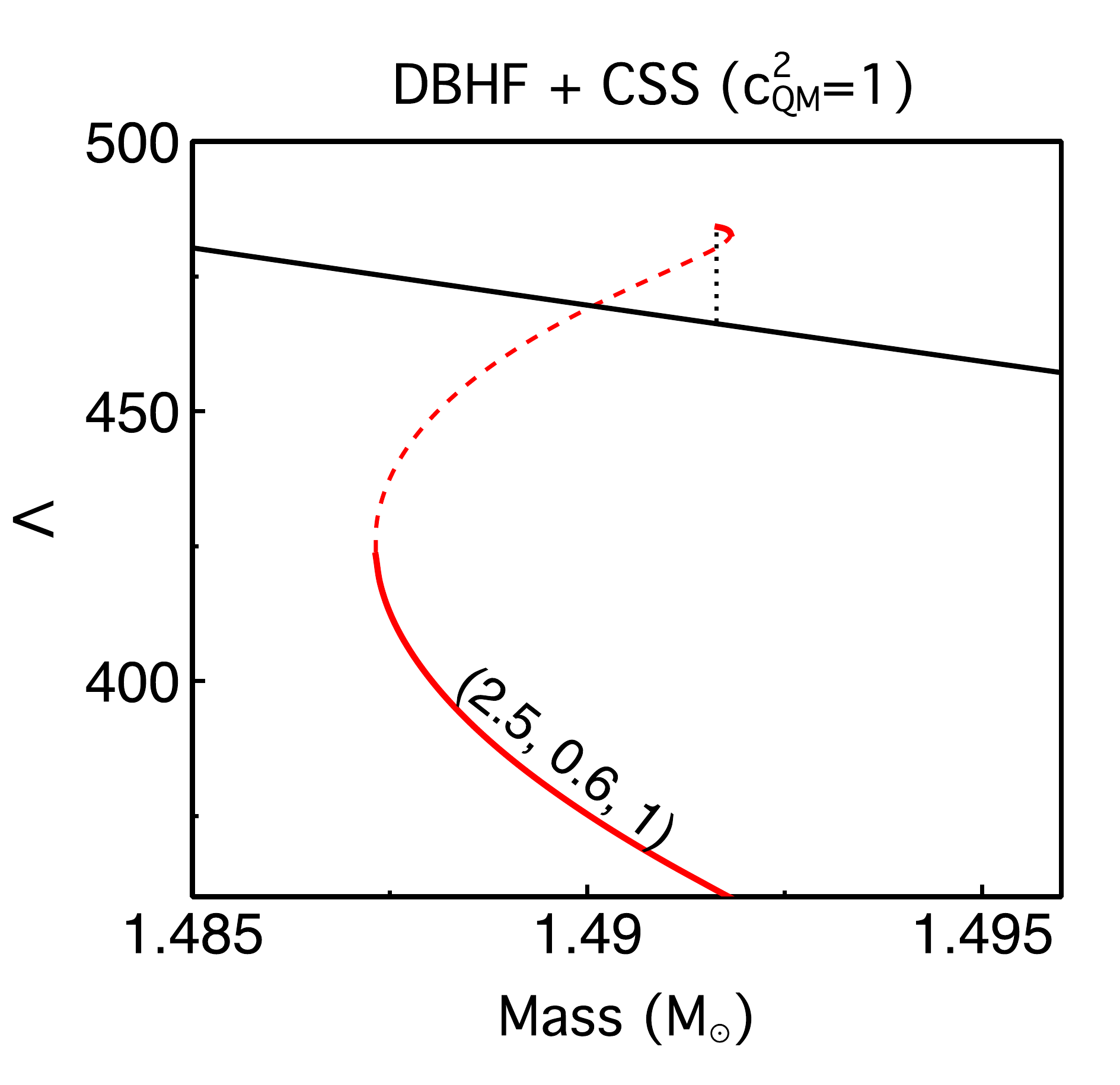}\\[-2ex]
}\parbox{0.26\hsize}{
\includegraphics[width=\hsize]{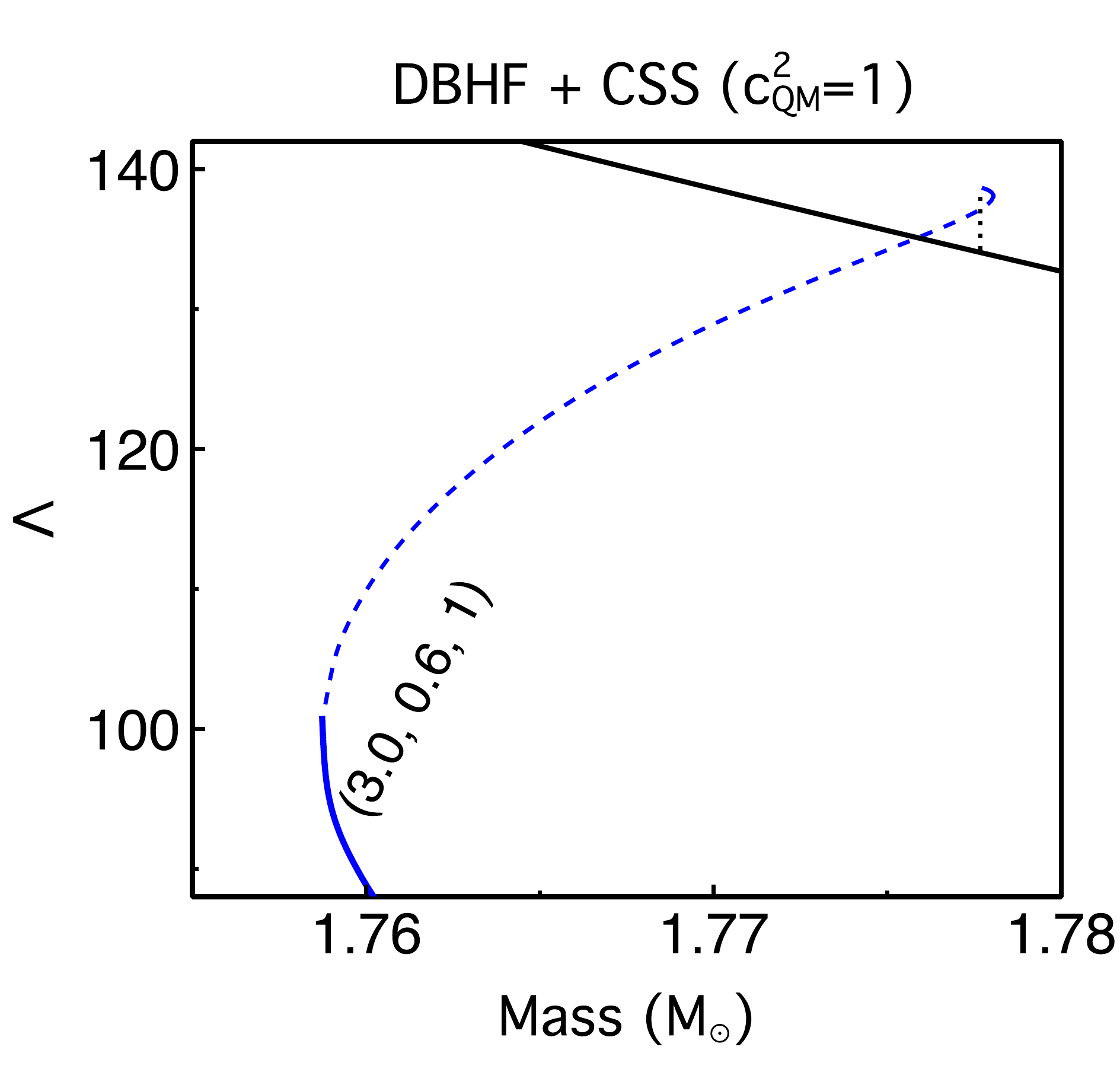}\\[-2ex]
}\parbox{0.26\hsize}{
\includegraphics[width=\hsize]{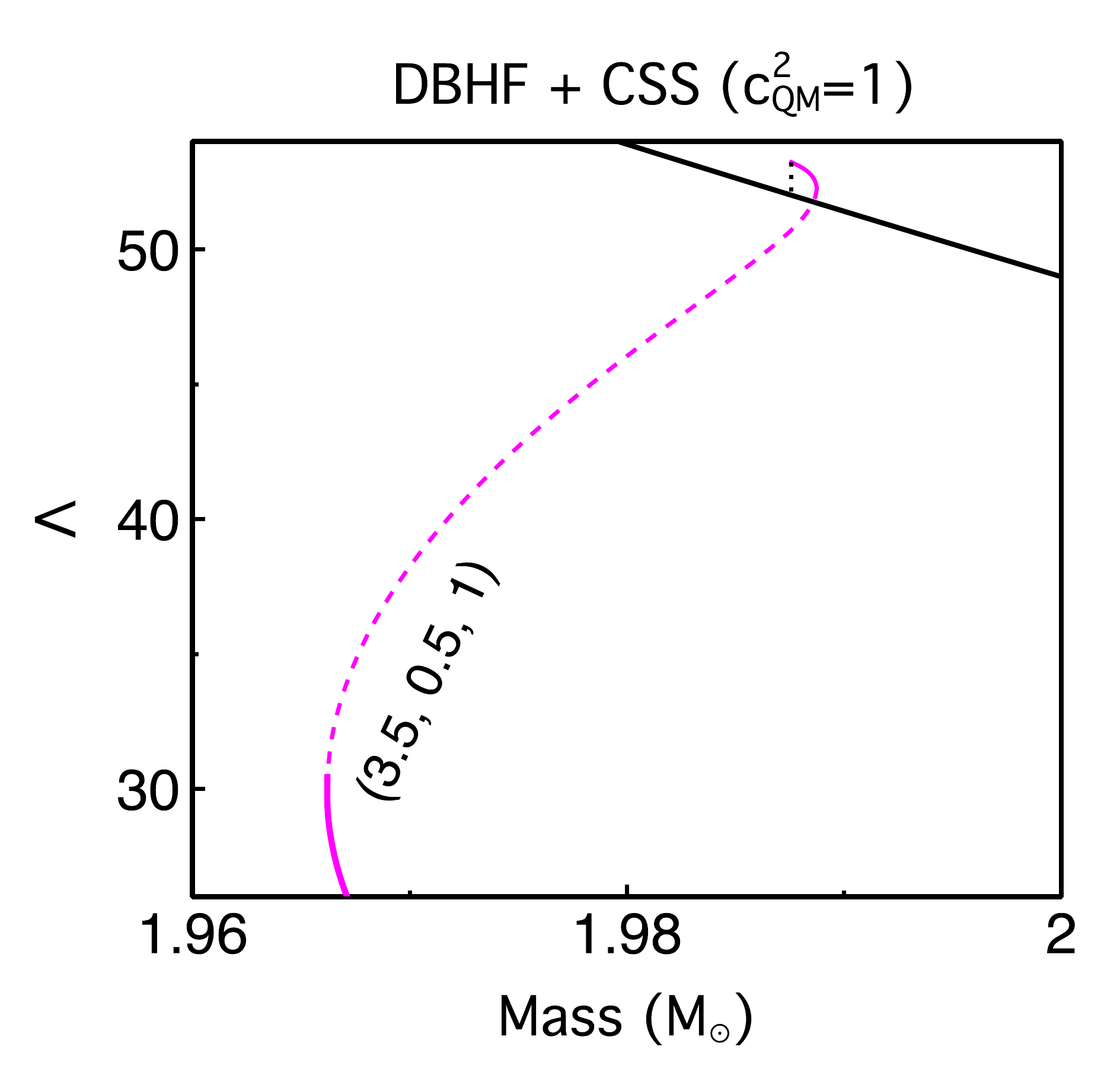}\\[-2ex]

}\\[1ex]

\parbox{0.26\hsize}{
\includegraphics[width=\hsize]{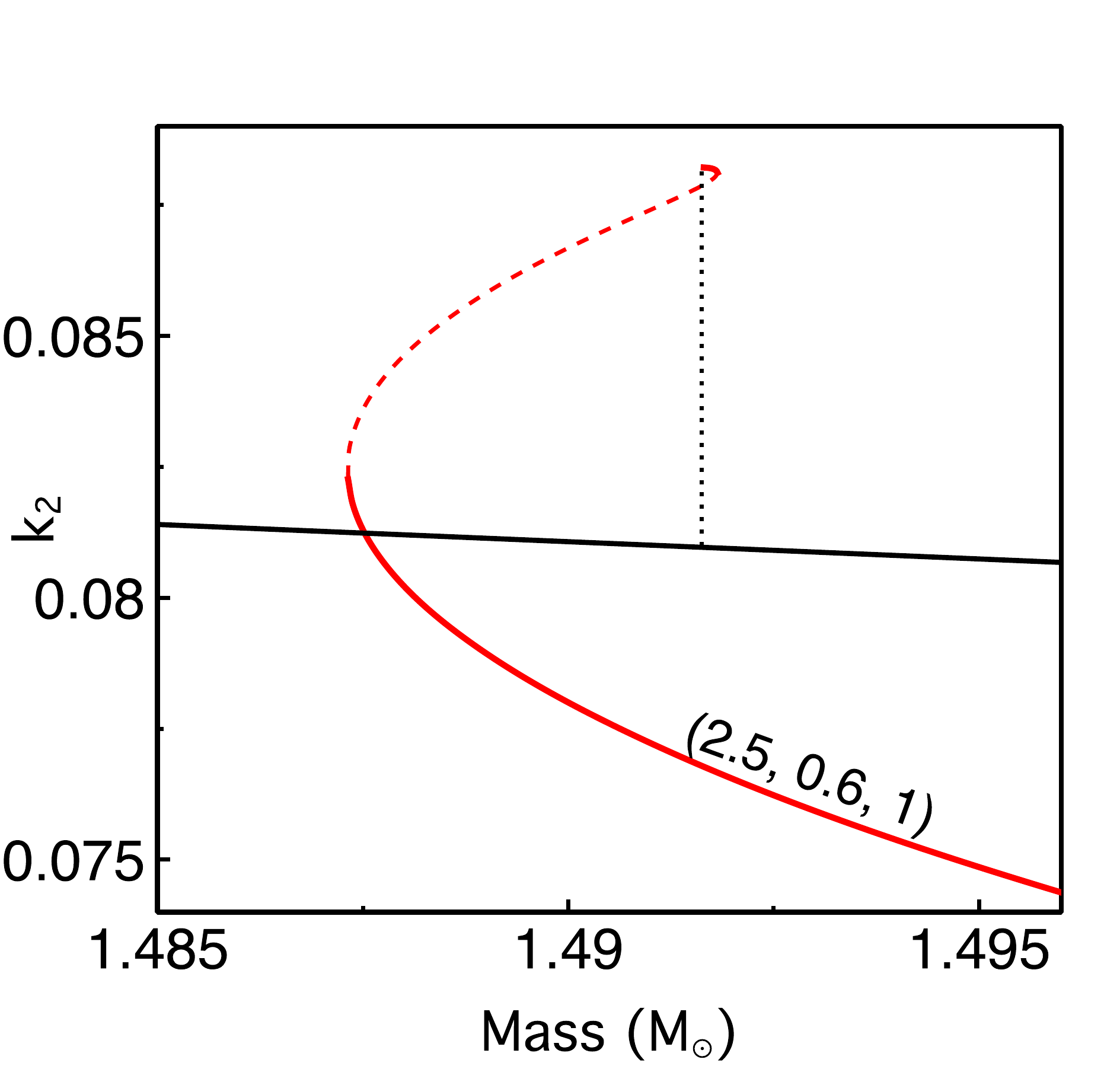}\\[-2ex]
}\parbox{0.26\hsize}{
\includegraphics[width=\hsize]{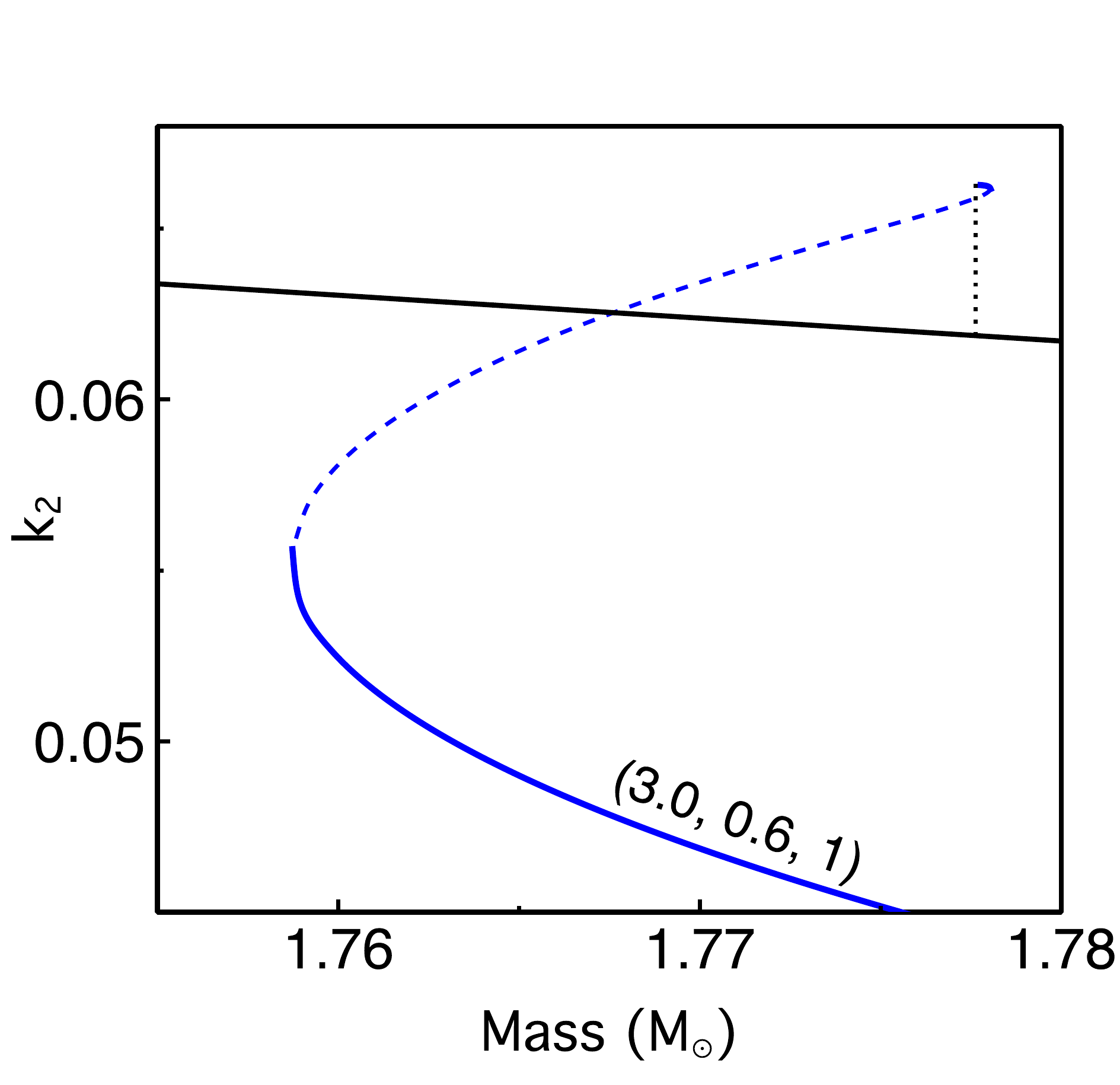}\\[-2ex]
}\parbox{0.26\hsize}{
\includegraphics[width=\hsize]{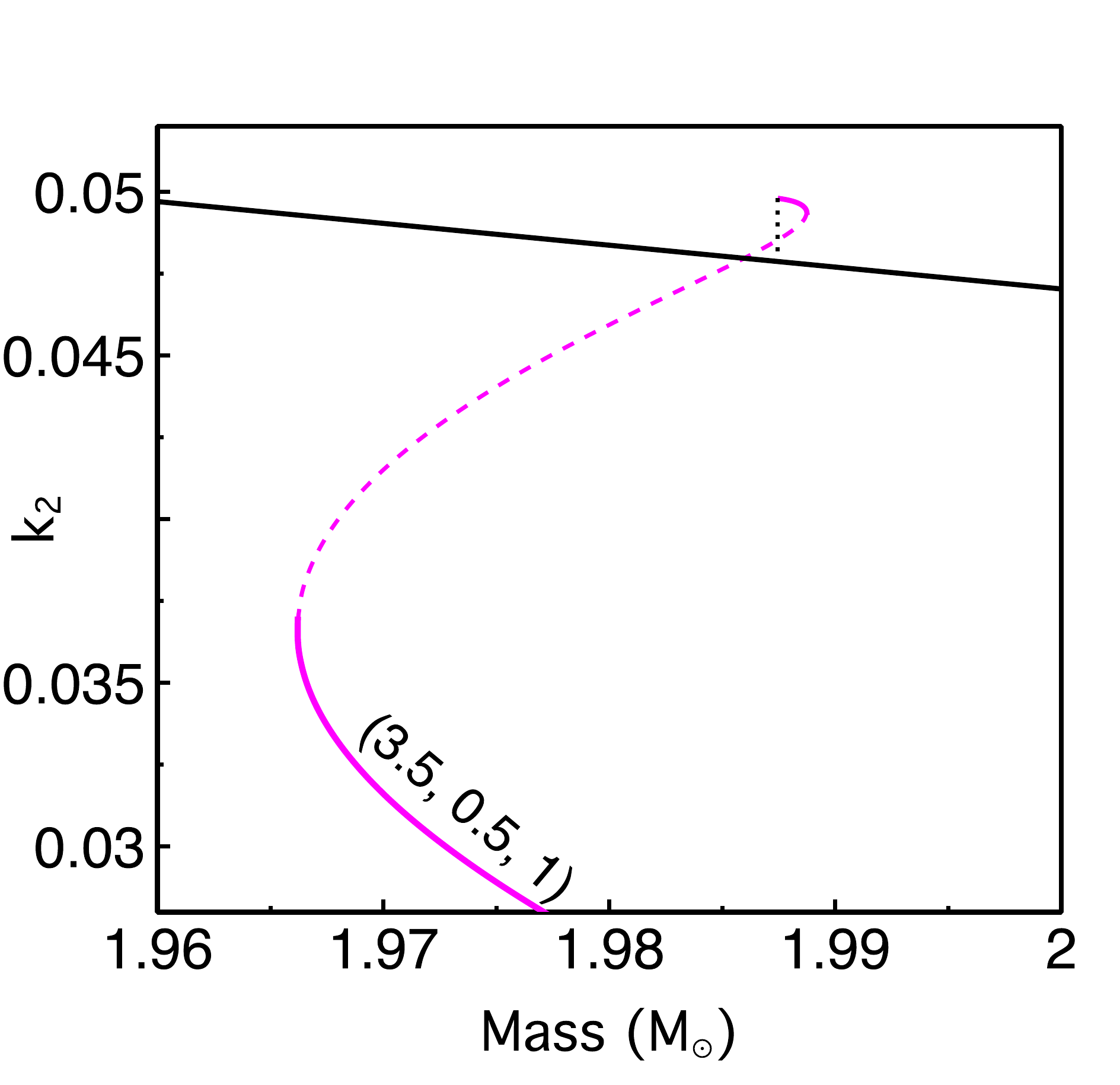}\\[-2ex]
}\\[1ex]
\caption{(Color online) Magnified plots of $\La (M)$ (upper) and $k_{2} (M)$ (lower) near the phase transition point $\Mtrans$ when $\pcent=\ptrans$, indicated by the vertical dot line. When the central pressure is below the transition pressure $\pcent<\ptrans$, purely-hadronic stellar configurations are determined by the nuclear matter EoS DBHF (black solid curve); at $\pcent=\ptrans$ ($M=\Mtrans$) depending on CSS phase transition parameters (taken from set III in Table~\ref{tab:hyb_EoS}; see also Fig.~\ref{fig:diag-dbhf-sfho-c2-1} right panel), due to the singularity in $1/c_s^2=d\ep/dp$, both $k_2$ and $\La$ are discontinuous; above $\ptrans$, solid (dashed) curves represent stable (unstable) hybrid configurations. All three EoSs give rise to the ``Both'' scenario of mass-radius in Fig.~\ref{fig:MR-De}.
}
\label{fig:lam-k2-M-dbhf-set3}
\end{figure*}

\begin{figure*}[htb]
\parbox{0.25\hsize}{
{\large ``Absent''}\\
\includegraphics[width=\hsize]{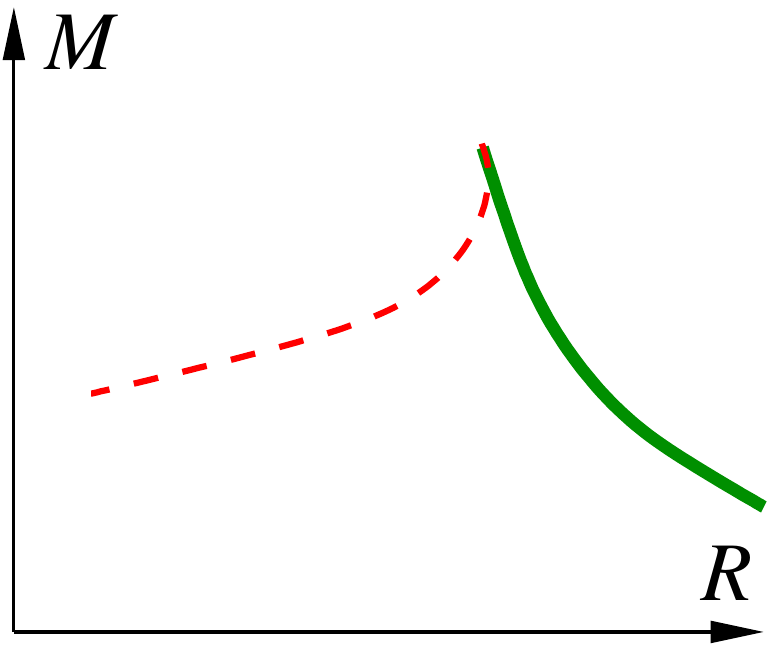}
\bc (a) \ec
}\parbox{0.25\hsize}{
{\large ``Both''}\\
\includegraphics[width=\hsize]{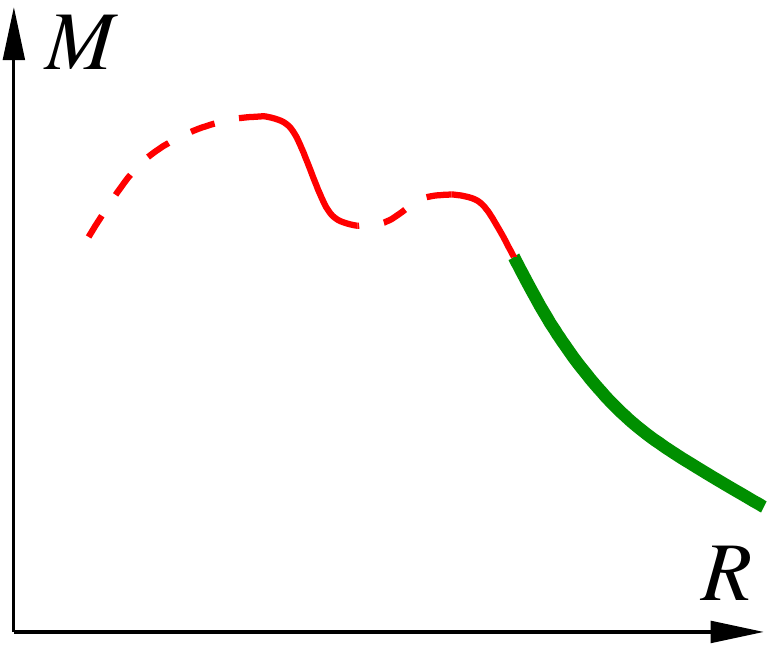}
\bc (b) \ec
}\parbox{0.25\hsize}{
\centerline{\large ``Connected''}
\includegraphics[width=\hsize]{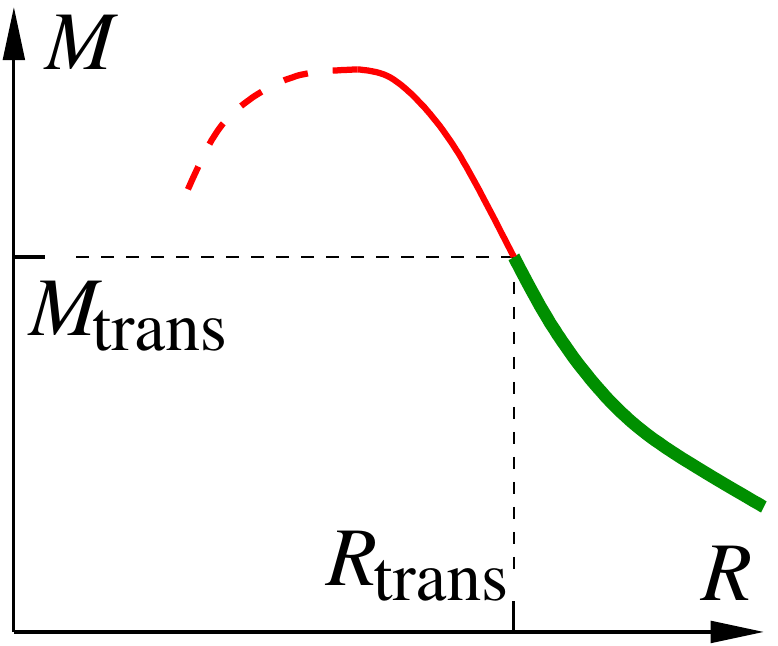}
\bc (c) \ec
}\parbox{0.25\hsize}{
\centerline{\large ``Disconnected''}
\includegraphics[width=\hsize]{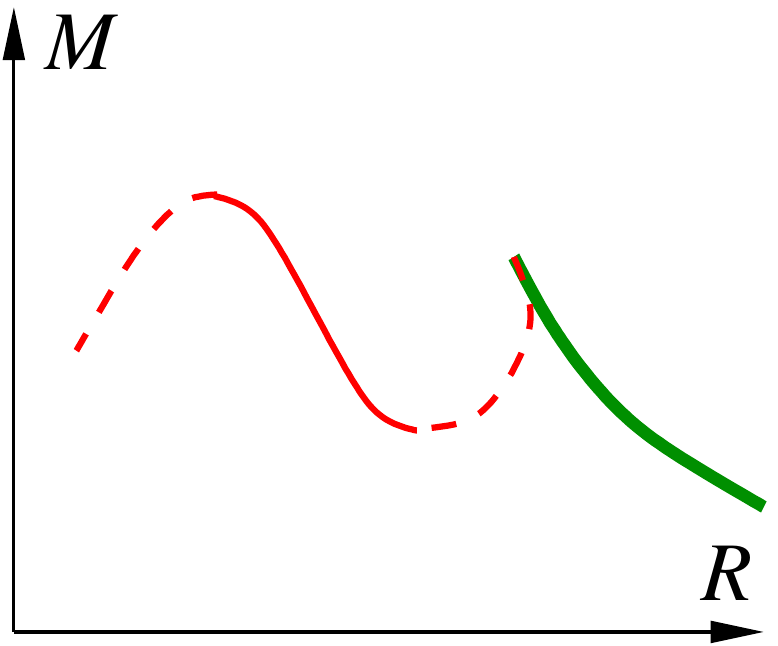}
\bc (d) \ec
}
\caption{
Four possible topologies of
the mass-radius relation for hybrid stars.
The thick (green) line is the hadronic branch. Thin solid (red) lines
are stable hybrid stars; thin dashed (red) lines are unstable hybrid stars.
In (a) the hybrid branch is absent. In (c) there is a connected branch.
In (d) there is a disconnected branch. In (b) there are both types of branch.
In realistic neutron star $M(R)$ curves,
the cusp that occurs in cases (a) and (d) is much smaller and
harder to see \cite{Haensel:1983,Lindblom:1998dp}. Adapted from Ref.~\cite{Alford:2013aca}.
}
\label{fig:MR-De}
\end{figure*}

Under certain circumstances there exist two separated stable hybrid branches, one connected to and the other disconnected from the hadronic branch. This ``both'' scenario illustrated in Fig.~\ref{fig:MR-De} is realized for EoSs in the ``B'' region of Fig.~\ref{fig:diag-dbhf-sfho-c2-1}, and it occurs more readily (i.e. those regions are larger) if the nuclear matter is stiff. We add three more plots showing tidal properties of such configurations in Fig.~\ref{fig:lam-k2-M-dbhf-set3}, with DBHF + CSS parameters in set III EoS of Table~\ref{tab:hyb_EoS}. 

For hybrid EoSs in the ``B'' or ``D'' region, at some mass close to the phase transition point there can be multiple values of the radius $R$, tidal Love number $k_2$ and tidal deformability $\La$, among which the lengths of intervals depend on the phase transition parameters. However, one should keep in mind that the range of masses that different branches overlap with each other are usually tiny, of order $10^{-2}\sim10^{-3}\,\Msolar$ (see discussion in Ref.~\cite{Alford:2013aca}), making the sophisticated change in tidal properties nearly impossible to detect. Heavier stars with a sizable quark core themselves, are in principle easily distinguishable from purely-hadronic ones due to the visible decrease in their tidal deformabilities (see Fig. \ref{fig:k2-Lam-set1} and Fig. \ref{fig:Lam-Mchirp-pt}).

\section{Constant-sound-speed (CSS) template for hybrid star configurations}
\label{app:css}

The CSS parametrization is applicable to high-density equations of
state for which (a) there is a sharp interface between nuclear matter and
a high-density phase which we will call quark matter, even when we do not make any assumptions about its physical nature; and (b) the speed of sound in the high-density
matter is pressure-independent for pressures ranging from the
first-order transition pressure up to the maximum central pressure
of neutron stars. 

Here we briefly recapitulate (see, e.g., Ref.~\cite{Zdunik:2012dj})
the construction of a
thermodynamically consistent equation of state
of the form in Eq.~\eqn{eqn:CSS_EoS}
\beq
\varepsilon\left(p\right) = \varepsilon_{0}+\frac{1}{c^2} p \ .
\label{eqn:ep_para}
\eeq
We start by writing the pressure in terms of the chemical potential
\beq
\ba{rcl}
p(\mu_{B}) &= &\dsp A \,\mu_{B}^{1+\beta} - B \ , \\[2ex]
\mu_{B}(p) &=&\dsp   \left(\frac{p+B}{A}\right)^{1/(1+\beta)}
\ .
\ea
\label{eqn:mup}
\eeq
Note that we have introduced an additional parameter $A$ with mass
dimension $3-\be$.
The value of $A$ can
be varied without affecting the energy-pressure relation \eqn{eqn:ep_para}. When constructing a first-order transition from some low-pressure EoS to
a high-pressure EoS of the form \eqn{eqn:ep_para}, we must
choose $A$ so that the pressure is a monotonically
increasing function of $\mu_B$ (i.e. so that the jump in $n_B$ at the transition
is not negative). The derivative with respect to $\mu_B$ yields
\beq
n_{B}(\mu_{B}) =  (1+\beta)\,A \,\mu_{B}^{\beta}
\label{eqn:nmu}
\eeq
and using $p = \mu_B n_B-\ep$, we obtain the energy density
\beq
\ep(\mu_{B})  =  B + \beta\,A\,\mu_{B}^{1+\beta} \ .
\label{eqn:emu}
\eeq
Then Eq.~\eqn{eqn:mup} gives energy density as a function of pressure
\beq
\varepsilon(p)  =  (1+\beta) B + \beta p 
\label{eqn:ep}
\eeq
which is equivalent to Eq.~\eqn{eqn:ep_para} with $1/c^2=\be$ and
$\ep_{0}=(1+\beta)B$.

\end{document}